\pdfoutput=1
\PassOptionsToPackage{dvipsnames}{xcolor}
\documentclass[sigconf]{acmart}
\usepackage{balance}

\copyrightyear{2025}
\acmYear{2025}
\setcopyright{rightsretained}
\acmConference[CCS '25]{Proceedings of the 2025 ACM SIGSAC Conference on Computer and Communications Security}{October 13--17, 2025}{Taipei, Taiwan}
\acmBooktitle{Proceedings of the 2025 ACM SIGSAC Conference on Computer and Communications Security (CCS '25), October 13--17, 2025, Taipei, Taiwan}
\acmDOI{10.1145/3719027.3765032}
\acmISBN{979-8-4007-1525-9/2025/10}

\ccsdesc[500]{Security and privacy~Distributed systems security}
\ccsdesc[500]{Theory of computation~Distributed algorithms}

\keywords{BFT Protocols; Blockchain; Accountability; Forensics}

\usepackage{subcaption}

\usepackage{import}
\usepackage[utf8]{inputenc}

\PassOptionsToPackage{hyphens}{url}%
\usepackage{hyperref}
\usepackage{graphicx}

\usepackage{amsmath}
\usepackage{amsthm}
\usepackage{amsfonts}
\usepackage{mathtools}

\usepackage{shortsym}

\usepackage{IEEEtrantools}
\allowdisplaybreaks

\usepackage[shortlabels,inline]{enumitem}
\setlist[itemize]{leftmargin=1.25em}
\setlist[enumerate]{leftmargin=1.75em}
\usepackage{xstring}   %
\usepackage[normalem]{ulem}

\usepackage{import}
\theoremstyle{plain}
\newtheorem{lemma}{Lemma}
\newtheorem{conjecture}{Conjecture}
\newtheorem*{theorem*}{Theorem}
\newtheorem*{conjecture*}{Conjecture}

\theoremstyle{definition}
\makeatletter
\renewcommand*\env@matrix[1][*\c@MaxMatrixCols c]{%
    \hskip -\arraycolsep
    \let\@ifnextchar\new@ifnextchar
    \array{#1}}
\makeatother

\DeclarePairedDelimiter\abs{\lvert}{\rvert}
\DeclarePairedDelimiter\len{\lvert}{\rvert}
\DeclarePairedDelimiter\norm{\lVert}{\rVert}

\makeatletter
\let\oldabs\abs
\def\abs{\@ifstar{\oldabs}{\oldabs*}}
\let\oldlen\len
\def\len{\@ifstar{\oldlen}{\oldlen*}}
\let\oldnorm\norm
\def\norm{\@ifstar{\oldnorm}{\oldnorm*}}
\makeatother

\DeclareMathOperator*{\argmax}{arg\,max}

\usepackage{pifont}

\usepackage{dsfont}

\usepackage{xspace}

\newcommand{\cf}[0]{cf.\xspace}
\newcommand{\ie}[0]{\emph{i.e.}\xspace}
\newcommand{\eg}[0]{\emph{e.g.}\xspace}

\usepackage[dvipsnames]{xcolor}

\definecolor{myA16zGrayLight}{RGB}{235,235,235}     %
\definecolor{myA16zGrayMedium}{RGB}{196,196,196}    %
\definecolor{myA16zGrayDark}{RGB}{44,34,34}         %
\definecolor{myA16zLavender}{RGB}{208,161,255}      %
\definecolor{myA16zMagenta}{RGB}{195,70,206}        %
\definecolor{myA16zMulberry}{RGB}{113,24,88}        %
\definecolor{myA16zLemonChiffon}{RGB}{250,234,157}  %
\definecolor{myA16zAmber}{RGB}{230,154,48}          %
\definecolor{myA16zRust}{RGB}{174,59,10}            %
\definecolor{myA16zLime}{RGB}{197,222,107}          %
\definecolor{myA16zAquamarine}{RGB}{82,216,145}     %
\definecolor{myA16zPine}{RGB}{60,87,44}             %
\definecolor{myA16zPacific}{RGB}{145,224,235}       %
\definecolor{myA16zTeal}{RGB}{36,197,201}           %
\definecolor{myA16zAzure}{RGB}{18,51,90}            %

\definecolor{jnSUCardinalRed}{HTML}{8c1515}
\definecolor{jnSUCardinalRedLight}{HTML}{B83A4B}
\definecolor{jnSUCardinalRedDark}{HTML}{820000}
\definecolor{jnSUWhite}{HTML}{ffffff}
\definecolor{jnSUCoolGrey}{HTML}{53565A}
\definecolor{jnSUBlack}{HTML}{2e2d29}
\definecolor{jnSUBlack100}{HTML}{2e2d29}
\definecolor{jnSUBlack90}{HTML}{43423E}
\definecolor{jnSUBlack80}{HTML}{585754}
\definecolor{jnSUBlack70}{HTML}{6D6C69}
\definecolor{jnSUBlack60}{HTML}{767674}
\definecolor{jnSUBlack50}{HTML}{979694}
\definecolor{jnSUBlack40}{HTML}{ABABA9}
\definecolor{jnSUBlack30}{HTML}{C0C0BF}
\definecolor{jnSUBlack20}{HTML}{D5D5D4}
\definecolor{jnSUBlack10}{HTML}{EAEAEA}

\definecolor{jnSUPaloAlto}{HTML}{175E54}
\definecolor{jnSUPaloAltoLight}{HTML}{2D716F}
\definecolor{jnSUPaloAltoDark}{HTML}{014240}
\definecolor{jnSUPaloVerde}{HTML}{279989}
\definecolor{jnSUPaloVerdeLight}{HTML}{59B3A9}
\definecolor{jnSUPaloVerdeDark}{HTML}{017E7C}
\definecolor{jnSUOlive}{HTML}{8F993E}
\definecolor{jnSUOliveLight}{HTML}{A6B168}
\definecolor{jnSUOliveDark}{HTML}{7A863B}
\definecolor{jnSUBay}{HTML}{6FA287}
\definecolor{jnSUBayLight}{HTML}{8AB8A7}
\definecolor{jnSUBayDark}{HTML}{417865}
\definecolor{jnSUSky}{HTML}{4298B5}
\definecolor{jnSUSkyLight}{HTML}{67AFD2}
\definecolor{jnSUSkyDark}{HTML}{016895}
\definecolor{jnSULagunita}{HTML}{007C92}
\definecolor{jnSULagunitaLight}{HTML}{009AB4}
\definecolor{jnSULagunitaDark}{HTML}{006B81}
\definecolor{jnSUPoppy}{HTML}{E98300}
\definecolor{jnSUPoppyLight}{HTML}{F9A44A}
\definecolor{jnSUPoppyDark}{HTML}{D1660F}
\definecolor{jnSUSpirited}{HTML}{E04F39}
\definecolor{jnSUSpiritedLight}{HTML}{F4795B}
\definecolor{jnSUSpiritedDark}{HTML}{C74632}
\definecolor{jnSUIlluminating}{HTML}{FEDD5C}
\definecolor{jnSUIlluminatingLight}{HTML}{FFE781}
\definecolor{jnSUIlluminatingDark}{HTML}{FEC51D}
\definecolor{jnSUPlum}{HTML}{620059}
\definecolor{jnSUPlumLight}{HTML}{734675}
\definecolor{jnSUPlumDark}{HTML}{350D36}
\definecolor{jnSUBrick}{HTML}{651C32}
\definecolor{jnSUBrickLight}{HTML}{7F2D48}
\definecolor{jnSUBrickDark}{HTML}{42081B}
\definecolor{jnSUArchway}{HTML}{5D4B3C}
\definecolor{jnSUArchwayLight}{HTML}{766253}
\definecolor{jnSUArchwayDark}{HTML}{2F2424}
\definecolor{jnSUStone}{HTML}{7F7776}
\definecolor{jnSUStoneLight}{HTML}{D4D1D1}
\definecolor{jnSUStoneDark}{HTML}{544948}
\definecolor{jnSUFog}{HTML}{DAD7CB}
\definecolor{jnSUFogLight}{HTML}{F4F4F4}
\definecolor{jnSUFogDark}{HTML}{B6B1A9}

\definecolor{jnSUDigitalRed}{HTML}{B1040E}
\definecolor{jnSUDigitalRedLight}{HTML}{E50808}
\definecolor{jnSUDigitalRedDark}{HTML}{820000}
\definecolor{jnSUDigitalBlue}{HTML}{006CB8}
\definecolor{jnSUDigitalBlueLight}{HTML}{6FC3FF}
\definecolor{jnSUDigitalBlueDark}{HTML}{00548f}
\definecolor{jnSUDigitalGreen}{HTML}{008566}
\definecolor{jnSUDigitalGreenLight}{HTML}{1AECBA}
\definecolor{jnSUDigitalGreenDark}{HTML}{006F54}

\definecolor{myParula01Blue}{RGB}{0,114,189}
\definecolor{myParula02Orange}{RGB}{217,83,25}
\definecolor{myParula03Yellow}{RGB}{237,177,32}
\definecolor{myParula04Purple}{RGB}{126,47,142}
\definecolor{myParula05Green}{RGB}{119,172,48}
\definecolor{myParula06LightBlue}{RGB}{77,190,238}
\definecolor{myParula07Red}{RGB}{162,20,47}

\usepackage{tikz}
\usetikzlibrary{calc}
\usetikzlibrary{arrows.meta}
\usetikzlibrary{patterns}
\usetikzlibrary{positioning}
\usetikzlibrary{decorations.pathreplacing}
\usetikzlibrary{decorations.pathmorphing}
\usetikzlibrary{calligraphy}
\usetikzlibrary{shapes.misc}
\usetikzlibrary{shapes.symbols}
\usetikzlibrary{shapes.arrows}
\usetikzlibrary{spy}
\usetikzlibrary{shapes.geometric}
\usetikzlibrary{intersections}
\usetikzlibrary{fit}

\usepackage{pgfplots}
\pgfplotsset{compat=1.17}
\usepgfplotslibrary{fillbetween}

\pgfplotsset{
    discard if not/.style 2 args={
            x filter/.code={
                    \edef\tempa{\thisrow{#1}}
                    \edef\tempb{#2}
                    \ifx\tempa\tempb
                    \else
                        
                    \fi
                }
        },
}

\pgfplotsset{
    mysimpleplot/.style = {
            every axis plot/.prefix style={thick},
            width=1.05\linewidth,
            height=0.75\linewidth,
            title style={font=\scriptsize,align=center},
            legend cell align=left,
            legend style={font=\scriptsize},
            legend columns=3,
            legend style={
                    at={(0.5,1)},
                    yshift=0.3em,
                    anchor=south,
                    draw=none,
                    /tikz/every even column/.append style={
                            column sep=0.3em
                        },
                    cells={
                            align=left
                        }
                },
            grid=both,
            minor tick num=9,
            major grid style={solid,very thin,draw=gray!50},
            minor grid style={solid,ultra thin,draw=gray!20},
            label style={font=\scriptsize,align=center},
            tick label style={font=\scriptsize},
        },
}

\pgfplotsset{
    mysimplefig1plot/.style = {
        mysimpleplot,
        xlabel={$\netX$},
        ylabel={$\tALident/n$},
        xmin=0.0, xmax=0.5,
        ymin=0.0, ymax=0.35,
        height=0.75\linewidth,
        width=\linewidth,
        yticklabel style={
                /pgf/number format/fixed,
                /pgf/number format/precision=2
            },
        scaled y ticks=false,
        xtick={0,0.1,0.2,0.25,0.33333,0.5},
        xticklabels={0,1/10,1/5,1/4,1/3,1/2},
        yticklabels={0,1/10,1/6,1/5,1/4,1/3},
        ytick={0,0.1,0.16666,0.2,0.25,0.33333},
    }
}

\tikzset{myparula11/.style={color=myParula01Blue,solid,mark=+,mark options={solid}}}
\tikzset{myparula12/.style={color=myParula01Blue,densely dashed,mark=x,mark options={solid}}}
\tikzset{myparula13/.style={color=myParula01Blue,densely dotted,mark=o,mark options={solid}}}
\tikzset{myparula14/.style={color=myParula01Blue,dashdotted,mark=triangle,mark options={solid}}}
\tikzset{myparula15/.style={color=myParula01Blue,dashdotdotted,mark=square,mark options={solid}}}

\tikzset{myparula21/.style={color=myParula02Orange,solid,mark=+,mark options={solid}}}
\tikzset{myparula22/.style={color=myParula02Orange,densely dashed,mark=x,mark options={solid}}}
\tikzset{myparula23/.style={color=myParula02Orange,densely dotted,mark=o,mark options={solid}}}
\tikzset{myparula24/.style={color=myParula02Orange,dashdotted,mark=triangle,mark options={solid}}}
\tikzset{myparula25/.style={color=myParula02Orange,dashdotdotted,mark=square,mark options={solid}}}

\tikzset{myparula31/.style={color=myParula03Yellow,solid,mark=+,mark options={solid}}}
\tikzset{myparula32/.style={color=myParula03Yellow,densely dashed,mark=x,mark options={solid}}}
\tikzset{myparula33/.style={color=myParula03Yellow,densely dotted,mark=o,mark options={solid}}}
\tikzset{myparula34/.style={color=myParula03Yellow,dashdotted,mark=triangle,mark options={solid}}}
\tikzset{myparula35/.style={color=myParula03Yellow,dashdotdotted,mark=square,mark options={solid}}}

\tikzset{myparula41/.style={color=myParula04Purple,solid,mark=+,mark options={solid}}}
\tikzset{myparula42/.style={color=myParula04Purple,densely dashed,mark=x,mark options={solid}}}
\tikzset{myparula43/.style={color=myParula04Purple,densely dotted,mark=o,mark options={solid}}}
\tikzset{myparula44/.style={color=myParula04Purple,dashdotted,mark=triangle,mark options={solid}}}
\tikzset{myparula45/.style={color=myParula04Purple,dashdotdotted,mark=square,mark options={solid}}}

\tikzset{myparula51/.style={color=myParula05Green,solid,mark=+,mark options={solid}}}
\tikzset{myparula52/.style={color=myParula05Green,densely dashed,mark=x,mark options={solid}}}
\tikzset{myparula53/.style={color=myParula05Green,densely dotted,mark=o,mark options={solid}}}
\tikzset{myparula54/.style={color=myParula05Green,dashdotted,mark=triangle,mark options={solid}}}
\tikzset{myparula55/.style={color=myParula05Green,dashdotdotted,mark=square,mark options={solid}}}

\tikzset{myparula61/.style={color=myParula06LightBlue,solid,mark=+,mark options={solid}}}
\tikzset{myparula62/.style={color=myParula06LightBlue,densely dashed,mark=x,mark options={solid}}}
\tikzset{myparula63/.style={color=myParula06LightBlue,densely dotted,mark=o,mark options={solid}}}
\tikzset{myparula64/.style={color=myParula06LightBlue,dashdotted,mark=triangle,mark options={solid}}}
\tikzset{myparula65/.style={color=myParula06LightBlue,dashdotdotted,mark=square,mark options={solid}}}

\tikzset{myparula71/.style={color=myParula07Red,solid,mark=+,mark options={solid}}}
\tikzset{myparula72/.style={color=myParula07Red,densely dashed,mark=x,mark options={solid}}}
\tikzset{myparula73/.style={color=myParula07Red,densely dotted,mark=o,mark options={solid}}}
\tikzset{myparula74/.style={color=myParula07Red,dashdotted,mark=triangle,mark options={solid}}}
\tikzset{myparula75/.style={color=myParula07Red,dashdotdotted,mark=square,mark options={solid}}}

\usepackage{multirow,booktabs}
\usepackage{makecell}

\usepackage{algorithm}
\usepackage{algorithmicx}
\usepackage[noend]{algpseudocode}
\makeatletter
\AddToHook{env/algorithmic/begin}{\def\@currentcounter{ALG@line}}
\makeatother

\algnewcommand{\LineComment}[1]{\State {\textcolor{gray}{// #1}}}

\newcommand{\alglocref}[2]{\cref{#1}, \cref{#2}}

\newcommand{\algloclabel}[1]{\phantomsection\label{#1}}

\algrenewcommand{\alglinenumber}[1]{\scriptsize\textcolor{gray}{\texttt{#1}}}
\algrenewcommand{\algorithmicindent}{1em}

\algnewcommand{\algfontsize}[0]{\footnotesize}

\algnewcommand{\algorithmicswitch}{\textbf{switch}}
\algdef{SE}[SWITCH]{Switch}{EndSwitch}[1]{\algorithmicswitch\ #1\ \algorithmicdo}{\algorithmicend\ \algorithmicswitch}%
\algtext*{EndSwitch}%

\algnewcommand{\algorithmiccase}{\textbf{case}}
\algdef{SE}[CASE]{Case}{EndCase}[1]{\algorithmiccase\ #1}{\algorithmicend\ \algorithmiccase}%
\algtext*{EndCase}%

\algnewcommand{\algorithmicon}{\textbf{on}}
\algdef{SE}[ON]{On}{EndOn}[1]{\algorithmicon\ #1\ \algorithmicdo}{\algorithmicend\ \algorithmicon}%
\algtext*{EndOn}%

\algnewcommand{\algorithmicat}{\textbf{at}}
\algdef{SE}[AT]{At}{EndAt}[1]{\algorithmicat\ #1\ \algorithmicdo}{\algorithmicend\ \algorithmicat}%
\algtext*{EndAt}%

\algnewcommand{\algorithmicrealfunction}{\textbf{function}}
\algdef{SE}[REALFUNCTION]{RealFunction}{EndRealFunction}[1]{\algorithmicrealfunction\ #1\ \algorithmicdo}{\algorithmicend\ \algorithmicrealfunction}%
\algtext*{EndRealFunction}%

\algnewcommand{\algorithmicthroughout}{\textbf{do throughout}}
\algdef{SE}[Throughout]{Throughout}{EndThroughout}[1]{\algorithmicthroughout\ #1\ \algorithmicdo}{\algorithmicend\ \algorithmicthroughout}%
\algtext*{EndThroughout}%

\algrenewcommand{\algorithmicdo}{}
\algrenewcommand{\algorithmicthen}{}

\algnewcommand{\algorithmicgoto}{\textbf{goto}}%
\algnewcommand{\Goto}[1]{\algorithmicgoto~\ref{#1}}%

\algnewcommand{\algorithmicassert}{\textbf{assert}}%
\algnewcommand{\Assert}[1]{\algorithmicassert~{#1}}%

\algnewcommand{\algorithmicbreak}{\textbf{break}}%
\algnewcommand{\Break}[0]{\algorithmicbreak}%

\algnewcommand{\algorithmicwaiton}{\textbf{wait on}}%
\algnewcommand{\WaitOn}[1]{\algorithmicwaiton~{#1}}%

\algnewcommand{\InlineRequire}[1]{\textbf{require} {#1}}

\algblock{ManualIndent}{EndManualIndent}
\algnotext{ManualIndent}
\algnotext{EndManualIndent}

\algdef{SE}[GENERICBLOCK]{GenericBlock}{EndGenericBlock}[1]{#1}{}%
\algtext*{EndGenericBlock}%

\subimport{./lib/}{defer.tex}

\usepackage[sort&compress,capitalize,nameinlink]{cleveref}

\crefalias{ALG@line}{line}

\crefname{figure}{Fig.}{Figs.}
\Crefname{figure}{Fig.}{Figs.}

\crefname{table}{Tab.}{Tabs.}
\Crefname{table}{Tab.}{Tabs.}

\crefname{section}{Sec.}{Secs.}
\Crefname{section}{Sec.}{Secs.}
\crefname{subsection}{Sec.}{Secs.}
\Crefname{subsection}{Sec.}{Secs.}
\crefname{subsubsection}{Sec.}{Secs.}
\Crefname{subsubsection}{Sec.}{Secs.}
\crefname{subsubsubsection}{Sec.}{Secs.}
\Crefname{subsubsubsection}{Sec.}{Secs.}
\crefname{appendix}{App.}{Apps.}
\Crefname{appendix}{App.}{Apps.}
\crefname{subappendix}{App.}{Apps.}
\Crefname{subappendix}{App.}{Apps.}
\crefname{subsubappendix}{App.}{Apps.}
\Crefname{subsubappendix}{App.}{Apps.}
\crefname{subsubsubappendix}{App.}{Apps.}
\Crefname{subsubsubappendix}{App.}{Apps.}

\crefname{algorithm}{Alg.}{Algs.}
\Crefname{algorithm}{Alg.}{Algs.}
\crefname{line}{ln.}{lns.}
\Crefname{line}{ln.}{lns.}

\crefname{proposition}{Prop.}{Props.}
\Crefname{proposition}{Prop.}{Props.}
\crefname{lemma}{Lem.}{Lems.}
\Crefname{lemma}{Lem.}{Lems.}
\crefname{theorem}{Thm.}{Thms.}
\Crefname{theorem}{Thm.}{Thms.}
\crefname{corollary}{Cor.}{Cors.}
\Crefname{corollary}{Cor.}{Cors.}
\crefname{definition}{Def.}{Defs.}
\Crefname{definition}{Def.}{Defs.}
\crefname{conjecture}{Conj.}{Conjs.}
\Crefname{conjecture}{Conj.}{Conjs.}
\crefname{remark}{Rem.}{Rems.}
\Crefname{remark}{Rem.}{Rems.}

\newcommand{\FFp}{\ensuremath{\mathsf{p}}}
\newcommand{\FFv}{\ensuremath{\mathsf{v}}}
\newcommand{\FFbmI}{\ensuremath{\mathsf{b}_{-1}}}
\newcommand{\FFb}{\ensuremath{\mathsf{b}}}
\newcommand{\FFtxs}{\ensuremath{\mathsf{txs}}}
\newcommand{\FFQC}{\ensuremath{\mathsf{Q}}}
\newcommand{\FFstage}{\ensuremath{\mathsf{s}}}

\newcommand{\Vote}{\ensuremath{\operatorname{Vote}}}
\newcommand{\VoteLive}{\ensuremath{\operatorname{VoteLive}}}
\newcommand{\Block}{\ensuremath{\operatorname{Block}}}
\newcommand{\MvalidB}{\ensuremath{\operatorname{valid}}}
\newcommand{\MvalidV}{\ensuremath{\operatorname{valid}}}
\newcommand{\MvalidQC}{\ensuremath{\operatorname{valid}}}
\newcommand{\StM}[2]{\ensuremath{\KM_{#1}^{#2}}}
\newcommand{\StMtxs}[2]{\ensuremath{{\KM\Kt}_{#1}^{#2}}}
\newcommand{\StQ}[2]{\ensuremath{\KQ_{#1}^{#2}}}
\newcommand{\StX}[2]{\ensuremath{\KX_{#1}^{#2}}}

\newcommand{\Accusation}{\ensuremath{\operatorname{Accusation}}}
\newcommand{\View}{\ensuremath{\operatorname{Transcript}}}
\newcommand{\FFt}{\ensuremath{\mathsf{t}}}
\newcommand{\FFM}{\ensuremath{\mathsf{M}}}

\newcommand{\GST}{\ensuremath{\mathsf{GST}}}
\newcommand{\netG}{\ensuremath{g}}
\newcommand{\netX}{\ensuremath{x}}
\newcommand{\netDeltaPeriod}{\ensuremath{\Delta'}}

\newcommand{\Tconf}{\ensuremath{T_{\mathrm{conf}}}}

\newcommand{\tS}{\ensuremath{\tau^{\mathrm{S}}}}
\newcommand{\tL}{\ensuremath{\tau^{\mathrm{L}}}}
\newcommand{\tALmax}{\ensuremath{\tau^{\mathrm{AL}}_{\mathrm{max}}}}
\newcommand{\tALident}{\ensuremath{\tau^{\mathrm{AL}}_{\mathrm{ident}}}}
\newcommand{\funLA}{\ensuremath{\psi}}
\newcommand{\epsAL}{\ensuremath{\varepsilon^{\mathrm{AL}}}}

\newcommand{\superviewLen}{\ensuremath{K_{\mathrm{views}}}}
\newcommand{\superviewEpsGap}{\ensuremath{\delta_{\mathrm{x}}}}
\newcommand{\superviewErrProb}{\ensuremath{\exp(-\superviewEpsGap g(\netDeltaPeriod)/6)}}

\newcommand{\critSubsetsCPA}{\ensuremath{\CP^{\mathrm{A}}}}
\newcommand{\critSubsetsG}{\ensuremath{G}}

\newcommand{\tx}{\ensuremath{\mathsf{tx}}}

\newcommand{\consistent}{\ensuremath{\asymp}}

\newcommand{\LOG}{\ensuremath{\Lambda}}
\newcommand{\Leader}{\ensuremath{L}}
\newcommand{\Blame}[3]{\ensuremath{\mathrm{Blame}_{#1,#2,#3}}}
\newcommand{\BlameNo}[3]{\ensuremath{\Blame{#1}{#2}{#3} \gets 0}}
\newcommand{\BlameYes}[3]{\ensuremath{\Blame{#1}{#2}{#3} \gets 1}}

\newcommand{\CPa}{\ensuremath{\CP_{\mathrm{a}}}}
\newcommand{\CPh}{\ensuremath{\CP_{\mathrm{h}}}}

\newcommand{\CUs}{\ensuremath{\CU_{\mathrm{s}}}}
\newcommand{\CUhl}{\ensuremath{\CU_{\mathrm{hl}}}}
\newcommand{\CUshl}{\ensuremath{\CU_{\mathrm{shl}}}}

\newtheorem{theorem}{Theorem}
\newtheorem{definition}{Definition}
\newtheorem{corollary}{Corollary}
\newtheorem{remark}{Remark}
\newtheorem{proposition}{Proposition}

\newcommand{\Prob}[1]{\ensuremath{\operatorname{Pr}\!\left[#1\right]}}
\newcommand{\Mean}[1]{\ensuremath{\operatorname{E}\!\left[#1\right]}}

\newcommand{\scriptspacing}[0]{\medmuskip=0mu\thickmuskip=0mu}

\newcommand{\myDrawScenario}[2]{
    \pgfmathsetmacro{\myrunningsum}{0};
    \foreach \name/\height/\label/\style in {#1} {
        \coordinate (\name top) at (0,{1-\myrunningsum});
        \coordinate (\name bot) at (0,{1-\myrunningsum-\height});
        \coordinate (\name toplet) at (0,{1-\myrunningsum});
        \coordinate (\name botlet) at (0,{1-\myrunningsum-\height});
        \coordinate (\name toprit) at (1,{1-\myrunningsum});
        \coordinate (\name botrit) at (1,{1-\myrunningsum-\height});
        \xdef\myrunningsum{(\myrunningsum + \height)};
        \draw [\style,shorten <=0.75pt,shorten >=0.75pt] ([xshift=-0.5em]\name top) -- ([xshift=-0.5em]\name bot);
        \draw (\name top) -- (\name bot) node [midway,left,xshift=-0.5em] (\name label) {\label};
    }
    \coordinate (top) at (0,1);
    \coordinate (bot) at (0,0);

    \pgfmathsetmacro{\myrunningsum}{0};
    \foreach \name/\width/\label/\style in {#2} {
        \coordinate (\name let) at ({\myrunningsum},0);
        \coordinate (\name rit) at ({\myrunningsum+\width},0);
        \coordinate (\name botlet) at ({\myrunningsum},0);
        \coordinate (\name botrit) at ({\myrunningsum+\width},0);
        \coordinate (\name toplet) at ({\myrunningsum},1);
        \coordinate (\name toprit) at ({\myrunningsum+\width},1);
        \xdef\myrunningsum{(\myrunningsum + \width)};
        \draw [\style,shorten <=0.75pt,shorten >=0.75pt] ([yshift=-0.5em]\name let) -- ([yshift=-0.5em]\name rit);
        \draw (\name let) -- (\name rit) node [midway,yshift=-1.5em] (\name label) {\label};
    }
    \coordinate (let) at (0,0);
    \coordinate (rit) at (1,0);

    \coordinate (toplet) at (0,1);
    \coordinate (botlet) at (0,0);
    \coordinate (toprit) at (1,1);
    \coordinate (botrit) at (1,0);

    \foreach \party/\height/\partylabel/\partystyle in {#1} {
        \foreach \view/\width/\viewlabel/\viewstyle in {#2} {
            \draw (\party top -| \view let) rectangle (\party bot -| \view rit);
            \coordinate (\party \view mid) at ($(\party top -| \view let)!0.5!(\party bot -| \view rit)$);
            \coordinate (\party \view toplet) at ($(\party top -| \view let)$);
            \coordinate (\party \view botlet) at ($(\party bot -| \view let)$);
            \coordinate (\party \view toprit) at ($(\party top -| \view rit)$);
            \coordinate (\party \view botrit) at ($(\party bot -| \view rit)$);
            \foreach \i in {1,2,3} {
                \coordinate (\party \view toplet\i) at ([xshift=\i*4pt,yshift=-4pt]$(\party top -| \view let)$);
                \coordinate (\party \view botlet\i) at ([xshift=\i*4pt,yshift=4pt]$(\party bot -| \view let)$);
                \coordinate (\party \view toprit\i) at ([xshift=\i*-4pt,yshift=-4pt]$(\party top -| \view rit)$);
                \coordinate (\party \view botrit\i) at ([xshift=\i*-4pt,yshift=4pt]$(\party bot -| \view rit)$);
            }
        }
    }
}%

\newcommand{\myIconAdvCrash}{\faAngry[regular]\faDizzy[regular]}
\newcommand{\myIconAdvPartition}{\faAngry[regular]\faCommentSlash[regular]}
\newcommand{\myIconHonPartition}{\faCommentSlash[regular]}

\newcommand{\myTextHighlighted}[1]{\textcolor{myA16zAmber}{#1}}
\newcommand{\myTextHighlightedDescription}[0]{in \myTextHighlighted{orange}}
\newcommand{\myTextHighlightedTwo}[1]{\textcolor{jnSUDigitalGreen}{#1}}
\newcommand{\myTextHighlightedTwoDescription}[0]{in \myTextHighlightedTwo{green}}

\usepackage{fontawesome5}
\usepackage{nicefrac}

\newif\ifcompilewithoutheavyfigures

\newif\ifcompilefullversion
\ifnum\pdfstrcmp{\jobname}{main-short}=0
    \compilefullversionfalse
\else
    \compilefullversiontrue
    \keywords{}
    \makeatletter
    \@ACM@nonacmtrue
    \makeatother
\fi

\newdefergroup{proofsblaming}[notdeferred]
\newdefergroup{proofsadjudication}[notdeferred]

\usepackage{xr}
\ifcompilefullversion\else
\input{appendix.aux}
\fi

\title{Accountable Liveness}%

\ifcompilefullversion\else
\thanks{See full version for appendix:
\url{https://eprint.iacr.org/2025/693}~\cite{fullversionofthispaper}}
\fi

\author{Andrew Lewis-Pye}%
\affiliation{%
    \institution{London School of Economics}%
    \city{London}%
    \country{UK}%
}%
\email{a.lewis7@lse.ac.uk}%
\author{Joachim Neu}%
\affiliation{%
    \institution{a16z Crypto Research}%
    \city{New York}%
    \state{NY}%
    \country{USA}%
}%
\email{jneu@a16z.com}%
\author{Tim Roughgarden}%
\affiliation{%
    \institution{Columbia University \& a16z Crypto Research}%
    \city{New York}%
    \state{NY}%
    \country{USA}%
}%
\email{tim.roughgarden@gmail.com}%
\author{Luca Zanolini}%
\affiliation{%
    \institution{Ethereum Foundation}%
    \city{London}%
    \country{UK}%
}%
\email{luca.zanolini@ethereum.org}%

\begin{document}%
\begin{abstract}%
    Safety and liveness are the two classical security properties of consensus protocols.
    Recent works have strengthened safety with \emph{accountability}: should any safety violation occur, a sizable fraction of adversary nodes can be proven to be protocol violators.
    This paper studies to what extent analogous accountability guarantees are achievable for \emph{liveness}.
    To reveal the full complexity of this question, we introduce an interpolation between the classical synchronous and partially-synchronous models that we call the \emph{$\netX$-partially-synchronous network model} in which, intuitively, at most an $\netX$ fraction of the time steps in any sufficiently long interval are asynchronous (and, as with a partially-synchronous network, all time steps are synchronous following the passage of an unknown ``global stablization time'').
    We prove a precise characterization of the parameter regime in which accountable liveness is achievable: if and only if $\netX < 1/2$ and $f < n/2$, where $n$ denotes the number of nodes and $f$ the number of nodes controlled by an adversary.
    We further refine the problem statement and our analysis by parameterizing by the number of violating nodes identified following a liveness violation, and provide evidence that the guarantees achieved by our protocol are near-optimal (as a function of $\netX$ and $f$).
    Our results provide rigorous foundations for liveness-accountability heuristics such as the  ``inactivity leaks'' employed in Ethereum.
\end{abstract}%
\maketitle%

\section{Introduction}
\label{sec:intro}

The atomic broadcast variant of Byzantine-fault tolerant (BFT) consensus is a fundamental problem in distributed computing, where a set of $n$ \emph{nodes} must agree on a total ordering of input \emph{transactions} into an output \emph{log}.
This problem has received renewed attention recently in the context of cryptocurrencies and blockchains because it is a fundamental primitive for 
such systems.
Specifically, each node in the protocol successively produces an output log of transactions, and the 
protocol must guarantee the two key security properties: \emph{safety}---ensuring that logs remain consistent across nodes and across time---and \emph{liveness}---ensuring that every input transaction is eventually included in the logs of nodes.
Despite some $f$ \emph{adversary} nodes acting arbitrarily (subject to being computationally bounded) in an effort to undermine consensus, these properties should hold for all non-adversary (a.k.a.\ \emph{honest}) nodes.
The typical guarantees for classical protocols state that if the fraction of adversary nodes is below a given threshold, the protocol remains \emph{safe} and \emph{live}.

However, if the fraction of adversary nodes exceeds this threshold, the protocol's security properties no longer hold. 
What then?
If there is a security violation, can we at least determine which nodes caused it? 
In data-center applications of consensus, this is useful for identifying and remedying faulty machines; 
and it becomes even more important in proof-of-stake blockchain applications of consensus, where self-interested nodes may deviate from the protocol for financial gain and may be willing to accept consensus security violations as collateral damage. Here, identifying such adversary nodes allows the system to confiscate their stake as a form of punishment, thereby incentivizing honest behavior, and to compensate damages from the violations.
Such aims are often referred to as \emph{crypto-economic security}.

Towards this goal, the literature has recently introduced the notion of \emph{accountable safety}~\cite{casper,bftforensics}, which is a strengthening~\cite{snapandchat,DBLP:conf/fc/NeuTT22} of safety that stipulates: if any two nodes at any two points in time ever have inconsistent output logs, 
\ie, a safety violation occurs,
then a substantial fraction
of nodes can be identified as having provably violated the protocol.
Specifically, if there is ever a safety violation between two honest nodes, then from their respective transcripts of the protocol's execution, a \emph{certificate of guilt} can be extracted for a set of nodes such that (a)~no honest node is ever falsely accused of having contributed to the safety violation, and (b)~the set of identified guilty nodes is guaranteed to be fairly large.

\subsection{Scope \& Contributions}
\label{sec:intro-scope}

This paper initiates the systematic study of
\emph{accountability for liveness}
(see \cref{sec:relatedwork,sec:addendum-relatedwork} for a discussion of related work).
More precisely, we
investigate under what circumstances, and through which protocol techniques, 
the following guarantee can be achieved:
should a liveness violation occur, 
then a certificate of guilt
is eventually produced
such that
(a)~%
no honest node is ever falsely accused,
and 
(b)~%
the set of identified (guilty) nodes is guaranteed to be fairly large.  
Note that, technically, traditional (\emph{eventual}) liveness
is such a weak guarantee that
violations
cannot be assessed by any one point in the execution~\cite{DBLP:journals/ipl/AlpernS85}
(as pending transactions may still get confirmed in the future).
We thus consider a slightly stronger (\emph{timely-})liveness notion
where transactions have to be confirmed within a deadline. 

Intuitively, because liveness violations would seem to typically involve the unexpected \emph{absence} of messages (\eg, votes) rather than the unexpected \emph{presence} of messages (\eg, double-voting to cause a safety violation), one might speculate that certificates of guilt (``proofs of misbehavior'') should be more difficult to guarantee for liveness violations than for safety violations.
We prove that accountable liveness is indeed harder to achieve than accountable safety, in two senses.

First, accountable safety is achievable in partial synchrony; indeed, any partially-synchronous protocol can be augmented to guarantee accountable safety without imposing any additional timing assumptions~\cite{cryptoeprint:2021/1169,accountabilityimpliesfinality}. 
(This statement assumes a computationally bounded adversary and secure digital signatures, assumptions we also make here.)
In contrast, we prove that accountable liveness \emph{cannot} be attained in a partially-synchronous network without additional timing assumptions 
(at least as long as the protocol is required to preserve safety under partial synchrony if all nodes are honest).
On the other hand, accountable liveness is vacuous in a (fully-)synchronous network,
where (at least in some variants~\cite{DBLP:journals/siamcomp/DolevS83,buterinblogpost,cryptoeprint:2024/1799,DBLP:conf/sp/HouYS22,DBLP:conf/sp/HouY23}) liveness can be guaranteed even if almost all nodes are adversarial.
To characterize the possibility--impossibility frontier, we introduce the \emph{$\netX$-partially-synchronous} network model in which, over any sufficiently long time interval, the network is asynchronous for at most an $\netX$ fraction of the time. This model generalizes both 
synchronous networks (when $\netX = 0$) and partially-synchronous networks (when $\netX = 1$), and 
may be of independent interest
in the context of recent work on
fine-grained network models 
(\eg,~\cite{DBLP:conf/wdag/GiridharanACN024}). 
Our goal is thus to design consensus protocols that satisfy standard security properties (safety and liveness) under partial synchrony but, \emph{additionally}, achieve accountable liveness in the $\netX$-partially-synchronous model (for $\netX$ as large as possible).

Second, accountable safety is achievable without any assumptions on the number of adversarial nodes---even if all but two of the nodes are adversarial, the honest nodes' transcripts will provide a certificate of guilt following any safety violation.\footnote{We note that~\cite{bftforensics} establishes lower bounds on the size of the adversary for which accountable safety can be provided for certain protocols. Nevertheless, appropriately designed protocols can provide accountability even if all but two nodes are adversarial (see~\cite{DBLP:conf/sigecom/BudishLR24,DBLP:conf/fc/NeuTT22,cryptoeprint:2021/1169,accountabilityimpliesfinality}, for example).}
By contrast, we show that no protocol offering standard guarantees in partial synchrony (as described above) can also provide accountable liveness in the case that there is an adversarial majority (even under synchrony).
Intuitively, this is because
if an honest minority could obtain a certificate of guilt implicating adversarial nodes in case of an adversarial-majority attack on liveness,
then, by symmetry, a minority adversary could construct certificates implicating honest nodes, which is a contradiction.
The interesting parameter regime is therefore an adversary that controls at least one-third of the nodes (as otherwise, standard protocols cannot suffer liveness violations and accountability is irrelevant) and less than one-half of the nodes. Our protocols guarantee accountable liveness in this parameter regime (in $\netX$-partial-synchrony with $\netX < 1/2$).

\subsection{Results}
\label{sec:intro-results}

\begin{figure*}[tbp]%
    \centering%
    \begin{subfigure}[b]{0.33\linewidth}
        \centering
        \begin{tikzpicture}[]
            \footnotesize
            \begin{axis}[mysimplefig1plot]

                \addplot [myparula11,no markers,mark=none,name path global=plt:results1-tALmax033-f033-Achievability1] table [x=x,y=fun] {figures/fig1/schemeAchievability1-tALmax0.33334-f0.33334.txt};
                \label{plt:results1-tALmax033-f033-Achievability1};
    
                \addplot [myparula21,densely dotted,no markers,mark=none,name path global=plt:results1-tALmax033-f033-Converse1] table [x=x,y=fun] {figures/fig1/schemeConverse1-tALmax0.33334-f0.33334.txt};
                \label{plt:results1-tALmax033-f033-Converse1};

                \addplot [draw=none,name path global=y0] {0};
                \addplot [draw=none,name path global=y1] {1};

                \addplot [myparula11,fill=myParula01Blue,fill opacity=0.1] fill between[of=plt:results1-tALmax033-f033-Achievability1 and y0];
                \addplot [myparula21,fill=myParula02Orange,fill opacity=0.1] fill between[of=plt:results1-tALmax033-f033-Converse1 and y1];
    
                \addplot [black,only marks,mark=o] coordinates {(0.33333,0.33333)};
                \label{plt:results1-tALmax033-f033-tight};
                
            \end{axis}
        \end{tikzpicture}%
        \caption{$\tALmax = \lfloor \frac{1}{3}(n-1) \rfloor + 1$}%
        \label{fig:results1-tALmax033-f033}
    \end{subfigure}
    \hfil
    \begin{subfigure}[b]{0.33\linewidth}
        \centering
        \begin{tikzpicture}[]
            \footnotesize
            \begin{axis}[mysimplefig1plot]

                \addplot [myparula11,no marks,name path global=plt:results1-tALmax040-f040-Achievability1] table [x=x,y=fun] {figures/fig1/schemeAchievability1-tALmax0.4-f0.4.txt};
                \label{plt:results1-tALmax040-f040-Achievability1};
    
                \addplot [myparula21,densely dotted,no marks,name path global=plt:results1-tALmax040-f040-Converse1] table [x=x,y=fun] {figures/fig1/schemeConverse1-tALmax0.4-f0.4.txt};
                \label{plt:results1-tALmax040-f040-Converse1};

                \addplot [draw=none,name path global=y0] {0};
                \addplot [draw=none,name path global=y1] {1};

                \addplot [myparula11,fill=myParula01Blue,fill opacity=0.1] fill between[of=plt:results1-tALmax040-f040-Achievability1 and y0];
                \addplot [myparula21,fill=myParula02Orange,fill opacity=0.1] fill between[of=plt:results1-tALmax040-f040-Converse1 and y1];
    
                \addplot [black,only marks,mark=o] coordinates {(0.33333,0.26666666666)};
                \label{plt:results1-tALmax040-f040-tight};
    
            \end{axis}
        \end{tikzpicture}%
        \caption{$\tALmax = \frac{4}{10}n$}%
        \label{fig:results1-tALmax040-f040}
    \end{subfigure}
    \hfil
    \begin{subfigure}[b]{0.33\linewidth}
        \centering
        \begin{tikzpicture}[]
            \footnotesize
            \begin{axis}[mysimplefig1plot]

                \addplot [myparula11,no marks,name path global=plt:results1-tALmax049-f049-Achievability1] table [x=x,y=fun] {figures/fig1/schemeAchievability1-tALmax0.49999-f0.49999.txt};
                \label{plt:results1-tALmax049-f049-Achievability1};
    
                \addplot [myparula21,densely dotted,no marks,name path global=plt:results1-tALmax049-f049-Converse1] table [x=x,y=fun] {figures/fig1/schemeConverse1-tALmax0.49999-f0.49999.txt};
                \label{plt:results1-tALmax049-f049-Converse1};

                \addplot [draw=none,name path global=y0] {0};
                \addplot [draw=none,name path global=y1] {1};

                \addplot [myparula11,fill=myParula01Blue,fill opacity=0.1] fill between[of=plt:results1-tALmax049-f049-Achievability1 and y0];
                \addplot [myparula21,fill=myParula02Orange,fill opacity=0.1] fill between[of=plt:results1-tALmax049-f049-Converse1 and y1];
    
                \addplot [black,only marks,mark=o] coordinates {(0.33333,0.1666666)};
                \label{plt:results1-tALmax049-f049-tight};
    
            \end{axis}
        \end{tikzpicture}%
        \caption{$\tALmax = \lfloor \frac{1}{2}(n-1) \rfloor$}%
        \label{fig:results1-tALmax049-f049}
    \end{subfigure}
    \caption[]{%
        Illustration of key results:
        Impossibility of accountable liveness for $\tALmax \geq \frac{1}{2}n$, provided in \cref{thm:impossibility-tALmax-geq-n2},
        and for $\netX\geq1/2$, provided in \cref{thm:impossibility-x-geq-12},
        determine choices of $\tALmax$ and range of $\netX$ plotted. 
        Plots show fraction of identified adversary nodes ($\tALident/n$) 
        achieved~(\ref{plt:results1-tALmax033-f033-Achievability1})
        by the scheme of \cref{sec:blaming,sec:adjudication,sec:consensus}
        (\cref{thm:adjudication-criticalsubsets1-acclive} with $\superviewEpsGap \approx 0$),
        and the $\tALident/n$ upper bound~(\ref{plt:results1-tALmax033-f033-Converse1}) of \cref{thm:impossibility-tALident1-ours,thm:impossibility-tALident1-pbft}.
        Achievability and impossibility are tight~(\ref{plt:results1-tALmax033-f033-tight})
        for $\netX=1/3$.
    }%
    \label{fig:results1}%
\end{figure*}%

We study atomic broadcast consensus protocols that, roughly speaking (see \cref{sec:modelprelims-goal} for a formal statement), guarantee two sets of requirements:
\begin{itemize}
    \item[(1)] As a \emph{baseline}, the protocol should satisfy the standard security and performance properties required of protocols for partial synchrony. Specifically, it should be safe and live up to $\lfloor (n-1)/3 \rfloor$ adversary nodes, with expected confirmation latency during synchrony on the order of the network delay bound $\Delta$ that is guaranteed to hold after the adversary-chosen ``global stabilization time'' (GST) of the partially-synchronous model.

    \item[(2)] \emph{Additionally}, for specified parameters $\netX$ and $\tALmax$,
    if the network happens to be $\netX$-partially-synchronous and there happen to be $f \leq \tALmax$ adversary nodes,
    then the protocol should be accountably live,
    \ie, produce certificates of guilt for ``many'' adversary nodes whenever a transaction is not confirmed in a timely fashion.
\end{itemize}

On the positive, 
achievability
side,
in \cref{sec:consensus,sec:blaming,sec:adjudication},
we present an accountably live protocol. %
Specifically, 
this protocol achieves:
\begin{theorem*}[Informal version of \cref{thm:adjudication-criticalsubsets1-acclive}]
    For any $\netX < 1/2$ and $n/3 < \tALmax < n/2$,
    the protocol of \cref{alg:tendermint-2,alg:tendermint-2-alproto-multishot,alg:adjudication-criticalsubsets1}
    is accountably live
    when run in $\netX$-partial-synchrony with $f \leq \tALmax$ adversary nodes,
    and identifies arbitrarily close to
    $\tALmax - \left\lfloor \frac{(1+x) (\tALmax - n/3)}{1-x} \right\rfloor$ adversary nodes when there is a liveness violation. 
\end{theorem*}
In addition, the protocol of \cref{alg:tendermint-2,alg:tendermint-2-alproto-multishot,alg:adjudication-criticalsubsets1} satisfies the aforementioned point~(1), namely the standard security and performance properties required of protocols for partial synchrony (\cf \cref{thm:tendermint-2-safety,thm:tendermint-2-liveness}).
We use the Tendermint protocol~\cite{tendermint} as the starting point for our construction, but
our techniques also readily apply to other PBFT-style protocols like HotStuff~\cite{hotstuff} or Streamlet~\cite{streamlet}.

On the negative, 
impossibility
side, we show in \cref{sec:impossibilities} that
the restrictions on $\netX$ and $\tALmax$ in the aforementioned theorem
are fundamental for accountable liveness:
\begin{theorem*}[Informal version of \cref{thm:impossibility-tALmax-geq-n2}]
    No protocol
    can simultaneously provide 
    safety under partial synchrony even when $f=0$,
    and 
    accountable liveness
    with $\tALmax \geq n/2$
    under synchrony.
\end{theorem*}
\begin{theorem*}[Informal version of \cref{thm:impossibility-x-geq-12,thm:impossibility-psync}]
    No optimally-resilient protocol
    can be accountably live
    under $\netX$-partial-synchrony
    for $\netX \geq 1/2$
    (and thus also not under partial synchrony).
\end{theorem*}
Here, ``optimally-resilient'' means that the numbers of adversary nodes the protocol can tolerate while remaining safe or live, respectively, are maximal.\footnote{We focus on optimally-resilient protocols since these are the most relevant, and because this also simplifies our analysis by ruling out certain complexities. For example, if we were not to focus on optimally-resilient protocols, a protocol might claim resilience that is sub-optimal and then appear to achieve non-trivial accountability simply by actually ruling out liveness violations beyond the claimed resilience.}

Finally, one can ask how good the protocol of \cref{sec:consensus,sec:blaming,sec:adjudication} is, in terms of the number of adversary nodes it can identify.
We show for classical PBFT-style protocols 
(including but not limited to
PBFT~\cite{pbft}, 
Tendermint~\cite{tendermint}, 
HotStuff~\cite{hotstuff}, 
CasperFFG~\cite{casper,casperethresearch}, 
and
Streamlet~\cite{streamlet})
an upper bound on the number of adversary nodes 
that can be identified,
and we conjecture this bound to hold for all (also non-PBFT-style) protocols:
\begin{theorem*}[Informal version of \cref{thm:impossibility-tALident1-ours,thm:impossibility-tALident1-pbft,thm:impossibility-tALident1-all}]
    For every $k\geq3$:
    Classical PBFT-style protocols
    (conjecture: all protocols!)
    that remain safe and live under $\lfloor (n-1)/3 \rfloor$ adversary nodes
    in partial synchrony,
    cannot be accountably live
    under $\netX$-partial-synchrony
    for $\netX \geq 1/k$
    and guarantee to identify
    $n/3 - \left\lfloor \frac{\tALmax - n/3}{k-2} \right\rfloor$
    (or more) adversary nodes when liveness is violated.
\end{theorem*}

\Cref{fig:results1} plots the number of adversary nodes the scheme of \cref{sec:blaming,sec:adjudication,sec:consensus} achieves to identify in case of a liveness violation (\ref{plt:results1-tALmax033-f033-Achievability1}),
and the aforementioned upper bound (\ref{plt:results1-tALmax033-f033-Converse1}),
for varying $\netX < 1/2$ and $n/3 < \tALmax < n/2$.
(Recall that $\tALmax \leq n/3$ is uninteresting because, by goal~(1), the protocol is then guaranteed to be live, so accountability is trivial.)
\Cref{fig:results1} shows that our impossibility result matches our achievability result 
closely.
Remarkably, they are tight for $\netX=1/3$.  

\subsection{Discussion}
\label{sec:intro-discussion}

\paragraph{Comparison to Synchronous Atomic Broadcast Protocols}
The definition of $\netX$-partial-synchrony for $\netX<1$ and with respect to a delay bound $\Delta$ implies synchrony with respect to a delay bound $\Delta^*$ that can be arbitrarily larger than $\Delta$; intuitively, $\Delta^*$ scales with the ``sufficiently large interval length''  under which the $\netX$-partially-synchronous condition is guaranteed to kick in. (See \cref{sec:modelprelims-discussion} for a more detailed discussion.)
In principle, therefore, one could employ a \emph{synchronous}
atomic broadcast protocol, such as 
a variant of the Dolev--Strong protocol~\cite{DBLP:journals/siamcomp/DolevS83,buterinblogpost,elaineshibook,cryptoeprint:2024/1799,DBLP:conf/sp/HouYS22,DBLP:conf/sp/HouY23}, with delay bound $\Delta^*$. However, such a protocol would always incur large confirmation latency (scaling with $\Delta^*$), and thus would not achieve our requirement of expected confirmation latency scaling with~$\Delta$ 
when
the network becomes synchronous. Such a protocol would also fail our requirement that it should be safe and live up to $\lfloor (n-1)/3 \rfloor$ adversary nodes in partial synchrony.

\paragraph{Add-On Features}

Known techniques, similar to those used for accountable safety, can be used to handle 
the reconfiguration process needed to remove adversary nodes once identified for violating liveness guarantees---such as manual intervention or externally triggered reconfiguration procedures~\cite{DBLP:journals/corr/abs-2310-06338}.
A more automated strategy, benefiting from the large implied synchrony bound $\Delta^*$ mentioned above, is described in~\cite{lewispye2025optimalfaulttolerance}, where the authors propose a wrapper that runs an accountably-safe state-machine replication protocol until a safety violation is detected. Upon detecting such a violation, the wrapper initiates a recovery procedure---requiring message delays to remain bounded by $\Delta^*$ during this recovery---to use certificates of guilt to reliably identify misbehaving nodes, remove them safely, and then restart the protocol without these adversarial nodes.
The same recovery procedure can be used with the certificates of guilt produced by liveness accountability in this work.

\subsection{Technical Overview}
\label{sec:tech-overview}

Once the model is formally defined in \cref{sec:modelprelims}, our first task towards establishing \cref{thm:adjudication-criticalsubsets1-acclive} 
is to describe 
a modification of Tendermint consensus that allows nodes to \emph{blame} others for a lack of liveness. Recall that an execution of Tendermint is partitioned into \emph{views}. Any node that does not see progress in a particular view will now blame a number of other nodes for lack of progress in that view.  
Here,
$\netX$-partial-synchrony 
ensures that (over a sufficiently long window) at least a $(1-\netX)$ fraction of views will be synchronous (\ie, message delivery will be reliable in those views), and for synchronous views we aim to ensure that the outcome of the blaming process has certain useful properties.
Specifically, the process of blaming other nodes 
is designed 
(in \cref{sec:blaming-simple})
so that:
\begin{enumerate}
    \item[(i)] No honest node is blamed by any honest node in any synchronous view.
    Since the honest nodes are a majority, this means that no honest node is ``majority blamed'' in any synchronous view. 
    \item[(ii)] In each synchronous view with an honest leader that does not confirm new transactions, there are at least $n/3$ adversary nodes that are blamed by all honest nodes (and so which are ``majority blamed'' in that view). 
\end{enumerate}
A weakness of (ii) above is that an adversarial leader may prevent progress in a synchronous view, removing the need for $n/3$ adversarial nodes to be majority blamed.
In a quantitative sense, this harms our ability to blame adversarial nodes, \ie, the fact that a single adversarial leader can get away with preventing progress during synchrony without many adversarial nodes getting blamed, limits the number of adversarial nodes that we can hold responsible in the event of a liveness violation.
For this reason, in \cref{sec:blaming-refined} we introduce the use of \emph{super-views}: each super-view consists of a number of views, with the number chosen so that (with random leader selection) each super-view is likely to have at least one honest leader. 
We can then consider an analogous notion of blame for super-views, which satisfies a corresponding version of (i) above, and which satisfies a version of (ii) with the weaker requirement that the super-view should have at least one honest leader.

Given this setup, we must then extract an ``adjudication rule'' for identifying guilty nodes in the event of an attack on liveness.  To demonstrate the basic idea behind the adjudication rule, consider (for now) a simplified setup in which we suppose all super-views have at least one honest leader. Suppose we see a sequence of super-views $\mathcal{U}$, all of which fail to make progress with liveness, and that we know a majority of super-views in $\mathcal{U}$ are synchronous (the number of super-views in $\mathcal{U}$ will be function of our formal definition of 
$\netX$-partial-synchrony
in \cref{sec:modelprelims}). 
In this section, we now describe an approach which suffices to identify at least \emph{one} adversary node in the case that there is a strict majority of honest nodes.  
Later, in \cref{sec:adjudication-complex}, we show how to generalize the method to identify a greater number of adversary nodes when a greater proportion of super-views are synchronous and/or a smaller number of nodes are adversary, and we also drop the assumption that every super-view has at least one honest leader.

\paragraph{A Trivial Heuristic}
If we can find a node that is majority blamed in a majority of super-views in $\mathcal{U}$, then we can identify that node as adversary. This holds since  an honest node is never majority blamed in a synchronous super-view and a majority of super-views in $\mathcal{U}$ are synchronous.  The following is therefore a \emph{sufficient} condition for being able to identify at least one adversary node: adversary nodes are on average majority blamed in over half of the super-views in $\mathcal{U}$.
This argument has one-sided error (no false positives, but may have false negatives).
In particular, this trivial heuristic is \emph{sound}, but unfortunately not \emph{complete}.

\paragraph{Extending the Trivial Heuristic}
The basic idea is that, if one chooses the \emph{right} subset of super-views  in $\mathcal{U}$ (efficiently computable and guaranteed to contain all synchronous super-views in $\mathcal{U}$), then the heuristic above remains \emph{sound} and is also \emph{complete} (\ie, applies whenever none of the views in $\mathcal{U}$ confirm new transactions).

\paragraph{The Details}
Recall that $\CU$ is a set of consecutive super-views.  
We assume that a majority of the super-views in $\CU$ are synchronous.
\begin{enumerate}
    \item Let $\critSubsetsCPA_u$ denote the set of nodes that are majority blamed in super-view $u\in \CU$.
    \item Form an undirected graph $G$ with $|\mathcal{U}|$ vertices, each  corresponding to a super-view in $\CU$, and an undirected edge $(u, u')$ whenever $|\critSubsetsCPA_u \cap \critSubsetsCPA_{u'} | \geq n/6$.
    \item  Let $\CU'$ denote the subset of super-views $u$ such that: 
    \begin{enumerate}
        \item $|\critSubsetsCPA_u|\geq n/3$; and
        \item $\operatorname{deg}_G(u)\geq |\mathcal{U}|/2$.
    \end{enumerate}
\end{enumerate}

\paragraph{Proof of Soundness}
The heuristic never outputs an honest node:
\begin{itemize} 
    \item If $u$ is synchronous, $|\critSubsetsCPA_u|\geq n/3$. 
    \item  If $u, u'$ are both synchronous, then $|\critSubsetsCPA_u \cap \critSubsetsCPA_{u'}| \geq n/6$ (since there are $<n/2$ adversary nodes).
    \item In the graph $G$ above, every vertex corresponding to a synchronous super-view has degree $\geq |\mathcal{U}|/2$ (by previous point and the assumption that a majority of super-views are synchronous).
    \item Every synchronous super-view belongs to $\CU'$ (by first and third points above).
    \item A majority of super-views in $\CU'$ are synchronous (by the previous point and the assumption that a majority of super-views in $\CU$ are synchronous).
    \item No honest node can be majority blamed in a majority of the super-views in $\CU'$ (because no honest node is majority blamed in any synchronous super-view).
\end{itemize} 

\paragraph{Proof of Completeness}
To show that the heuristic is guaranteed to output a (necessarily adversary) node, we argue as follows: 
\begin{itemize}
    \item Consider a $0$--$1$ matrix $M$ with rows indexed by adversary nodes $p$ and columns indexed by super-views $u$ of $\CU'$. Define $M_{p,u} = 1$ if $p\in \critSubsetsCPA_u$ and $M_{p,u} = 0$ otherwise.
    \item For every synchronous super-view in $\CU'$, the corresponding column sum is at least $n/3$.
    \item By properties~(a) and~(b) of super-views in $\CU'$, for every asynchronous super-view in $\CU'$, the corresponding column sum is at least $n/6$. This is because, by~(b), every asynchronous super-view $u\in\CU'$ must have overlap $|\critSubsetsCPA_u \cap \critSubsetsCPA_{u'} | \geq n/6$ with some synchronous super-view $u$, but in synchronous super-views only adversary nodes are majority blamed, so the overlap counts into the column sum of $u$ in $M$ despite $M$ having rows only for adversary nodes.
    \item Because over half the super-views of $\CU'$ are synchronous (see above), the sum of $M$'s entries is $\geq \frac{|\CU'|}{2}\cdot \frac{n}{3} + \frac{|\CU'|}{2}\cdot \frac{n}{6} = \frac{n|\CU'|}{4}$.
     \item  Because there are fewer than $n/2$ rows (by assumption on the number of adversary nodes), the average row sum is $> \frac{|\CU'|}{2}$.
     \item  Thus, there exists a row with more than $\frac{|\CU'|}{2}$ 1's---an adversary node that is majority blamed in a majority of the views in $\CU'$.
\end{itemize}

\paragraph{Outline}

\Cref{sec:modelprelims} introduces the $\netX$-partial-synchrony model
and our notion of accountable liveness, and also covers various preliminaries.
\Cref{sec:consensus} reviews the starting point of our protocol, a variant of Tendermint consensus.
This protocol subsequently serves as our running example, 
in \cref{sec:blaming,sec:adjudication}, for how to achieve accountable liveness.
\Cref{sec:impossibilities} proves the two impossibility results for accountable liveness that are mentioned above.
\Cref{sec:relatedwork} discusses related work. 
\Cref{sec:conclusion} concludes.
\Cref{sec:addendum-relatedwork} discusses 
additional 
related works.

\section{Model \& Preliminaries}
\label{sec:modelprelims}

We model protocol execution and define consensus security properties
in a standard way, except for our use of $\netX$-partially-synchronous networks (\cref{sec:modelprelims-oftensync})
and our definition of accountable liveness (\cref{sec:modelprelims-accountableliveness}).
There are $n$ \emph{nodes}, denoted $p\in\CP$, each of which has a cryptographic identity
(public/secret key pair for signatures) that is 
commonly known
(\emph{public-key infrastructure, PKI}).
There is an \emph{environment} that, over time, inputs \emph{transactions} to the nodes,
and to which, over time, each node outputs a \emph{log} of transactions.
The nodes' objective is \emph{atomic broadcast},
\ie, to reach agreement on an ordering of their input transactions into their output logs.
For this purpose, nodes can send each other messages over a \emph{network}.
A 
computationally bounded \emph{adversary} seeks to disrupt consensus
and for this purpose can corrupt nodes and delay network messages.
For ease of exposition, we treat cryptographic signatures as \emph{ideal}, \ie, we assume that the adversary cannot forge signatures of honest nodes.

Time proceeds 
in discrete \emph{rounds} and, for simplicity, nodes are assumed to have synchronized clocks.
In each round, each node receives messages from the network and possibly transactions from the environment.
The node then updates its internal state, produces messages to send to other nodes via the network, and outputs a log of confirmed transactions to the environment.
The adversary chooses $f$ nodes, denoted $\CPa\subseteq\CP$, to corrupt at the beginning of the execution and in particular before any protocol randomness is drawn (\emph{static corruption}).
The adversary learns the internal state of \emph{adversary} nodes, and can make them deviate from the protocol arbitrarily for the entire execution (\emph{permanent Byzantine faults}).
Non-adversary \emph{honest} nodes ($\CPh\subseteq\CP$) follow the protocol.

\subsection{\texorpdfstring{$\netX$}{x}-Partially-Synchronous Networks}
\label{sec:modelprelims-oftensync}

When nodes send messages to each other,
the messages are \emph{delayed} by the adversary subject to certain constraints.
We propose the  \emph{$\netX$-partially-synchronous network model} 
as an interpolation between two classic network models, 
\emph{synchronous}~\cite{sync-model}
and \emph{partially-synchronous}~\cite{model-psync} (also called \emph{eventually-synchronous}) networks.

Specifically, under synchrony, there is a known delay upper-bound of $\Delta > 0$ rounds.
It will be convenient to view the synchronous network model as follows. When a node instructs the network to send a message to another node,
the message is enqueued in the recipient's pending message queue together with a \emph{countdown}
initialized to $\Delta$.
The adversary can decrease the countdown at will.
The countdown also decreases by $1$ with each round.
Once the countdown hits $0$, the message is delivered to the recipient at the beginning of the next round.
Partially-synchronous networks extend the above with a \emph{global stabilization time} $\GST < \infty$,
a round
which the adversary chooses adaptively.
The countdown of 
pending messages
is guaranteed to decrement 
each round only after $\GST$.

Our \emph{$\netX$-partially-synchronous setting} inherits the assumptions of partial synchrony.
Furthermore, for every round before $\GST$, the adversary adaptively chooses 
whether the round is synchronous (\ie, all countdowns decrement) or not. (All rounds after $\GST$ are synchronous.)
Call an interval {\em $\Delta$-synchronous} if each of its rounds is synchronous in this sense. The restriction on the adversary is
parameterized by (in addition to $\Delta$)
a known function $\netG$ and a known value $\netX \in [0,1]$.
The restriction is then that, 
for every partition of the execution into \emph{periods} of length $\netDeltaPeriod$ rounds,
and for every interval of $\netG(\netDeltaPeriod)$ periods, at least a $(1-\netX)$ fraction of the periods are 
$\Delta$-synchronous.
For example, $\netG(1)$ specifies the minimum length of an interval (in rounds) for which it is guaranteed that at most an $\netX$ fraction of the interval's rounds are asynchronous.
We generally assume that $\netG$ grows unboundedly with $\netDeltaPeriod$---the longer the period of synchrony~$\netDeltaPeriod$ desired, the longer one may have to wait for it.
For every function~$\netG$, the $\netX$-partially-synchronous model generalizes the synchronous ($\netX=0$) and partially synchronous ($\netX=1$) models.\footnote{Note that safe \emph{and live} consensus is achievable in $\netX$-partial-synchrony even for $\GST=\infty$. As such, the model is meaningful also without $\GST$.
We chose to keep $\GST$ to retain synchrony and partial synchrony as special cases of $\netX$-partial-synchrony.}
Where precision about the network parameters is required, we write \emph{$(\Delta, \netG, \netX)$-partial-synchrony}.
Since $\netX$ is the dominant quantitative parameter
affecting accountable liveness,
while $\Delta$ and $\netG$ are often clear from context,
we also often write \emph{$\netX$-partial-synchrony}.
\Cref{sec:modelprelims-discussion} discusses the plausibility of the $\netX$-partially-synchronous model from a practical perspective.

\subsection{Atomic Broadcast}
\label{sec:modelprelims-atomicbroadcast}

For logs, \ie, 
transaction sequences,
we write 
$\LOG \preceq \LOG'$
iff
$\LOG$ is a prefix of or equal to $\LOG'$,
and $\LOG \consistent \LOG'$ (``$\LOG$ is consistent with $\LOG'$'')
iff $\LOG \preceq \LOG'$ or $\LOG' \preceq \LOG$.
We follow the usual definition of atomic broadcast:
\begin{definition}
    \label{def:atomic-broadcast}
    A protocol $\Pi$
    where node $p$ at round $t$
    \emph{confirms}
    the output log $\LOG_t^p$,
    achieves \emph{atomic broadcast} 
    with
    \emph{safety resilience} $\tS$
    and
    \emph{liveness resilience} $\tL$,
    iff
    in every execution satisfying the desired network model:
    \begin{itemize}
        \item \textbf{Safety}:
                If $f \leq \tS$,
                then:
                $\forall p, q \in \CPh:
                \forall t, t':
                \LOG_t^p \consistent \LOG_{t'}^q$.
        \item \textbf{Liveness}:
                If $f \leq \tL$,
                then:
                for every transaction $\tx$ input to all $\CPh$ by $t_0^{\tx}$:
                $\exists t_1^{\tx} < \infty:
                \forall t \geq t_1^{\tx}:
                \forall p \in \CPh:
                \tx \in \LOG_t^p$.
    \end{itemize} 
    For the minimum $t_0^{*\tx}, t_1^{*\tx}$
    that satisfy the liveness condition,
    called the \emph{input round} and \emph{confirmation round} of $\tx$, respectively,
    we define $\Tconf^\tx \triangleq t_1^{*\tx} - t_0^{*\tx}$ as the \emph{confirmation latency} of $\tx$.
\end{definition}
We say a protocol satisfying \cref{def:atomic-broadcast}
is \emph{$\tS$-safe} and \emph{$\tL$-live}.

\subsection{Accountable Liveness}
\label{sec:modelprelims-accountableliveness}

In addition to the parameters $(\Delta, \netG, \netX)$ of the $\netX$-partially-synchronous 
model,
the notion of accountable liveness is parametric in
\emph{period length} $\netDeltaPeriod$,
\emph{accountable-liveness resilience} $\tALmax$,
\emph{sensitivity} $\tALident$,
and \emph{failure probability} $\epsAL$,
which will become clear below.
Given these parameters,
we build up the definition of accountable liveness in three steps.
First, we define what it means for a \emph{timely-liveness violation} to occur.
Then, we define what it means for a protocol message to constitute a \emph{certificate of guilt}.
Finally, we define what it means for an atomic broadcast protocol to be \emph{accountably live}.

Traditional (\emph{eventual}) liveness (as in \cref{def:atomic-broadcast})
is a weaker property in some sense, 
and a stronger property in another sense,
than what we need.
Specifically, 
traditional liveness is 
so weak
that violations of it cannot be assessed by any one point in time~\cite{DBLP:journals/ipl/AlpernS85}
(any transaction in question may still get confirmed in the future).
But $(\Delta, \netG, \netX)$-partial-synchrony
with $\netX<1$
allows us to demand a more ambitious liveness guarantee
where transactions need to be confirmed 
within $\netDeltaPeriod \netG(\netDeltaPeriod)$ rounds.
On the other hand,
traditional liveness is strong in 
that it requires \emph{all} honest nodes
to confirm a transaction.
For our purposes, a slightly weaker variant suffices,
that requires only \emph{all except ``a few'' honest nodes} to confirm.
This is particularly meaningful under $(\Delta, \netG, \netX)$-partial-synchrony with $\netX<1$,
because
that setting
implies a (large) delay upper-bound of $\Delta^* \triangleq \netDeltaPeriod \netG(\netDeltaPeriod)$ 
so that
if \emph{some} honest node has confirmed a transaction,
then \emph{all} honest nodes will have confirmed that transaction $\Delta^*$ later.

Concretely, we say a protocol with maximum liveness resilience~$\tL$ satisfies \emph{timely-liveness} if all transactions
are confirmed within $\netDeltaPeriod \netG(\netDeltaPeriod)$ rounds
by all except $\tL$ honest nodes.
We concern ourselves with accountability for timely-liveness violations:
\begin{definition}
    \label{def:liveness-violation}
    For a protocol with maximum liveness resilience~$\tL$, and for period length $\Delta'$, an execution has a \emph{timely-liveness violation} at round $t$
    iff there exists a transaction $\tx$
    with input round $t_0^{\tx} \leq t - \netDeltaPeriod \netG(\netDeltaPeriod)$
    that \emph{(strictly) more than $\tL$} honest nodes have not confirmed by $t$,
    \ie,
    $\exists\text{$\tx$ input to all $\CPh$ by $t_0^{\tx} \leq t - \netDeltaPeriod \netG(\netDeltaPeriod)$}:
    \exists \CP'\subseteq\CPh: (|\CP'| > \tL) \land
    (\forall p \in \CP': \tx \not\in \LOG_t^p)$. 
\end{definition}

Accountability is carried out through certificates of guilt:\footnote{%
Note that \cref{def:certificate-of-guilt} is a \emph{conditional} variant
(where only $(\Delta, \netG, \netX)$-partially-synchronous executions with $f \leq \tALmax$ are considered)
of the notion of a certificate of guilt that appeared in the accountable-safety literature~\cite[p.\ 5]{lewispye2025optimalfaulttolerance} (where no such restrictions are imposed on the executions considered).
As shown in \cref{sec:intro-scope,sec:intro-results,thm:impossibility-x-geq-12,thm:impossibility-tALmax-geq-n2}, the two restrictions on the executions in \cref{def:certificate-of-guilt}, namely on network timing and on a maximum adversary strength,
are necessary for the notion of accountable liveness to be achievable.
\Cref{sec:modelprelims-discussion} discusses that for suitably chosen parameters, restricting to $\netX$-partially-synchronous executions may practically not be very severe. 
}
\begin{definition}
    \label{def:certificate-of-guilt}
    A protocol message $m$ constitutes a \emph{certificate of guilt} for a node $p$
    iff $m$ is never obtained by any node in any 
    $(\Delta, \netG, \netX)$-partially-synchronous execution
    with $f \leq \tALmax$
    in which $p\in\CPh$.
\end{definition}

Intuitively,
accountable liveness 
means that, whenever
an input transaction remains unconfirmed for 
``sufficienty long'' by ``many'' honest nodes,
``eye-witness evidence'' 
(\eg, many attestations to the unexpected absence of certain messages) emerges
that accuses at least $\tALident$ nodes
of misbehavior.
This eye-witness evidence
is guaranteed to implicate only adversarial nodes provided
the network was 
$(\Delta, \netG, \netX)$-partially-synchronous
and  $f \leq \tALmax$. 
\begin{definition}
    \label{def:accountable-liveness}
    An atomic broadcast protocol $\Pi$
    is \emph{accountably live}
    in $(\Delta, \netG, \netX)$-partially-synchronous networks 
    with
    \emph{period length} $\netDeltaPeriod$,
    \emph{accountable-liveness resilience} $\tALmax$,
    \emph{sensitivity} $\tALident$,
    and \emph{failure probability} $\epsAL$,
    iff for any fixed round $t$:
    For every adversary with $f \leq \tALmax$
    in $(\Delta, \netG, \netX)$-partially-synchronous networks,
    with probability at least $(1-\epsAL)$,
    if a timely-liveness
    violation occurs at round $t$,
    then, eventually, some honest node broadcasts a 
    certificate of guilt for some set $\CP'\subseteq\CP$ of nodes
    with $|\CP'| \geq \tALident$.
\end{definition}

For simplicity, and as this captures the crux of the problem, we state our (accountable) liveness claims for 
liveness violations that occur at a \emph{fixed round}.
(Accountable) liveness properties for \emph{all} transactions 
\emph{over an entire appropriately-bounded execution horizon} are readily obtained with a union bound.

\subsection{Overall Protocol Design Goal}
\label{sec:modelprelims-goal}

We now formally state our protocol design objective.
We aim to design atomic broadcast protocols that guarantee two requirements:
\begin{itemize}
    \item[(1)] As a \emph{baseline}, for any given $\tau, \Delta$, when run among $n=3\tau+1$ nodes in a \emph{partially-synchronous} network (\ie, $\netX$-partially-synchronous with $\netX=1$), the protocol should be $(\tS=\tau)$-safe and $(\tL=\tau)$-live, and guarantee expected confirmation latency $O(\Delta)$, independent of $\netG$, for transactions input after $\GST$.
    In other words, the protocol should satisfy the standard security and performance properties required of protocols for partial synchrony.

    \item[(2)] \emph{Additionally}, for any given $\netG, \netX, \tALmax, \epsAL$,
    the protocol should be accountably live
    for some $\netDeltaPeriod, \tALident$
    if the network happens to be $(\Delta, \netG, \netX)$-partially-synchronous and if $f \leq \tALmax$.
    Note that $\tALmax > \tL, \epsAL < 1, \tALident > 0$ constitutes the non-trivial regime.
    Smaller $\netDeltaPeriod$ and larger $\tALident$ are better,
    since it 
    shortens the duration of non-confirmation after which accountability is required,
    and increases the number of adversary nodes identified.
\end{itemize}

\subsection{Discussion}
\label{sec:modelprelims-discussion}

Recall that under $(\Delta, \netG, \netX)$-partial-synchrony,
the assumptions of partial synchrony hold regarding $\Delta$ and $\GST$.
In addition, there is a known function $\netG$ and a known value $\netX \in [0,1]$, 
such that, 
before $\GST$,
for any partition of time into periods of $\netDeltaPeriod$ rounds,
any interval of $\netG(\netDeltaPeriod)$ periods
has at most $\netX$ fraction of the periods not be $\Delta$-synchronous.
Is such an assumption practically plausible?
Consider that typical global round-trip times in the Internet are in the order of hundreds of milliseconds,
and that Internet connectivity outages lasting longer than hours are exceedingly rare.
Furthermore, network service-level agreements 
commonly promise
at most a certain latency 
for at least a certain fraction of 
every long-enough period of time.
This leads us to believe that
with
$\netX$ of tens of percentage points,
$\Delta$ of seconds,
$\netDeltaPeriod$ of tens of seconds,
and $\netDeltaPeriod \netG(\netDeltaPeriod)$ of tens of hours,
$\netX$-partial-synchrony is at least a plausible assumption.
It does not appear much less plausible than
assumptions of Internet delay upper-bounds of seconds,
implicit in systems like Ethereum~\cite{gasper} or Cardano~\cite{DBLP:conf/crypto/KiayiasRDO17}.
At the same time, the longer periods of synchrony one demands,
the longer one plausibly has to wait, justifying that
$\netG(\netDeltaPeriod)$ should grow unboundedly with $\netDeltaPeriod$.

Note that $\netX$-partial-synchrony for $\netX<1$
implies a large delay upper-bound of $\Delta^* \triangleq \netDeltaPeriod \netG(\netDeltaPeriod) \gg \Delta$.
The reader may then ask, why not run a synchronous atomic broadcast protocol
with that delay bound $\Delta^*$?
Given that some such protocols 
are safe and live 
even if almost all nodes are adversary~\cite{DBLP:journals/siamcomp/DolevS83,cryptoeprint:2024/1799,buterinblogpost,DBLP:conf/sp/HouYS22,DBLP:conf/sp/HouY23},
they would trivially be accountably live as well, since there are no timely-liveness violations (let alone liveness violations) to begin with.
However, such protocols suffer from high latency in the order of $\Delta^* \gg \Delta$, and thus do not satisfy the target expected confirmation latency of $O(\Delta)$ after $\GST$.
Furthermore, such protocols do not guarantee safety under partial synchrony.
Thus, such protocols do not satisfy the baseline goal~(1) set out in \cref{sec:modelprelims-goal}.

The reader may wonder why we focus on the atomic broadcast variant of consensus,
rather than, for instance, on the state-machine replication variant, which in addition to nodes also models the system's clients~\cite{cryptoeprint:2024/1799}.
Three reasons:
(1)~Impossibility results for atomic broadcast are stronger. 
Yet, our protocol is straightforwardly extended to state-machine replication.
(2)~Our $\netX$-partially-synchronous network model
incorporates periods of asynchrony, and earlier works suggest~\cite{cryptoeprint:2024/1799}
that while there is
a considerable difference between state-machine replication and atomic broadcast
under synchrony, this may not be the case 
when safety during periods of asynchrony is required.
(3)~Accountable liveness inherently arises from the interplay of nodes and their network delay (\cf \cref{sec:intro-scope}). Unlike in accountable safety, clients play no particular role in accountable liveness, so, for simplicity, we leave them aside.

We focus exclusively on accountability of the liveness property here,
and otherwise stick to ``regular'' unaccountable safety.
This is because accountability of the safety property
has been studied extensively already in earlier works~\cite{bftforensics,DBLP:conf/fc/NeuTT22,DBLP:conf/sigecom/BudishLR24,casper,gasper,snapandchat}, and those techniques are orthogonal and can readily be applied independently to make an accountably live protocol accountably safe.

\section{Consensus Protocol}
\label{sec:consensus}

\begin{algorithm}[tb]
    \caption{Tendermint consensus variant, code for node $p$
    (based on \cite{tendermint}, \cite[Sec.\ 9.1]{DBLP:conf/sigecom/BudishLR24};
    with extra delay highlighted \myTextHighlightedDescription{}, and extra round of liveness voting highlighted \myTextHighlightedTwoDescription{}; \cf \cref{alg:tendermint-original})}
    \label{alg:tendermint-2}
    \begin{algorithmic}[1]
        \algfontsize%
        \State
            \algloclabel{loc:tendermint-2-blocks}
            Blocks:
            \begin{itemize}[itemsep=0pt,topsep=0pt,parsep=0pt,partopsep=0pt]
                \item $b \triangleq \Block(\FFp, \FFv, \FFbmI, \FFQC, \FFtxs)$ consists of:
                    creator $\FFp$,
                    view $\FFv$,
                    parent block $\FFbmI$,
                    quorum certificate $\FFQC$,
                    transactions $\FFtxs$.
                \item Genesis block: $b_0 \triangleq \Block(0, 0, \bot, \emptyset, \emptyset)$.
                \item Validity:
                    \begin{IEEEeqnarray}{rCCCl}
                        \MvalidB(\StM{}{}, b)
                        &\triangleq&
                        && (b = b_0) \nonumber\\
                        &&\lor&& ((b \in \StM{}{}) \nonumber\\ %
                        &&&\land& \MvalidQC(\StM{}{}, b.\FFQC, b.\FFbmI, 1) \nonumber\\
                        &&&\land& (b.\FFp = \Leader_{b.\FFv}) \nonumber\\
                        &&&\land& (b.\FFv > b.\FFbmI.\FFv))
                        \IEEEeqnarraynumspace
                    \end{IEEEeqnarray}
            \end{itemize}
        \State
            \algloclabel{loc:tendermint-2-votes}
            Votes:
            \begin{itemize}[itemsep=0pt,topsep=0pt,parsep=0pt,partopsep=0pt]
                \item $w \triangleq \Vote(\FFp, \FFb, \FFstage)$ consists of:
                    creator $\FFp$,
                    target block $\FFb$,
                    vote stage $\FFstage$.
                \item Validity:
                    \begin{IEEEeqnarray}{rCCCl}
                        \MvalidV(\StM{}{}, w)
                        &\triangleq&
                        && (w \in \StM{}{}) \land \MvalidB(\StM{}{}, w.\FFb)
                        \IEEEeqnarraynumspace
                    \end{IEEEeqnarray}
                \item \myTextHighlightedTwo{$w' \triangleq \VoteLive(\FFp, \FFv)$ consists of:
                    creator $\FFp$,
                    view $\FFv$.}
                \item \myTextHighlightedTwo{Validity:}
                    \begin{IEEEeqnarray}{rCCCl}
                        \myTextHighlightedTwo{\MvalidV(\StM{}{}, w')}
                        &\myTextHighlightedTwo{{}\triangleq{}}&
                        && \myTextHighlightedTwo{(w' \in \StM{}{})}
                        \IEEEeqnarraynumspace
                    \end{IEEEeqnarray}
            \end{itemize}
        \State
            \algloclabel{loc:tendermint-2-qcs}
            Quorum certificates (QCs):
            \begin{itemize}[itemsep=0pt,topsep=0pt,parsep=0pt,partopsep=0pt]
                \item Validity:
                    \begin{IEEEeqnarray}{rCCCl}
                        \MvalidQC(\StM{}{}, Q, b, s)
                        &\triangleq&
                        && ((Q = \emptyset) \nonumber\\ 
                        &&&\land& (b = b_0)) \nonumber\\
                        &&\lor&& ((\forall w \in Q: \MvalidV(\StM{}{}, w) \land (w.\FFb = b) \land (w.\FFstage = s)) \nonumber\\
                        &&&\land& (|\{ w.\FFp \mid w \in Q \}| > 2n/3)) 
                        \IEEEeqnarraynumspace
                    \end{IEEEeqnarray}
                \item Notation: If $\MvalidQC(\StM{}{}, Q, b, s)$, then $\FFv(Q) \triangleq b.\FFv$.
            \end{itemize}
        \State $\StM{}{} \gets \{ b_0 \}$%
            \algloclabel{loc:tendermint-2-stateM}%
        \State $\StQ{}{} \gets \emptyset$
        \State
        At all times, re-broadcast all messages and transactions received from the network or as input.
        \State
        At all times, 
        add to $\StM{}{}$ any message (block or vote) received from the network, 
        upon verification and stripping of the message's signature (ensuring that message $m$ was indeed created by $m.\FFp$),
        and any transaction received from the network or as input.
        Denote the set of transactions in $\StM{}{}$ as $\StMtxs{}{}$.
        Record when elements are added to $\StM{}{}$,
        to allow access to $\StM{}{}$ and $\StMtxs{}{}$ ``as of'' time $t$
        as $\StM{t}{}$ and $\StMtxs{t}{}$.%
        \algloclabel{loc:tendermint-2-stateM-explain}
        \State
        At all times, confirm as log $\LOG_t^p$ the sequence of transactions on the path from $b_0$ to $b$ iff
        $\exists Q_1, Q_2 \subseteq \StM{}{}: \MvalidQC(\StM{}{}, Q_1, b, 1) \land \MvalidQC(\StM{}{}, Q_2, b, 2)$.%
        \algloclabel{loc:tendermint-2-confirm}
        \For{$v=1,2,3,...$}
            \At{\myTextHighlighted{$t=12 \Delta v + 2 \Delta$}}%
                \algloclabel{loc:tendermint-2-propose}
                \If{$\Leader_v = p$}
                    \State $(b, Q) \gets \argmax_{(b, Q): \MvalidQC(\StM{}{}, Q, b, 1)} b.\FFv$%
                        \algloclabel{loc:tendermint-2-Lvchooseparent}
                    \State Sign and broadcast $\Block(\FFp \gets p, \FFv \gets v, \FFbmI \gets b, \FFQC \gets Q, \FFtxs \gets 
                        \{ \tx \in \StMtxs{}{} \mid \tx \not\in b_0.\FFtxs \| ... \| b.\FFtxs, \text{ for the path from $b_0$ to $b$} \})$%
                        \algloclabel{loc:tendermint-2-Lvproduceblock}
                \EndIf
            \EndAt
            \At{\myTextHighlighted{$t=12 \Delta v + 4 \Delta$}}%
                \algloclabel{loc:tendermint-2-vote1}
                \If{$\exists b: \MvalidB(\StM{}{}, b) \land (b.\FFv = v)$}%
                        \algloclabel{loc:tendermint-2-validprop}%
                        \Comment{$\MvalidB(\StM{}{}, b) \implies (b.\FFp = \Leader_v)$. Proceed only with one $b$ satisfying the condition.}
                    \If{$\FFv(\StQ{}{}) \leq \FFv(b.\FFQC)$}%
                            \algloclabel{loc:tendermint-2-admissibleprop}
                        \State Sign and broadcast $\Vote(\FFp \gets p, \FFb \gets b, \FFstage \gets 1)$%
                            \algloclabel{loc:tendermint-2-stage1vote}
                    \EndIf
                \EndIf
            \EndAt
            \At{\myTextHighlighted{$t=12 \Delta v + 7 \Delta$}}%
                \algloclabel{loc:tendermint-2-vote2}
                \If{$\exists b, Q: (b.\FFv = v) \land \MvalidQC(\StM{}{}, Q, b, 1)$}%
                    \algloclabel{loc:tendermint-2-validstage1qc}
                    \Comment{Proceed only with one $(b,Q)$ satisfying the condition.}
                    \State $\StQ{}{} \gets Q$%
                        \algloclabel{loc:tendermint-2-updatelock}
                    \State Sign and broadcast $\Vote(\FFp \gets p, \FFb \gets b, \FFstage \gets 2)$%
                        \algloclabel{loc:tendermint-2-stage2vote}
                \EndIf
            \EndAt
            \myTextHighlightedTwo{
            \At{$t=12 \Delta v + 10 \Delta$}%
                \algloclabel{loc:tendermint-2-vote3}
                \If{$\StMtxs{12 \Delta v}{} \subseteq \LOG_{12 \Delta v + 10 \Delta}^p$}%
                    \algloclabel{loc:tendermint-2-liveness}
                    \State Sign and broadcast $\VoteLive(\FFp \gets p, \FFv \gets v)$%
                        \algloclabel{loc:tendermint-2-stage3vote}
                \EndIf
            \EndAt
            }
        \EndFor
    \end{algorithmic}
\end{algorithm}

A variant of the Tendermint consensus protocol~\cite{tendermint} (inspired by~\cite[Sec.\ 9.1]{DBLP:conf/sigecom/BudishLR24}) is provided in \cref{alg:tendermint-2}, described as pseudo-code from the perspective of any honest node $p$.
Like earlier Tendermint versions, the protocol proceeds in \emph{views}.
Each view $v$ is associated with a randomly selected \emph{leader} node $\Leader_v$, known to all nodes.
Conceptually, each view consists of 
a proposal by the leader (\alglocref{alg:tendermint-2}{loc:tendermint-2-propose}), 
followed by rounds of voting (\alglocref{alg:tendermint-2}{loc:tendermint-2-vote1,loc:tendermint-2-vote2,loc:tendermint-2-vote3}).

There are two main differences compared to traditional Tendermint~\cite{tendermint} (\cref{alg:tendermint-original}):
More time is allotted (highlighted \myTextHighlightedDescription{} in \cref{alg:tendermint-2}) for each proposal and voting phase,
and there is an extra third round of voting
(highlighted \myTextHighlightedTwoDescription{} in \cref{alg:tendermint-2}).

Let us discuss the first main difference.
For simplicity, the extra delay before a view's proposal allows
to analyze each view's liveness in isolation:
if the network is synchronous \emph{for the duration of that view $v$} (\emph{synchronous view}),
\ie, messages sent during $12 \Delta v$ to $12 \Delta (v+1) - \Delta$ arrive within $\Delta$ time,
and $\Leader_v$ is honest,
and ``enough'' votes are cast, then a new consensus decision is reached (by all honest nodes).
Furthermore, for liveness alone,
simple $\Delta$ delay would suffice before the proposal,
after the proposal, and after each of the votes.
Our delays of $2\Delta$, $2\Delta$, $3\Delta$, $3\Delta$, $2\Delta$, respectively, are chosen to enable accountable liveness,
as will become clear in \cref{sec:blaming}.

Regarding the second main difference,
note that the extra third round of voting
is different from the earlier two,
in that nodes do not vote for a block,
but indicate whether all transactions recently seen as pending have been confirmed.
These $\VoteLive$ votes are used in \cref{sec:blaming,sec:adjudication}
to detect liveness violations
and trigger the production of certificates of guilt.
The $\VoteLive$ votes are produced but not consumed in \cref{alg:tendermint-2},
and thus 
have no influence on
proposing, voting, or confirming in \cref{alg:tendermint-2},
and can therefore be neglected in the traditional safety and liveness analyses of \cref{alg:tendermint-2} (\cref{thm:tendermint-2-safety,thm:tendermint-2-liveness}).

Regarding notation,
for any protocol state variable $\StX{}{}$, we denote by $\StX{t}{p}$ the state as viewed by (honest) node $p$ at time $t$.
If we set $t=\infty$, we mean the state as viewed at the end of the execution.
We may omit the node if clear from context, such as in \cref{alg:tendermint-2}.
Specifically, 
we denote by $\StM{t}{p}$ the set of message received from the network, upon verification and stripping of the messages' signatures (\alglocref{alg:tendermint-2}{loc:tendermint-2-stateM,loc:tendermint-2-stateM-explain}), as viewed by node $p$ at time $t$,
and
for any $t$ including $\infty$, we define $\StM{t}{\cup} \triangleq \bigcup_{p\in\CPh} \StM{t}{p}$ to denote the union of $\StM{t}{p}$ across all honest nodes.
By $t^-$ we mean the time \emph{just before} any honest node executes its code for time $t$.
By $t^+$ we mean the time \emph{just after} all honest nodes have executed their code for time $t$.

\begin{lemma}
    \label{thm:tendermint-2-safety}
    Assuming $n>3f$, \cref{alg:tendermint-2} is safe
    in partial synchrony 
    (\ie, $\tS = \lfloor (n-1)/3 \rfloor$).
\end{lemma}
\begin{lemma}
    \label{thm:tendermint-2-liveness}
    Assuming $n>3f$, \cref{alg:tendermint-2} is live
    in partial synchrony 
    (\ie, $\tL = \lfloor (n-1)/3 \rfloor$), 
    with expected confirmation latency $O(\Delta)$
    after $\GST$.
\end{lemma}
The proofs are analogous to those for earlier Tendermint variants~\cite{tendermint}~\cite[Sec.\ 9.1]{DBLP:conf/sigecom/BudishLR24} and are therefore relegated to \cref{sec:addendum-consensus}.

Note that \cref{thm:tendermint-2-safety,thm:tendermint-2-liveness} are both under partial synchrony and under the assumption $n>3f$, as is part of our design goal.
In fact, \cref{thm:tendermint-2-safety,thm:tendermint-2-liveness} show that \cref{alg:tendermint-2} satisfies goal~(1) in  \cref{sec:modelprelims-goal}.
In \cref{sec:blaming}, we analyze what we can learn from \emph{individual} synchronous views.
In \cref{sec:adjudication}, we finally leverage the $\netX$-partially-synchronous model, and the \emph{aggregate} combinatorial structure of synchronous views it implies, to arrive at accountable liveness.

\section{Blame Accounting}
\label{sec:blaming}

\begin{figure*}[tbp]%
    \centering%
    \begin{tikzpicture}[x=2cm,y=1cm]
        \footnotesize

        \coordinate (txenters) at (0,0);
        \coordinate (livenessviol) at (1,0);
        \coordinate (viewsknown) at (2,0);
        \coordinate (certifproduced) at (3,0);

        \coordinate (timestart) at ([xshift=-0.5cm]txenters);
        \coordinate (timeend) at ([xshift=1cm]certifproduced);

        \draw [-Latex] (timestart) -- (timeend) node [pos=1,above] {\textsc{Time}};
        \draw ([yshift=3pt]txenters) -- ([yshift=-3pt]txenters) node [below] {$t_0$};
        \draw ([yshift=3pt]livenessviol) -- ([yshift=-3pt]livenessviol) node [below] {$t_1$};
        \draw ([yshift=3pt]viewsknown) -- ([yshift=-3pt]viewsknown) node [below] {$t_2$};
        \draw ([yshift=3pt]certifproduced) -- ([yshift=-3pt]certifproduced) node [below] {$t_3$};

        \draw [-Latex,shorten >=2pt,shorten <=2pt] (livenessviol) to [bend right=35] node[below, sloped,align=center] {Propagate\\transcripts} (viewsknown);
        \draw [-Latex,shorten >=2pt,shorten <=2pt] (viewsknown) to [bend right=35] node[below, sloped,align=center] {Propagate\\accusations} (certifproduced);

        \draw [|-|,shorten <=0.5pt,shorten >=0.5pt] ([yshift=-1.25cm]txenters) -- ([yshift=-1.25cm]livenessviol) node [midway,below] {$\netDeltaPeriod \netG(\netDeltaPeriod)$};
        \draw [|-|,shorten <=0.5pt,shorten >=0.5pt] ([yshift=-1.25cm]livenessviol) -- ([yshift=-1.25cm]viewsknown) node [midway,below] {$\netDeltaPeriod \netG(\netDeltaPeriod)$};
        \draw [|-|,shorten <=0.5pt,shorten >=0.5pt] ([yshift=-1.25cm]viewsknown) -- ([yshift=-1.25cm]certifproduced) node [midway,below] {$\netDeltaPeriod \netG(\netDeltaPeriod)$};

        \draw [Latex-,shorten <=5pt] (certifproduced) -- ++(0,1) -- ++(0.1,0) node [pos=1,right,anchor=north west,yshift=0.8em,align=left] {%
        (4)~Produce certificate of guilt from propagated accusations:\\%
        $\exists p': \exists \Sigma\subseteq\StM{}{}: |\{ p'' \mid \Accusation(\FFp=p'', \FFp'=p') \in \Sigma \}| > n/2$};

        \draw [Latex-,shorten <=5pt] (txenters) -- ++(0,1) -- ++(0.1,0) node [pos=1,right,anchor=north west,yshift=0.8em,align=left] {%
        (1)~Input $\tx$};

        \draw [Latex-,shorten <=5pt] (livenessviol) -- ++(0,3.9) -- ++(0.1,0) node [pos=1,right,anchor=north west,yshift=1em,align=left] {%
        (2)~(Continuously) sign and broadcast $\View(\FFp=p,\FFt=t_1,\FFM=\StM{t_1}{p})$.\\%
        If $\lnot \exists v \in \CV: \exists \CP' \subseteq \CP: (|\CP'| > 2n/3) \land (\forall p'\in\CP': \VoteLive(\FFp=p', \FFv=v) \in \StM{}{})$, then take note of a potential timely-liveness violation at $t_1$.};

        \draw [Latex-,shorten <=5pt] (viewsknown) -- ++(0,2.85) -- ++(0.1,0) node [pos=1,right,anchor=north west,yshift=0.8em,align=left] {%
        (3)~If note was taken of a potential timely-liveness violation at $t_1$, then:\\%
        Among propagated transcripts, for each $p''\in\CP$, retain one $\StM{t_1}{p''}$.\\%
        $\CP' \gets \funLA(\StM{t_1}{p_1}, ..., \StM{t_1}{p_n})$\quad(use $\bot$ as $\StM{t_1}{p_i}$ for any $i$ where no $\StM{t_1}{p_i}$ was received)\\%
        For each $p' \in \CP'$, sign and broadcast $\Accusation(\FFp=p, \FFp'=p')$.};

        \node [align=left,anchor=west] (certif) at ([yshift=-0.5cm]timeend) {Certificate of guilt $\Sigma$};
        \draw [-Latex,shorten >=1pt,shorten <=2pt] (certifproduced) to [bend right=20] (certif.west);

        \draw [decoration=snake,decorate] (txenters) -- (livenessviol) node [midway,below,yshift=-2pt] {$\tx$ unconfirmed};
        
    \end{tikzpicture}%
    \caption[]{%
        Overview of how certificates of guilt for \cref{alg:tendermint-2} are produced,
        from the perspective of a node $p$:
        (2)~%
        Nodes continuously share their transcripts of \cref{alg:tendermint-2} with each other.
        If $p$ has not received $\VoteLive$ from
        $>2n/3$ nodes for 
        the most recent views $\CV$ of \cref{alg:tendermint-2} of the 
        last
        $\netDeltaPeriod\netG(\netDeltaPeriod)$ rounds
        ($\CV$ and $\netDeltaPeriod$ are determined in \cref{sec:blaming-refined,sec:adjudication}),
        then $p$ takes note of a potential timely-liveness violation.
        (3)~%
        Each node retains one transcript for each other node. Since honest transcripts are unique and guaranteed to propagate within $\netDeltaPeriod\netG(\netDeltaPeriod)$ rounds, honest nodes always have the correct transcript for other honest nodes.
        If $p$ has taken note of a potential timely-liveness violation, then:
        A function $\funLA$, the centerpiece of liveness accountability developed in \cref{sec:blaming,sec:adjudication}, is used to identify guilty nodes (and never mis-identifies honest nodes).
        Accusations for these nodes are shared among honest nodes.
        (4)~%
        Accusations against a particular node by a majority of nodes constitute a certificate of guilt.
        (1)~%
        If a transaction remains unconfirmed for $\netDeltaPeriod\netG(\netDeltaPeriod)$ rounds, then $2\netDeltaPeriod\netG(\netDeltaPeriod)$ rounds later a certificate of guilt for ``many'' adversary nodes is produced.
    }%
    \label{fig:overview1}%
\end{figure*}%

In \cref{sec:blaming,sec:adjudication}, 
we describe how certificates of guilt are produced
so as to render the consensus protocol of \cref{alg:tendermint-2} accountably live
according to \cref{def:accountable-liveness}.
A high-level overview of the accountability process, which nodes run in parallel to \cref{alg:tendermint-2}, is given in \cref{fig:overview1}.

At all times, 
nodes take note of potential timely-liveness violations (\cref{fig:overview1}, step~(2)),
based on the lack of $2n/3$-quorums of $\VoteLive$ messages
for the most recent views $\CV$ of \cref{alg:tendermint-2} 
of the recent $\netDeltaPeriod\netG(\netDeltaPeriod)$ rounds
(final determination of $\netDeltaPeriod$ and $\CV$ is made in \cref{sec:blaming-refined,sec:adjudication};
for now, think of them as $12 \Delta$ and as the most recent $\netG(\netDeltaPeriod)$ views in \cref{alg:tendermint-2}, respectively).
Honest nodes sign and broadcast their \emph{transcripts} of \cref{alg:tendermint-2}, which means their $\StM{}{}$, 
on a continuously ongoing basis.\footnote{There are various ways to reduce the communication overhead of this step, which we leave to future work, to retain simplicity, and since our primary focus here is to show the achievability of liveness accountability, not its most efficient implementation.}

Assuming $\netX$-partial-synchrony, these transcripts propagate to all honest nodes within $\netDeltaPeriod\netG(\netDeltaPeriod)$ time (\cref{fig:overview1}, step~(3)).
Honest nodes discard equivocating, invalidly-signed, and syntactically-malformed transcripts,
and use a default of $\bot$ for missing transcripts,
so that for each node and time they retain exactly one transcript.
If an honest node has taken note of a potential timely-liveness violation,
the node applies a function $\funLA$
that is the centerpiece of liveness accountability developed in \cref{sec:blaming,sec:adjudication}.
Namely, $\funLA$ is applied to the sanitized transcripts,
to obtain a set $\CP'$ of seemingly guilty nodes, for which the node then signs and broadcasts accusations.

Again assuming $\netX$-partial-synchrony, these accusations propagate to all honest nodes within $\netDeltaPeriod\netG(\netDeltaPeriod)$ time (\cref{fig:overview1}, step~(4)).
If a node was accused for a particular point in time by a majority of nodes,
then that constitutes a certificate of guilt for that node.

Note that in the above process, the adversary may cook up or alter its transcripts, subject to not being able to forge signatures.
Honest nodes do not attempt to filter transcripts \emph{semantically}, \ie, based on the information contained---rather, $\funLA$ will take care of that.
Note that when $\funLA$ is invoked by an honest node, it is guaranteed that the proper transcripts of all honest nodes are provided to it.
\Cref{sec:blaming,sec:adjudication} are all about how to design $\funLA$ such that,
under those circumstances, $\funLA$ never outputs an honest node,
and $\funLA$ outputs a ``large'' set of adversary nodes whenever
a transaction has remained unconfirmed for more than $\netDeltaPeriod\netG(\netDeltaPeriod)$ time
(\cref{fig:overview1}, step~(1)).

The $\funLA$ we construct 
proceeds in two steps:
(1) \emph{Blame accounting} (\cref{sec:blaming}):
First, $\funLA$ uses the $\{ \StM{t}{p} \}$ to 
count how frequently a node $p$ did not see a vote from node $p'$
that $p'$ ``ought to'' have sent, where ``ought to'' is determined 
based on $\StM{t}{p}$ and the assumption that the network was synchronous.
We say that $p$ \emph{blames} $p'$ for unexplained missing votes
that may have led to a timely-liveness violation.
Note that these blame counts in isolation may be inaccurate,
but form the basis for the next step.
(2) \emph{Adjudication rule} (\cref{sec:adjudication}):
Second, $\funLA$ leverages the assumptions that, roughly,
``not too many'' nodes are adversary
and that the network is ``often synchronous'' ($\netX$-partial-synchrony),
and as a result, adversary nodes draw more blame than honest nodes,
and an \emph{adjudication rule} can be used
to reliably identify adversary nodes based on the blame counts.

\subsection{Simple Blame Accounting}
\label{sec:blaming-simple}

\begin{algorithm}[tbp]
    \caption{Blame accounting for \cref{alg:tendermint-2}, given views $\CV$}
    \label{alg:tendermint-2-alproto}
    \begin{algorithmic}[1]
        \algfontsize%
        \State $\StM{t}{p} \gets \StM{}{}$ of \alglocref{alg:tendermint-2}{loc:tendermint-2-stateM}, for node $p$ and ``receive-time-annotated'', \ie, supports access to $\StM{}{}$ in the view of $p$ of \cref{alg:tendermint-2} ``as of'' any time $t$
        \State $\forall p\in\CP, v\in\CV, p'\in\CP: \BlameNo{p}{v}{p'}$%
            \Comment{Default: no blame}
            \For{$p\in\CP, v\in\CV, p'\in\CP$}
                \If{$\exists b \in \StM{12 \Delta v + 3\Delta}{p}: \MvalidB(\StM{12 \Delta v + 3\Delta}{p}, b) \land (b.\FFv = v)$}%
                        \algloclabel{loc:tendermint-2-alproto-validprop}%
                        \algloclabel{loc:tendermint-2-alproto-l4}%
                        \Comment{Assuming synchrony, if $\Leader_v$ has sent a valid proposal $b$ for view $v$ in time ...}
                    \If{$\lnot \exists Q \subseteq \StM{12 \Delta v + 1 \Delta}{p}, b' \in \StM{12 \Delta v + 1 \Delta}{p}:
                    \MvalidQC(\StM{12 \Delta v + 1 \Delta}{p}, Q, b', 1) \land (\FFv(Q) > \FFv(b.\FFQC))$}
                            \algloclabel{loc:tendermint-2-alproto-admissibleprop}%
                            \algloclabel{loc:tendermint-2-alproto-l5}%
                            \Comment{... and $p'$ cannot possibly have had an excuse not to stage-$1$ vote for the proposal ...}
                        \If{$\lnot \exists w \in \StM{12 \Delta v + 5 \Delta}{p}, b'' \in \StM{12 \Delta v + 5 \Delta}{p}: \MvalidV(\StM{12 \Delta v + 5 \Delta}{p}, w) \land (w.\FFp = p') \land (w.\FFb = b'') \land (w.\FFstage = 1) \land (b''.\FFv = v)$}%
                                \algloclabel{loc:tendermint-2-alproto-didntvote-1}%
                                \algloclabel{loc:tendermint-2-alproto-l6}%
                                \Comment{... and yet $p'$ did not cast any valid stage-$1$ vote for any valid proposal for view $v$ ...}
                            \State 
                                $\BlameYes{p}{v}{p'}$%
                                \algloclabel{loc:tendermint-2-alproto-blame-1}%
                                \Comment{... then $p'$ deserves blame!}
                        \EndIf
                    \EndIf
                \EndIf
                \If{$\exists b \in \StM{12 \Delta v + 6 \Delta}{p}, Q \subseteq \StM{12 \Delta v + 6 \Delta}{p}:
                            (b.\FFv = v)
                            \land \MvalidQC(\StM{12 \Delta v + 6 \Delta}{p}, Q, b, 1)$}%
                        \algloclabel{loc:tendermint-2-alproto-validqc}%
                        \algloclabel{loc:tendermint-2-alproto-l8}%
                        \Comment{Assuming synchrony, if there was a valid proposal $b$ for $v$, and there was a valid stage-$1$ QC for $b$ ...}
                    \If{$\lnot \exists w \in \StM{12 \Delta v + 8 \Delta}{p}, b' \in \StM{12 \Delta v + 8 \Delta}{p}, Q' \subseteq \StM{12 \Delta v + 8 \Delta}{p}: \MvalidV(\StM{12 \Delta v + 8 \Delta}{p}, w) \land (w.\FFp = p') \land (w.\FFb = b') \land (w.\FFstage = 2) \land (b.\FFv = v) \land \MvalidQC(\StM{12 \Delta v + 8 \Delta}{p}, Q, b, 1)$}%
                            \algloclabel{loc:tendermint-2-alproto-didntvote-2}%
                            \algloclabel{loc:tendermint-2-alproto-l9}%
                            \Comment{... and yet $p'$ did not cast any valid stage-$2$ vote for any valid proposal $b'$ for view $v$ for which there was a valid stage-$1$ QC ...}
                        \State 
                            $\BlameYes{p}{v}{p'}$%
                            \algloclabel{loc:tendermint-2-alproto-blame-2}%
                            \Comment{... then $p'$ deserves blame!}
                    \EndIf
                \EndIf
                \If{$\StMtxs{12 \Delta v + 1}{p} \subseteq \LOG_{12 \Delta v + 9 \Delta}^p$}%
                        \algloclabel{loc:tendermint-2-alproto-liveness}%
                        \algloclabel{loc:tendermint-2-alproto-l11}%
                        \Comment{Assuming synchrony, if all transactions pending at the beginning of the view were confirmed within the view ...}
                    \If{$\lnot \exists w' \in \StM{12 \Delta v + 11 \Delta}{p}: (w'.\FFp = p') \land (w'.\FFv = v)$}
                            \algloclabel{loc:tendermint-2-alproto-didntvote-3}%
                            \algloclabel{loc:tendermint-2-alproto-l12}%
                            \Comment{... and yet $p'$ did not cast a $\VoteLive$ vote for view $v$ ...}
                        \State 
                            $\BlameYes{p}{v}{p'}$%
                            \algloclabel{loc:tendermint-2-alproto-blame-3}%
                            \Comment{... then $p'$ deserves blame!}
                    \EndIf
                \EndIf
            \EndFor
    \end{algorithmic}
\end{algorithm}

Simple \emph{blame accounting} for \cref{alg:tendermint-2} is provided in \cref{alg:tendermint-2-alproto}.
It is invoked with a set of views $\CV$.
The final choice of $\CV$ is made in \cref{sec:blaming-refined,sec:adjudication}.
For the remainder of \cref{sec:blaming-simple},
think of $\CV$ 
as the set of the $\netG(\netDeltaPeriod)$ most recent views of \cref{alg:tendermint-2},
with $\netDeltaPeriod = 12 \Delta$.
This choice and $\netX$-partial-synchrony gives us that at most $\netX$ fraction of views in $\CV$ are not synchronous 
(\ie, violate the network-delay upper-bound $\Delta$).
In the following, we observe how \cref{alg:tendermint-2-alproto} behaves for individual views $v\in\CV$, irrespective of the choice of $\CV$.

\begin{lemma}
    \label{lem:tendermint-2-advaccountability}
    With \cref{alg:tendermint-2,alg:tendermint-2-alproto},
    for every $v\in\CV$,
    if the network is synchronous for view $v$, and $\Leader_v$ is honest,
    then:
    unless the block $b$ proposed by $\Leader_v$ is confirmed by all honest nodes by $12 \Delta v + 9 \Delta$
    \emph{and} all honest nodes see $\VoteLive$ messages for $v$ from $>2n/3$ nodes by $12 \Delta v + 12 \Delta$,
    there is a set $\CP'$ with $|\CP'| \geq n/3$,
    such that
    for every $p\in\CPh$,
    for all $p'\in\CP'$,
    $\Blame{p}{v}{p'} = 1$ in \cref{alg:tendermint-2-alproto}.
\end{lemma}

Intuitively, this means that if conditions are ``very good'' during view $v$,
but there is (or appears to be) no liveness, then there is a large set $\CP'$ of nodes that all honest nodes blame simultaneously.

\begin{lemma}
    \label{lem:tendermint-2-honacquittal}
    With \cref{alg:tendermint-2,alg:tendermint-2-alproto},
    for every $v\in\CV$,
    if the network is synchronous for view $v$,
    then 
    for every $p\in\CPh$,
    for every $p'\in\CPh$,
    $\Blame{p}{v}{p'} = 0$ in \cref{alg:tendermint-2-alproto}.
\end{lemma}

Intuitively, this means that if conditions are ``good'' during view $v$,
then honest nodes do not blame each other.

Before we proceed to the proofs of \cref{lem:tendermint-2-advaccountability,lem:tendermint-2-honacquittal},
let us explain the intuition behind the design of \cref{alg:tendermint-2-alproto} and its relation to \cref{alg:tendermint-2}.
Recall that compared to traditional Tendermint variants~\cite{tendermint}, \cite[Sec.\ 9.1]{DBLP:conf/sigecom/BudishLR24}, which have $\Delta$ delay after the proposal and after each round of voting,
\cref{alg:tendermint-2} also has delay \emph{before} the proposal.
This pedagogical simplification, which can be discharged, 
allows to treat the liveness of each view independently;
it entails that the timeframes of the decision processes of different views do not overlap: 
If the network is synchronous for a view and the view's leader is honest,
then the delay before the proposal ensures that the proposal is such that it is viewed as admissible (\alglocref{alg:tendermint-2}{loc:tendermint-2-admissibleprop}) to vote for by all honest nodes.
Subsequent delays after the proposal and after each round of voting ensure that, assuming synchrony, all honestly produced messages reach all honest nodes in time to proceed with the next protocol phase and to ultimately lead to a new confirmed block.

\Cref{alg:tendermint-2} increases the phase-separating delays from $\Delta$, $\Delta$, $\Delta$, $\Delta$ to $2\Delta$, $2\Delta$, $3\Delta$, $3\Delta$, $2\Delta$.
The key reason for increasing the delays 
is:
Assuming synchrony, if an honest node $p$ sees a message $m$ by time $t$,
then it knows that all nodes must see $m$ by time $t+\Delta$, due to $p$ re-broadcasting $m$ at $t$.
Vice versa, if $p$ does not see $m$ by time $t$, it knows that no honest node can have seen $m$ by time $t-\Delta$.
Thus, an honest node's $\StM{t}{p}$ serves as an inner and outer bound (in the sense of subsets) on the message sets of any other (purportedly honest) node shortly after or before, respectively.
Thus, from $\StM{t}{p}$ of an honest node $p$,
assuming synchronous view $v$,
one can infer (part of) the internal state of another (purportedly honest) node $p'$ as of $t-\Delta$, to the extent that the state affects the sending/not-sending of messages, which is observable in $\StM{t}{p}$.
For instance, \alglocref{alg:tendermint-2-alproto}{loc:tendermint-2-alproto-admissibleprop} infers the ``highest'' (by view) lock $\StQ{}{p'}$ any honest $p'$ could have had by $12 \Delta v$ (and, since locks are updated only at \alglocref{alg:tendermint-2}{loc:tendermint-2-updatelock}, by $12 \Delta v + 4\Delta$, which is critical for the check of \alglocref{alg:tendermint-2}{loc:tendermint-2-admissibleprop} to pass).

One can also, assuming synchronous view $v$, infer from $\StM{t}{p}$ for honest $p$ what another (purportedly honest) $p'$ ``ought to'' do at $t+\Delta$, to the extent that such action is affected by the receipt/non-receipt of messages.
For instance, \alglocref{alg:tendermint-2-alproto}{loc:tendermint-2-alproto-validprop} infers that $p'$ receives a valid timely proposal for view $v$ (\alglocref{alg:tendermint-2}{loc:tendermint-2-validprop}) and thus ``ought to'' stage-$1$ vote for it in \alglocref{alg:tendermint-2}{loc:tendermint-2-stage1vote}.

Since there is $\Delta$ time lag in the backward and forward direction in the arguments above, respectively, this explains why most of the $\Delta$ delays in the original Tendermint protocol turn into $2\Delta$ delays in \cref{alg:tendermint-2} when liveness accountability is sought.
The $3\Delta$ delay between the first and second round of voting in \cref{alg:tendermint-2} is needed because there is ``an extra round of indirection'' between one honest node seeing a stage-$1$ QC and all honest nodes seeing it.
Specifically, the check of \alglocref{alg:tendermint-2-alproto}{loc:tendermint-2-alproto-validqc} needs to be $\Delta$ after the check of \alglocref{alg:tendermint-2-alproto}{loc:tendermint-2-alproto-didntvote-1} (to ensure that if \alglocref{alg:tendermint-2-alproto}{loc:tendermint-2-alproto-validqc} fails, there is a large set of nodes that are blamed by \emph{all} honest nodes in \alglocref{alg:tendermint-2-alproto}{loc:tendermint-2-alproto-didntvote-1}),
which in turn must be $\Delta$ after \alglocref{alg:tendermint-2}{loc:tendermint-2-stage1vote}, and
\alglocref{alg:tendermint-2-alproto}{loc:tendermint-2-alproto-validqc}
needs to be $\Delta$ before \alglocref{alg:tendermint-2}{loc:tendermint-2-stage2vote} (to ensure that if \alglocref{alg:tendermint-2-alproto}{loc:tendermint-2-alproto-validqc} passes, all nodes ``ought to'' stage-$2$ vote in \alglocref{alg:tendermint-2}{loc:tendermint-2-stage2vote}).
The case of block confirmation (based on stage-$1$ and stage-$2$ QCs) and a quorum of $\VoteLive$ votes from $>2n/3$ nodes is analogous;
hence the $3\Delta$ delay there as well.

\deferredproof{proofsblaming}{lem:tendermint-2-advaccountability}
\begin{defer}{proofsblaming}
\begin{proof}[Proof of \cref{lem:tendermint-2-advaccountability}]
    Assume the network is synchronous for view $v$
    and $\Leader_v$ is honest and proposes $b$.
    We consider two cases:
    (A)~Some honest node does not view $b$ as confirmed by $12 \Delta v + 9 \Delta$.
    (B)~Every honest node views $b$ as confirmed by $12 \Delta v + 9 \Delta$,
    but some honest nodes does not see $\VoteLive$ messages for $v$ from $>2n/3$ nodes by $12 \Delta v + 12 \Delta$.
    We show that in both cases there is a set $\CP'$ with $|\CP'| \geq n/3$ such that
    for every honest node $p$,
    for all $p\in\CP'$,
    $\Blame{p}{v}{p'} = 1$ in \cref{alg:tendermint-2-alproto}.

    First, case~(A). 
    Let $q$ be an honest node that does not view $b$ as confirmed by $12 \Delta v + 9 \Delta$.
    Since the network is synchronous and $\Leader_v$ is honest,
    the condition of \alglocref{alg:tendermint-2-alproto}{loc:tendermint-2-alproto-validprop}
    is met 
    for
    every honest node $p$.
    The condition of \alglocref{alg:tendermint-2-alproto}{loc:tendermint-2-alproto-admissibleprop} must be met 
    for
    every honest node because $\Leader_v$ is honest and chooses $(b, Q)$ according to \alglocref{alg:tendermint-2}{loc:tendermint-2-Lvchooseparent}, and the network is synchronous.
    Since $q$ does not view $b$ as confirmed by $12 \Delta v + 9 \Delta$,
    but $b$ is viewed as valid by every honest node
    after $12 \Delta v + 3 \Delta$ due to network synchrony and $\Leader_v$ being honest,
    either (a)
    there must be a set $\CP'$ with $|\CP'| \geq n/3$ such that
    $q$ has not seen a valid stage-$1$ vote for $b$ from any $p' \in \CP'$ by $12 \Delta v + 9 \Delta$, 
    and thus
    no honest node has seen a valid stage-$1$ vote for $b$ from any $p' \in \CP'$ by $12 \Delta v + 8 \Delta$ (and thus also not by $12 \Delta v + 6 \Delta$ or by $12 \Delta v + 5 \Delta$),
    or (b)
    there must be a set $\CP'$ with $|\CP'| \geq n/3$ such that
    $q$ has not seen a valid stage-$2$ vote for $b$ from any $p' \in \CP'$ by $12 \Delta v + 9 \Delta$,
    and thus
    no honest node has seen a valid stage-$2$ vote for $b$ from any $p' \in \CP'$ by $12 \Delta v + 8 \Delta$.

    Suppose (a) holds.
    Then, since $\Leader_v$ is honest and $b$ is its unique proposal for view $v$, the condition of
    \alglocref{alg:tendermint-2-alproto}{loc:tendermint-2-alproto-didntvote-1}
    is met 
    for
    every honest node $p$ for all $p' \in \CP'$.
    Then,
    for every honest node $p$,
    $\BlameYes{p}{v}{p'}$
    in \alglocref{alg:tendermint-2-alproto}{loc:tendermint-2-alproto-blame-1} for all those $p' \in \CP'$, as desired.

    Suppose (b) holds.
    Suppose there is some honest node $r$ 
    for which
    the condition of \alglocref{alg:tendermint-2-alproto}{loc:tendermint-2-alproto-validqc} 
    is 
    not satisfied.
    Then, by synchrony and since $\Leader_v$ is honest, 
    there must be a set $\CP'$ with $|\CP'| \geq n/3$ such that
    no honest node has seen a valid stage-$1$ vote for $b$ from any $p' \in \CP'$ by $12 \Delta v + 5 \Delta$,
    and we are done by the same argument as for (a).
    We thus assume that the condition of \alglocref{alg:tendermint-2-alproto}{loc:tendermint-2-alproto-validqc} holds for every honest node.
    Then, by (b),
    there must be a set $\CP'$ with $|\CP'| \geq n/3$ such that
    the condition of \alglocref{alg:tendermint-2-alproto}{loc:tendermint-2-alproto-didntvote-2} is met for all $p' \in \CP'$ 
    for
    every honest node $p$, so that 
    $\BlameYes{p}{v}{p'}$
    in \alglocref{alg:tendermint-2-alproto}{loc:tendermint-2-alproto-blame-2} for all $p' \in \CP'$, as desired.

    Now consider case~(B).
    Suppose $q$ is an honest node that does not see $\VoteLive$ messages for $v$ from $>2n/3$ nodes by $12 \Delta v + 12 \Delta$,
    but all honest nodes view $b$ as confirmed by $12 \Delta v + 9 \Delta$.
    By synchrony and since $\Leader_v$ is honest and since all honest nodes view $b$ as confirmed by $12 \Delta v + 9 \Delta$ by assumption,
    for every honest node $p$,
    $\StMtxs{12 \Delta v + 1}{p} \subseteq \StMtxs{12 \Delta v + 2}{\Leader_v} \subseteq \LOG_{12 \Delta v + 9 \Delta}^p$.
    (Note that a node $r$ can determine the confirmed output log of another node $s$ as per any round $t$ based on the transcript of $s$.)
    Thus, 
    \alglocref{alg:tendermint-2-alproto}{loc:tendermint-2-alproto-liveness}
    is met for every honest node $p$.
    Since $q$ does not see $\VoteLive$ messages for $v$ from $>2n/3$ nodes by $12 \Delta v + 12 \Delta$,
    by synchrony,
    there must be a set $\CP'$ with $|\CP'| \geq n/3$ such that
    no honest node has seen a $\VoteLive$ messages for $v$ from any $p' \in \CP'$ by $12 \Delta v + 11 \Delta$.
    Thus, for every honest node $p$ and every $p' \in \CP'$,
    the condition of
    \alglocref{alg:tendermint-2-alproto}{loc:tendermint-2-alproto-didntvote-3}
    holds, so that
    $\BlameYes{p}{v}{p'}$
    in \alglocref{alg:tendermint-2-alproto}{loc:tendermint-2-alproto-blame-3}, as desired.
\end{proof}
\end{defer}

\deferredproof{proofsblaming}{lem:tendermint-2-honacquittal}
\begin{defer}{proofsblaming}
\begin{proof}[Proof of \cref{lem:tendermint-2-honacquittal}]
    Suppose, for contradiction, that the network is synchronous for view $v$
    but 
    for some honest $p$ and for some honest $p'$,
    $\Blame{p}{v}{p'} = 1$ in \cref{alg:tendermint-2-alproto}.
    Then, 
    $\BlameYes{p}{v}{p'}$
    either (a) in \alglocref{alg:tendermint-2-alproto}{loc:tendermint-2-alproto-blame-1},
    (b) in \alglocref{alg:tendermint-2-alproto}{loc:tendermint-2-alproto-blame-2},
    or (c) in \alglocref{alg:tendermint-2-alproto}{loc:tendermint-2-alproto-blame-3}.

    Suppose (a) holds.
    Then, it must be that the conditions of
    \alglocref{alg:tendermint-2-alproto}{loc:tendermint-2-alproto-validprop} and
    \alglocref{alg:tendermint-2-alproto}{loc:tendermint-2-alproto-admissibleprop}
    are met 
    for
    $p$.
    But then, by the assumption of synchrony,
    in the view of $p'$,
    the conditions of
    \alglocref{alg:tendermint-2}{loc:tendermint-2-validprop} and
    \alglocref{alg:tendermint-2}{loc:tendermint-2-admissibleprop}
    were met, so that $p'$ cast a valid stage-$1$ vote
    for a block proposed in view $v$,
    and thus, again by synchrony,
    the condition of 
    \alglocref{alg:tendermint-2-alproto}{loc:tendermint-2-alproto-didntvote-1}
    cannot be met 
    for
    $p$,
    a contradiction to 
    $\BlameYes{p}{v}{p'}$
    in \alglocref{alg:tendermint-2-alproto}{loc:tendermint-2-alproto-blame-1}.

    Suppose (b) holds.
    Then, it must be that the condition of
    \alglocref{alg:tendermint-2-alproto}{loc:tendermint-2-alproto-validqc}
    is met 
    for
    $p$.
    But then, by the assumption of synchrony,
    the condition of \alglocref{alg:tendermint-2}{loc:tendermint-2-validstage1qc}
    was met in the view of $p'$,
    so that $p'$ cast a valid stage-$2$ vote
    for a valid block $b$
    by $12 \Delta v + 7 \Delta$, which, again by synchrony, $p$ must have seen by $12 \Delta v + 8 \Delta$,
    so that the condition of
    \alglocref{alg:tendermint-2-alproto}{loc:tendermint-2-alproto-didntvote-2}
    cannot be met 
    for
    $p$,
    a contradiction to 
    $\BlameYes{p}{v}{p'}$
    in \alglocref{alg:tendermint-2-alproto}{loc:tendermint-2-alproto-blame-2}.

    Suppose (c) holds.
    Then, it must be that the condition of
    \alglocref{alg:tendermint-2-alproto}{loc:tendermint-2-alproto-liveness}
    is met 
    for
    $p$.
    But then, by the assumption of synchrony,
    $\StMtxs{12 \Delta v}{p'} \subseteq \StMtxs{12 \Delta v + 1}{p} \subseteq \LOG_{12 \Delta v + 9 \Delta}^p \subseteq \LOG_{12 \Delta v + 10 \Delta}^{p'}$,
    so that
    the condition of \alglocref{alg:tendermint-2}{loc:tendermint-2-liveness}
    was met in the view of $p'$,
    and $p'$ cast a $\VoteLive$ vote
    for view $v$
    by $12 \Delta v + 10 \Delta$, which, again by synchrony, $p$ must have seen by $12 \Delta v + 11 \Delta$,
    so that the condition of
    \alglocref{alg:tendermint-2-alproto}{loc:tendermint-2-alproto-didntvote-3}
    cannot be met 
    for
    $p$,
    a contradiction to 
    $\BlameYes{p}{v}{p'}$
    in \alglocref{alg:tendermint-2-alproto}{loc:tendermint-2-alproto-blame-3}.
\end{proof}
\end{defer}

\subsection{Refined Blame Accounting}
\label{sec:blaming-refined}

\begin{algorithm}[tbp]
    \caption{Refined blame accounting for \cref{alg:tendermint-2}, given super-views $\{ \CV_u \}_{u\in\CU}$ (changes over \cref{alg:tendermint-2-alproto} highlighted \myTextHighlightedDescription{})}
    \label{alg:tendermint-2-alproto-multishot}
    \begin{algorithmic}[1]
        \algfontsize%
        \State $\StM{t}{p} \gets \StM{}{}$ of \alglocref{alg:tendermint-2}{loc:tendermint-2-stateM}, for node $p$ and ``receive-time-annotated'', \ie, supports access to $\StM{}{}$ in the view of $p$ of \cref{alg:tendermint-2} ``as of'' any time $t$
        \State $\forall p\in\CP, \myTextHighlighted{u\in\CU}, p'\in\CP: \BlameNo{p}{\myTextHighlighted{u}}{p'}$%
            \For{$p\in\CP, \myTextHighlighted{u\in\CU, v\in\CV_u}, p'\in\CP$}
                \GenericBlock{\emph{... same as \alglocref{alg:tendermint-2-alproto}{loc:tendermint-2-alproto-l4} ...}}%
                        \algloclabel{loc:tendermint-2-alproto-multishot-l4}
                    \GenericBlock{\emph{... same as \alglocref{alg:tendermint-2-alproto}{loc:tendermint-2-alproto-l5} ...}}%
                            \algloclabel{loc:tendermint-2-alproto-multishot-l5}
                        \GenericBlock{\emph{... same as \alglocref{alg:tendermint-2-alproto}{loc:tendermint-2-alproto-l6} ...}}%
                                \algloclabel{loc:tendermint-2-alproto-multishot-l6}
                            \State 
                                $\BlameYes{p}{\myTextHighlighted{u}}{p'}$%
                                \algloclabel{loc:tendermint-2-alproto-multishot-blame-1}%
                                \algloclabel{loc:tendermint-2-alproto-multishot-l7}
                        \EndGenericBlock
                    \EndGenericBlock
                \EndGenericBlock
                \GenericBlock{\emph{... same as \alglocref{alg:tendermint-2-alproto}{loc:tendermint-2-alproto-l8} ...}}%
                        \algloclabel{loc:tendermint-2-alproto-multishot-l8}
                    \GenericBlock{\emph{... same as \alglocref{alg:tendermint-2-alproto}{loc:tendermint-2-alproto-l9} ...}}%
                            \algloclabel{loc:tendermint-2-alproto-multishot-l9}
                        \State 
                            $\BlameYes{p}{\myTextHighlighted{u}}{p'}$%
                            \algloclabel{loc:tendermint-2-alproto-multishot-blame-2}%
                            \algloclabel{loc:tendermint-2-alproto-multishot-l10}
                    \EndGenericBlock
                \EndGenericBlock
                \GenericBlock{\emph{... same as \alglocref{alg:tendermint-2-alproto}{loc:tendermint-2-alproto-l11} ...}}%
                        \algloclabel{loc:tendermint-2-alproto-multishot-l11}
                    \GenericBlock{\emph{... same as \alglocref{alg:tendermint-2-alproto}{loc:tendermint-2-alproto-l12} ...}}%
                            \algloclabel{loc:tendermint-2-alproto-multishot-l12}
                        \State 
                            $\BlameYes{p}{\myTextHighlighted{u}}{p'}$%
                            \algloclabel{loc:tendermint-2-alproto-multishot-blame-3}%
                            \algloclabel{loc:tendermint-2-alproto-multishot-l13}
                    \EndGenericBlock
                \EndGenericBlock
            \EndFor
    \end{algorithmic}
\end{algorithm}

Looking at \cref{lem:tendermint-2-advaccountability,lem:tendermint-2-honacquittal},
the blame accounting of \cref{alg:tendermint-2-alproto}
has a drawback.
Based on the network model, views of \cref{alg:tendermint-2} can be categorized into three groups.
Either (a) the network is asynchronous for the view,
or (b) the network is synchronous for the view but the view's leader $\Leader_v$ is adversary,
or (c) the network is synchronous for the view and the view's leader $\Leader_v$ is honest.
\Cref{lem:tendermint-2-advaccountability,lem:tendermint-2-honacquittal} make no guarantee for (a)---which seems unavoidable.
\Cref{lem:tendermint-2-honacquittal} protects honest nodes from blame during (b) and (c)---which seems the best we can hope for.
But \cref{lem:tendermint-2-advaccountability} ensures blame for adversary nodes, if there is no liveness, only during (c).
During (b), the adversary ``gets away'' without blame.
While, qualitatively, this still works, quantitatively, it reduces the number of adversarial nodes we can hold responsible in the event of a liveness violation.

The refined blame accounting of \cref{alg:tendermint-2-alproto-multishot} 
improves this.
It assumes a partition $\{ \CV_u \}_{u\in\CU}$ of views into \emph{super-views}. 
If the network is synchronous for a super-view $u\in\CU$,
then 
honest nodes are protected from blame,
and
as long as \emph{any} view $v\in\CV_u$ has an honest leader $\Leader_v$,
``many'' adversary nodes are blamed by all honest nodes for the super-view $u$ if there is no liveness.
Foreshadowing \cref{sec:adjudication},
$\{ \CV_u \}_{u\in\CU}$ is chosen as follows:
For $\netDeltaPeriod = 12 \Delta \superviewLen$ for some fixed parameter $\superviewLen$,
choose the $\netG(\netDeltaPeriod)$ most recent periods of length $\netDeltaPeriod$ of the execution, aligned to view boundaries in \cref{alg:tendermint-2}, as super-views $\CU$. Each of these periods $u$ corresponds to $\superviewLen$ views of \cref{alg:tendermint-2}, which make $\CV_u$.
Finally, $\CV \triangleq \bigcup_{u\in\CU} \CV_u$.
Note that even for moderate $\superviewLen$, \emph{most} super-views $u$
have at least \emph{some} view $v\in\CV_u$ with honest $\Leader_v$,
due to the random choice of $\Leader_v$.
This and $\netX$-partial-synchrony will give us that almost $(1-x)$ fraction of super-views in $\CU$ are synchronous \emph{and} have some honest leader.
Revisiting the earlier categorization of views, for super-views, this eliminates the problematic group (b) and leaves us to deal only with (a) and (c).
In the following, we observe how \cref{alg:tendermint-2-alproto-multishot} behaves for individual super-views $u\in\CU$, irrespective of the choice of $\{ \CV_u \}_{u\in\CU}$.
Changes in \cref{lem:tendermint-2-advaccountability-multishot,lem:tendermint-2-honacquittal-multishot} over \cref{lem:tendermint-2-advaccountability,lem:tendermint-2-honacquittal} are 
highlighted 
\myTextHighlightedDescription{}.

\begin{lemma}
    \label{lem:tendermint-2-advaccountability-multishot}
    With \cref{alg:tendermint-2,alg:tendermint-2-alproto-multishot},
    for every \myTextHighlighted{super-view $u\in\CU$},
    if 
    \myTextHighlighted{the network is synchronous for \emph{all} views $\CV_u$,
    and $\Leader_v$ is honest for \emph{some} view $v\in\CV_u$,}
    then:
    unless the block $b$ proposed by $\Leader_v$ is confirmed by all honest nodes by $12 \Delta v + 9 \Delta$
    \emph{and} all honest nodes see $\VoteLive$ messages for $v$ from $>2n/3$ nodes by $12 \Delta v + 12 \Delta$,
    there is a set $\CP'$ with $|\CP'| \geq n/3$,
    such that
    for every $p\in\CPh$,
    for all $p'\in\CP'$,
    $\Blame{p}{\myTextHighlighted{u}}{p'} = 1$ in \cref{alg:tendermint-2-alproto-multishot}.
\end{lemma}
\begin{proof}
    Follows the same steps as the proof of \cref{lem:tendermint-2-advaccountability}.
\end{proof}

\begin{lemma}
    \label{lem:tendermint-2-honacquittal-multishot}
    With \cref{alg:tendermint-2,alg:tendermint-2-alproto-multishot},
    for every \myTextHighlighted{super-view $u\in\CU$},
    if
    \myTextHighlighted{the network is synchronous for \emph{all} views $\CV_u$,}
    then 
    for every $p\in\CPh$,
    for every $p'\in\CPh$,
    $\Blame{p}{\myTextHighlighted{u}}{p'} = 0$ in \cref{alg:tendermint-2-alproto-multishot}.
\end{lemma}
\begin{proof}
    Follows directly from \cref{lem:tendermint-2-honacquittal}.
\end{proof}

\section{Adjudication Rule}
\label{sec:adjudication}

\label{sec:adjudication-complex}

\begin{algorithm}[tbp]
    \caption{Critical-Subsets adjudication rule for \cref{alg:tendermint-2,alg:tendermint-2-alproto-multishot}, given super-views $\{ \CV_u \}_{u\in\CU}$, $\tALmax < n/2$, $\netX < 1/2$}
    \label{alg:adjudication-criticalsubsets1}
    \begin{algorithmic}[1]
        \algfontsize%
        \State Compute $\Blame{p}{u}{p'}$ for all $p\in\CP,p'\in\CP,u\in\CU$ using \cref{alg:tendermint-2-alproto-multishot}
        \State $\forall u\in\CU: \critSubsetsCPA_u \gets \{ p'\in\CP \mid \sum_{p\in\CP} \Blame{p}{u}{p'} \geq n-\tALmax \}$
        \State Undirected graph $\critSubsetsG \gets (\CU, \{ \{u,u'\} \mid |\critSubsetsCPA_u \cap \critSubsetsCPA_{u'}| \geq 2n/3 - \tALmax \})$
        \State $\CU' \gets \{ u\in\CU \mid (|\critSubsetsCPA_u| \geq n/3) \land (\deg_{\critSubsetsG}(u) > (x+\superviewEpsGap)|\CU|) \}$
        \State \Return{$\{ p\in\CP \mid \exists \CU''\subseteq\CU': (\forall u\in\CU'': p\in\critSubsetsCPA_u) \land (|\CU''| > (x+\superviewEpsGap)|\CU'|) \}$}
    \end{algorithmic}
\end{algorithm}

In \cref{sec:tech-overview}, we described an adjudication rule that suffices to determine at least one adversary node in the event of a timely-liveness violation. A precise version of this rule is specified in \cref{alg:adjudication-criticalsubsets1}. 
We now tackle the analysis for 
this rule
in the general case for parameters $\netX$, $\tALmax$, and when, due to the random leader election, not all super-views may have some honest leader.
Recall that
$\{ \CV_u \}_{u\in\CU}$ are chosen as follows:
For $\netDeltaPeriod = 12 \Delta \superviewLen$ for some fixed parameter $\superviewLen$,
we choose the $\netG(\netDeltaPeriod)$ most recent periods of length $\netDeltaPeriod$ of the execution, aligned to view boundaries in \cref{alg:tendermint-2}, as super-views $\CU$. Each of these periods $u$ corresponds to $\superviewLen$ views of \cref{alg:tendermint-2}, which make up $\CV_u$.
Also, $\CV \triangleq \bigcup_{u\in\CU} \CV_u$.
A super-view is \emph{synchronous} if the network respects a message-delay upper-bound $\Delta$ during view $v$ of \cref{alg:tendermint-2} for all $v\in\CV_u$.

Note that even with moderate $\superviewLen$, \emph{most} super-views $u$
have at least \emph{some} view $v\in\CV_u$ with honest $\Leader_v$,
due to the random choice of $\Leader_v$.
This and $\netX$-partial-synchrony give us that, for a fixed point in time where $\funLA$ is invoked, 
and appropriate choice of $\superviewEpsGap$,
except with 
some small probability,
at least $(1-\netX-\superviewEpsGap)$ fraction of super-views in $\CU$ are synchronous \emph{and} have some honest leader; vice versa, at most $(\netX+\superviewEpsGap)$ fraction of super-views in $\CU$ are either not synchronous \emph{or} have no honest leader.
Denote by $\CUs$ the subset of $\CU$ containing all synchronous super-views,
by $\CUhl$ the subset of $\CU$ containing all super-views with some honest leader,
and by $\CUshl \triangleq \CUs \cap \CUhl$.
Complementation of these sets is taken with respect to $\CU$.
Note that with this notation we expect $|\CUshl| \geq (1-\netX-\superviewEpsGap) |\CU|$,
which \cref{lem:adjudication-criticalsubsets1-overwhelminglysomehonestleader}
establishes based on a well-known Chernoff bound.

\begin{defer}{proofsadjudication}
\begin{proposition}[Chernoff bound]
    \label{prop:chernoff-indepbernoulli}
    Let $Z_1, ..., Z_m$ be independent Bernoulli random variables with $p_i \triangleq \Prob{Z_i = 1}$.
    Let $\mu \triangleq \Mean{\sum_{i=1}^m Z_i}$.
    Then, $\forall c\geq0$:
    \begin{IEEEeqnarray}{rCl}
        \Prob{\left(\sum_{i=1}^m Z_i\right) \geq (1+c)\mu}
        &\leq&
        \exp\left(\frac{- c^2 \mu}{2+c} \right).
        \IEEEeqnarraynumspace
    \end{IEEEeqnarray}
\end{proposition}
\end{defer}

\begin{lemma}
    \label{lem:adjudication-criticalsubsets1-overwhelminglysomehonestleader}
    For a fixed 
    round,
    and fixed $\superviewEpsGap > 0$,
    assuming $(\Delta, \netG, \netX)$-partial-synchrony
    and $f \leq \tALmax < n/2$,
    and choosing $\superviewLen = \lceil \log_2(\frac{2}{\superviewEpsGap}) \rceil$,
    except with probability $\superviewErrProb$,
    we get $|\CUshl| \geq (1-\netX-\superviewEpsGap) |\CU|$.
\end{lemma}
\deferredproof{proofsadjudication}{lem:adjudication-criticalsubsets1-overwhelminglysomehonestleader}
\begin{defer}{proofsadjudication}
\begin{proof}%
    For each super-view $u\in\CU$,
    let $Z_u$ be the random variable with 
    $Z_u = 0$ if $\exists v\in\CV_u: \Leader_v \in \CPh$, and
    $Z_u = 1$ otherwise, \ie, if $\forall v\in\CV_u: \Leader_v \in \CPa$.
    Since $|\CPa| = f \leq \tALmax < n/2$, and leaders per view are chosen independently and uniformly at random,
    and each super-view has $\superviewLen$ views,
    $\Mean{Z_u} = \Prob{Z_u = 1} \leq 2^{-\superviewLen}$.
    Then, from \cref{prop:chernoff-indepbernoulli},
    \begin{IEEEeqnarray}{rCl}
        \Prob{\left(\sum_{u\in\CU} Z_i\right) \geq 2 \cdot 2^{-\superviewLen} \netG(\netDeltaPeriod)}
        &\leq&
        \exp\left(- 2^{-\superviewLen} \netG(\netDeltaPeriod)/3\right).
        \IEEEeqnarraynumspace
    \end{IEEEeqnarray}
    For any target $\superviewEpsGap > 0$,
    by choosing $\superviewLen = \lceil \log_2(\frac{2}{\superviewEpsGap}) \rceil$,
    except with probability $\superviewErrProb$,
    at most $\superviewEpsGap$ fraction of super-views in $\CU$ will not have \emph{some} honest leader:
    $|\CUhl^c| \leq \superviewEpsGap |\CU|$.
    
    Furthermore, due to $(\Delta, \netG, \netX)$-partial-synchrony
    and our choice of $\netDeltaPeriod$,
    we get $|\CUs^c| \leq x |\CU|$.
    Thus, from a union bound,
    $|\CUshl| = |\CUs \cap \CUhl| \geq |\CU| - |\CUs^c| - |\CUhl^c| = (1 - x - \superviewEpsGap)|\CU|$. 
\end{proof}
\end{defer}

Since we consider the regime where $\netX<1/2$,
we may choose $\superviewEpsGap < 1/2 - x$
by choosing $\superviewLen = \lceil \log_2(\frac{4}{1-2x}) \rceil + 1$,
to ensure $1-\netX-\superviewEpsGap > \netX+\superviewEpsGap$,
which we subsequently make use of.
\begin{remark}
    \label{rem:assumptions}
    For \cref{lem:adjudication-criticalsubsets1-CUshlInCUprime,lem:adjudication-criticalsubsets1-soundness,lem:adjudication-criticalsubsets1-completeness},
    we assume that
    some honest node took note of a potential timely-liveness violation in \cref{fig:overview1},
    that $(\Delta, \netG, \netX)$-partial-synchrony holds,
    and that $1-\netX-\superviewEpsGap > \netX+\superviewEpsGap$.
\end{remark}

We first show that $\CUshl \subseteq \CU'$:
\begin{lemma}
    \label{lem:adjudication-criticalsubsets1-CUshlInCUprime}
    For a fixed 
    round,
    under the assumptions of \cref{rem:assumptions},
    except with probability $\superviewErrProb$,
    for $\CU'$ of \cref{alg:adjudication-criticalsubsets1},
    $\CUshl \subseteq \CU'$.
\end{lemma}
\deferredproof{proofsadjudication}{lem:adjudication-criticalsubsets1-CUshlInCUprime}
\begin{defer}{proofsadjudication}
\begin{proof}%
    Let $u\in\CUshl$.
    Assuming the blame counts are based on all honest transcripts
    (which they are, as we have argued at the beginning of \cref{sec:blaming} and in \cref{fig:overview1}), 
    and using the assumption that 
    some honest node took note of a potential timely-liveness violation,
    and using $|\CPa| \leq \tALmax$, so that $|\CPh| \geq n-\tALmax$, by \cref{lem:tendermint-2-advaccountability-multishot},
    $|\critSubsetsCPA_u| \geq n/3$.

    Let $u,u'\in\CUshl$,
    so per the above,
    $|\critSubsetsCPA_u| \geq n/3$,
    $|\critSubsetsCPA_{u'}| \geq n/3$.
    Due to \cref{lem:tendermint-2-honacquittal-multishot},
    $\critSubsetsCPA_u \subseteq \CPa$,
    $\critSubsetsCPA_{u'} \subseteq \CPa$.
    Since $|\CPa| \leq \tALmax$,
    by an intersection argument,
    $|\critSubsetsCPA_u \cap \critSubsetsCPA_{u'}| \geq 2n/3 - \tALmax$.

    Since any two $u,u'\in\CUshl$
    have an edge in $\critSubsetsG$ of \cref{alg:adjudication-criticalsubsets1} per the above,
    and $|\CUshl| \geq (1-\netX-\superviewEpsGap) |\CU| > (\netX+\superviewEpsGap) |\CU|$ due to $\netX$-partial-synchrony
    and $1-\netX-\superviewEpsGap > \netX+\superviewEpsGap$ by assumption,
    for every $u\in\CUshl$,
    $\deg_{\critSubsetsG}(u) > (\netX+\superviewEpsGap)|\CU|$.

    Thus, for every $u\in\CUshl$,
    $|\critSubsetsCPA_u| \geq n/3$ and $\deg_{\critSubsetsG}(u) > (\netX+\superviewEpsGap)|\CU|$,
    so $u\in\CU'$, as desired.
\end{proof}
\end{defer}

We then show that \cref{alg:adjudication-criticalsubsets1} never outputs an honest node:
\begin{lemma}[Soundness of $\funLA$]
    \label{lem:adjudication-criticalsubsets1-soundness}
    For a fixed 
    round,
    under the assumptions of \cref{rem:assumptions},
    except with probability $\superviewErrProb$,
    no $p\in\CPh$ is returned by \cref{alg:adjudication-criticalsubsets1}.
\end{lemma}
\deferredproof{proofsadjudication}{lem:adjudication-criticalsubsets1-soundness}
\begin{defer}{proofsadjudication}
\begin{proof}%
    Recall that $\CUshl \subseteq \CU'$ by \cref{lem:adjudication-criticalsubsets1-CUshlInCUprime},
    $\CU' \subseteq \CU$ by definition in \cref{alg:adjudication-criticalsubsets1},
    and $|\CUshl| \geq (1-\netX-\superviewEpsGap) |\CU|$.
    Thus,
    $|\CU' \cap \CUshl| = |\CUshl| \geq (1-\netX-\superviewEpsGap) |\CU| \geq (1-\netX-\superviewEpsGap) |\CU'|$,
    \ie, $\CUshl$ makes up at least $(1-\netX-\superviewEpsGap)$ fraction of $\CU'$,
    \ie, at most $\netX+\superviewEpsGap$ fraction of $\CU'$ are not in $\CUshl$.

    Suppose for contradiction that $p\in\CPh$ is returned by \cref{alg:adjudication-criticalsubsets1}.
    Then there is a subset $\CU''$ of $\CU'$
    such that $p\in\critSubsetsCPA_u$ for all $u\in\CU''$
    and $|\CU''| > (\netX+\superviewEpsGap) |\CU'|$.
    From the latter, there must be $u\in\CU''\cap\CUshl$
    so that $p\in\critSubsetsCPA_u$,
    \ie, $p$ is blamed by at least $n - \tALmax$ nodes in $u$,
    but, together with $\tALmax < n/2$,
    \cref{lem:tendermint-2-honacquittal-multishot}
    rules such $u\in\CUshl$ out.
    This is the desired contradiction.
\end{proof}
\end{defer}

\begin{defer}{proofsadjudication}
A counting lemma is used in the proof of \cref{lem:adjudication-criticalsubsets1-completeness}:
\begin{lemma}
    \label{lem:generalized-pidgeonhole}
    Let $\Omega$ be some set with $|\Omega| < \infty$,
    $h: \Omega \to \{0,...,m\}$ be some function, for some $m\in\IN$,
    and $c < \frac{1}{|\Omega|} \sum_{\omega\in\Omega} h(\omega) \triangleq \mu$.
    Then,
    $|\{ \omega \mid h(\omega) \leq c \}| \leq \lfloor |\Omega| \frac{m - \mu}{m - c} \rfloor$.
\end{lemma}
\begin{proof}
    Let $\Omega_- \triangleq \{ \omega \mid h(\omega) \leq c \}$,
    $\Omega_+ \triangleq \{ \omega \mid h(\omega) > c \}$.
    Then,
    $\mu = \frac{1}{|\Omega|} \sum_{\omega\in\Omega} h(\omega) = \frac{1}{|\Omega|} \left( \sum_{\omega\in\Omega_-} h(\omega) + \sum_{\omega\in\Omega_+} h(\omega) \right)$.
    Since $h(\omega) \leq c$ for all $\omega\in\Omega_-$,
    and $h(\omega) \leq m$ for all $\omega\in\Omega_+$,
    and also $|\Omega| = |\Omega_-| + |\Omega_+|$,
    $|\Omega| \mu \leq |\Omega_-| c + (|\Omega| - |\Omega_-|) m$.
    So, rearranging,
    $|\Omega| \mu - |\Omega| m \leq |\Omega_-| (c - m)$.
    Finally, multiplying both sides by $(-1)$,
    and dividing,
    $|\Omega|\frac{m - \mu}{m - c} \geq |\Omega_-|$.
    Since $|\Omega_-| \in \IN$,
    $|\{ \omega \mid h(\omega) \leq c \}| = |\Omega_-| \leq \lfloor |\Omega| \frac{m - \mu}{m - c} \rfloor$.
\end{proof}
\end{defer}

Finally, we show that \cref{alg:adjudication-criticalsubsets1} outputs a certain number of (necessarily adversary) nodes:
\begin{lemma}[Completeness of $\funLA$]
    \label{lem:adjudication-criticalsubsets1-completeness}
    For a fixed 
    round,
    under the assumptions of \cref{rem:assumptions},
    except with probability $\superviewErrProb$,
    \cref{alg:adjudication-criticalsubsets1} returns $\CP'$
    with $|\CP'| \geq f - \left\lfloor \frac{(f - n/3) + (\netX+\superviewEpsGap) (\tALmax - n/3)}{1-\netX-\superviewEpsGap} \right\rfloor$.
\end{lemma}
\deferredproof{proofsadjudication}{lem:adjudication-criticalsubsets1-completeness}
\begin{defer}{proofsadjudication}
\begin{proof}%
    Consider a $|\CPa| \times |\CU'|$ matrix $M$ with rows indexed by $p\in\CPa$ and columns indexed by $u\in\CU'$ of \cref{alg:adjudication-criticalsubsets1}.
    Let $M_{p,u} = 1$ if $p\in\critSubsetsCPA_u$, and $0$ otherwise.
    Recall that $\CUshl \subseteq \CU'$ (\cref{lem:adjudication-criticalsubsets1-CUshlInCUprime}).
    Consider $u\in\CU'\cap\CUshl=\CUshl$.
    By \cref{lem:tendermint-2-advaccountability-multishot,lem:tendermint-2-honacquittal-multishot},
    $\sum_{p\in\CPa} M_{p,u} \geq n/3$.
    Now consider $u\in\CU'\cap\CUshl^c$.
    By definition of $\CU'$ in \cref{alg:adjudication-criticalsubsets1},
    $\deg_{\critSubsetsG}(u) > (\netX+\superviewEpsGap)|\CU|$.
    Since $\CUshl$ makes up at least $(1-\netX-\superviewEpsGap)$ fraction of $\CU'$
    (proof of \cref{lem:adjudication-criticalsubsets1-soundness}),
    \ie, at most $\netX+\superviewEpsGap$ fraction of $\CU'$ are not in $\CUshl$,
    there is some $u'\in\CUshl$ such that
    $\{u,u'\} \in \critSubsetsG$
    and thus $|\critSubsetsCPA_u \cap \critSubsetsCPA_{u'}| \geq 2n/3 - \tALmax$.
    By \cref{lem:tendermint-2-honacquittal-multishot},
    $\critSubsetsCPA_{u'} \subseteq \CPa$,
    thus $\critSubsetsCPA_u \cap \critSubsetsCPA_{u'} \subseteq \CPa$,
    thus $|\critSubsetsCPA_u \cap \CPa| \geq 2n/3 - \tALmax$,
    thus $\sum_{p\in\CPa} M_{p,u} \geq 2n/3 - \tALmax$.

    Since $|\CU' \cap \CUshl| \geq (1-\netX-\superviewEpsGap) |\CU'|$ (proof of \cref{lem:adjudication-criticalsubsets1-soundness}),
    and thus $|\CU' \cap \CUshl^c| \leq (\netX+\superviewEpsGap) |\CU'|$,
    and $2n/3 - \tALmax \leq n/3$ due to $n/3 \leq \tALmax < n/2$,
    $\sum_{p\in\CPa} \sum_{u\in\CU'} M_{p,u} \geq (1-\netX-\superviewEpsGap) |\CU'| n/3 + (\netX+\superviewEpsGap) |\CU'| (2n/3 - \tALmax) = |\CU'| (n/3 + (\netX+\superviewEpsGap) (n/3 - \tALmax))$.

    Since there are $f = |\CPa|$
    rows in $M$,
    the average row weight $\mu$ of $M$ is 
    $\mu \triangleq |\CU'| (n/3 + (\netX+\superviewEpsGap) (n/3 - \tALmax)) / f$.
    Note that for every $f \leq \tALmax$,
    $\mu > (\netX+\superviewEpsGap) |\CU'|$, since $\netX+\superviewEpsGap < 1/2$.
    Consider the function $h: \CPa \to \{0,...,|\CU'|\}, p \mapsto \sum_{u\in\CU'} M_{p,u}$.
    Using \cref{lem:generalized-pidgeonhole},
    $|\{ p \mid h(p) \leq (\netX+\superviewEpsGap) |\CU'| \}| \leq 
    \left\lfloor \frac{f - n/3 - (\netX+\superviewEpsGap) (n/3 - \tALmax)}{1-\netX-\superviewEpsGap} \right\rfloor$.
    
    Thus, the number of rows in $M$ with weight more than $(\netX+\superviewEpsGap) |\CU'|$
    is at least $f - \left\lfloor \frac{f - n/3 - (\netX+\superviewEpsGap) (n/3 - \tALmax)}{1-\netX-\superviewEpsGap} \right\rfloor$.
    The $p$ corresponding to these rows satisfy the return condition of \cref{alg:adjudication-criticalsubsets1},
    $\exists \CU''\subseteq\CU': (\forall u\in\CU'': p\in\critSubsetsCPA_u) \land (|\CU''| > (\netX+\superviewEpsGap)|\CU'|)$,
    where we recall that $M_{p,u} = 1$ if $p\in\critSubsetsCPA_u$, and $0$ otherwise.
\end{proof}
\end{defer}

\begin{remark}
    \label{rem:adjudication-criticalsubsets1-completeness-coreset}
    It is easy to verify that when \cref{alg:adjudication-criticalsubsets1} is used in the context of $\funLA$ in \cref{fig:overview1}, that $\CP'$ obtained from $\funLA$ when input all honest transcripts and $\bot$ for all adversary transcripts,
    satisfies the bound on $|\CP'|$ of \cref{lem:adjudication-criticalsubsets1-completeness},
    and is a subset of any $\CP''$ obtained from $\funLA$ when input all honest transcripts
    and any other adversary-produced transcripts. This property is crucial to ensuring that $\CP'$ is accused by \emph{all} honest nodes in the context of \cref{fig:overview1},
    so that subsequently a certificate of guilt for $\CP'$ is formed.
\end{remark}

We combine the above results to assert accountable liveness:
\begin{theorem}
    \label{thm:adjudication-criticalsubsets1-acclive}
    For any given $\netG, \netX < 1/2, n/3 < \tALmax < n/2, \epsAL$,
    the atomic broadcast protocol of \cref{alg:tendermint-2}
    together with \cref{alg:tendermint-2-alproto-multishot,alg:adjudication-criticalsubsets1},
    with design parameter $\superviewEpsGap$,
    when 
    instantiated with $\superviewLen = \lceil \log_2(\frac{2}{\superviewEpsGap}) \rceil + C$
    and $\netDeltaPeriod = 12 \Delta \superviewLen$,
    where $C$ is chosen such that $\superviewErrProb \leq \epsAL$,
    is accountably live
    for 
    $\tALident = \tALmax - \left\lfloor \frac{(1+\netX+\superviewEpsGap) (\tALmax - n/3)}{1-\netX-\superviewEpsGap} \right\rfloor$,
    when run in $(\Delta, \netG, \netX)$-partial-synchrony with $f \leq \tALmax$.
\end{theorem}
\deferredproof{proofsadjudication}{thm:adjudication-criticalsubsets1-acclive}
\begin{defer}{proofsadjudication}
\begin{proof}%
    \emph{Soundness.}
    Suppose a certificate of guilt is produced for node $p$ and time $t$,
    according to the process of \cref{fig:overview1,sec:blaming,sec:adjudication}.
    And towards a contradiction, suppose $p\in\CPh$.
    Because $f \leq \tALmax < n/2$,
    an honest node $p'$ has accused $p$.
    Thus, $p'$ obtained $p\in\CP'$ for $\CP'$ returned by $\funLA$.
    Since $p'$ is honest, $p'$ must have 
    taken note of a potential timely-liveness violation in \cref{fig:overview1}.
    But by \cref{lem:adjudication-criticalsubsets1-soundness},
    under the given circumstances,
    no $p\in\CPh$ is returned by \cref{alg:adjudication-criticalsubsets1}.
    This is a contradiction, so $p\in\CPa$.

    \emph{Completeness.}
    Suppose an execution with a timely-liveness
    violation at $t$.
    Then more than $\lfloor (n-1)/3 \rfloor$ honest nodes have not voted $\VoteLive$ for any of the views in $\CV$, and thus all honest nodes have taken note of a potential timely-liveness violation at $t$ in \cref{fig:overview1}.
    Thus, all honest nodes apply $\funLA$,
    and
    by \cref{lem:adjudication-criticalsubsets1-completeness,rem:adjudication-criticalsubsets1-completeness-coreset},
    and considering that $f \leq \tALmax$,
    there is a set $\CP'$
    with $|\CP'| \geq \tALmax - \left\lfloor \frac{(1+\netX+\superviewEpsGap) (\tALmax - n/3)}{1-\netX-\superviewEpsGap} \right\rfloor$
    so that for every honest node,
    \cref{alg:adjudication-criticalsubsets1} as part of $\funLA$ returns a superset of $\CP'$.
    All honest nodes obtain (a superset of) $\CP'$ from $\funLA$ by time $t + \netDeltaPeriod\netG(\netDeltaPeriod)$, according to the process of \cref{fig:overview1,sec:blaming,sec:adjudication}, and accuse all $\CP'$.
    By $t + 2\netDeltaPeriod\netG(\netDeltaPeriod)$, these accusations have propagated
    to all honest nodes, and since honest nodes are a majority,
    they have produced a certificate of guilt for $\CP'$.
\end{proof}
\end{defer}

\Cref{thm:adjudication-criticalsubsets1-acclive} shows that
the protocol consisting of the combination of \cref{alg:tendermint-2,alg:tendermint-2-alproto-multishot,alg:adjudication-criticalsubsets1}
satisfies goal~(2) in \cref{sec:modelprelims-goal}.
Together with \cref{thm:tendermint-2-safety,thm:tendermint-2-liveness},
this completes the picture that the protocol satisfies the goals laid out in \cref{sec:modelprelims-goal}.

It is important to observe that, as required by \cref{def:accountable-liveness},
$\tALident$ of \cref{thm:adjudication-criticalsubsets1-acclive} provides a lower bound \emph{across \emph{all} executions with $f \leq \tALmax$}, on the number of identifiable adversary nodes.
How does the number of identifiable adversary nodes depend on the actual number of adversary nodes $f$?
This is equivalent to asking for a variant of \cref{def:accountable-liveness} with a function $\tALident(f)$ instead of a constant $\tALident$.
The answer is provided by \cref{lem:adjudication-criticalsubsets1-completeness}.

\section{Impossibility Results}
\label{sec:impossibilities}

It is well-known~\cite{model-psync} that atomic broadcast protocols
that are $\tS$-safe and $\tL$-live under partial synchrony,
must satisfy $n > 2\tL + \tS$.
Since protocols with these properties are the baseline
laid out in our design goals in \cref{sec:modelprelims-goal},
we focus subsequently on protocols that achieve 
that bound
in an optimal fashion:
\begin{definition}
    \label{def:optresilient}
    An atomic broadcast protocol
    with safety resilience $\tS$
    and liveness resilience $\tL$
    is \emph{optimally resilient}
    iff $n = 2\tL + \tS + 1$. 
\end{definition}

\subsection{Impossibility for \texorpdfstring{$\tALmax \geq n/2$}{tALmax >= n/2}}
\label{sec:impossibility-tALmax-geq-n2}

A perhaps intuitive consequence of the requirements 
that 
``enough'' honest nodes alone should be able to produce certificates of guilt,
while ``few'' adversary nodes alone should not,
is 
that
non-trivial accountable liveness 
is impossible
for $\tALmax \geq n/2$:
\begin{theorem}
    \label{thm:impossibility-tALmax-geq-n2}
    No atomic broadcast protocol 
    can be $0$-safe under partial synchrony
    and achieve non-trivial accountable liveness
    with
    $\tALmax \geq n/2$ and $\tALident > 0$
    under $(\Delta, \netG, \netX)$-partial-synchrony 
    (even if $\netX=0$, \ie, even under synchrony).
\end{theorem}
The proof is relegated to \cref{sec:proof-impossibility-tALmax-geq-n2}.
(Note that for any $\tau>0$, $\tau$-safety implies $(\tau-1)$-safety, so $0$-safety is the weakest non-trivial such threshold safety property.)

\subsection{Impossibility for \texorpdfstring{$\netX \geq 1/2$}{x >= 1/2}}
\label{sec:impossibility-x-geq-12}

For the scheme of \cref{sec:blaming,sec:adjudication},
it is intuitive that non-trivial accountable liveness cannot be achieved if the network is ``more asynchronous than synchronous'', since honest nodes can then be framed by the adversary more often than they can reliably detect adversary behavior.
Perhaps interestingly, this impossibility holds for \emph{all} consensus protocols and \emph{all} liveness accountability mechanisms:
\begin{theorem}
    \label{thm:impossibility-x-geq-12}
    No optimally-resilient atomic broadcast protocol
    with $\tL > 0$
    that is $0$-safe under partial synchrony
    can achieve non-trivial accountable liveness
    with
    $\tALmax > \tL$ and $\tALident > 0$
    under $(\Delta, \netG, \netX)$-partial-synchrony
    for $\netX \geq 1/2$.
\end{theorem}
The proof is relegated to \cref{sec:proof-impossibility-x-geq-12}.

\begin{corollary}
    \label{thm:impossibility-psync}
    No optimally-resilient atomic broadcast protocol 
    with $\tL>0$ 
    that is $0$-safe under partial synchrony
    can achieve non-trivial accountable liveness
    with
    $\tALmax > \tL$ and $\tALident > 0$
    under partial synchrony.
\end{corollary}
\begin{proof}
    Follows from \cref{thm:impossibility-x-geq-12} and
    the fact that partial synchrony is a special case of $\netX$-partial-synchrony with $\netX=1$.
\end{proof}

\subsection{Impossibility for \texorpdfstring{$\tALident$}{tALident}}
\label{sec:impossibility-tALident1}

\Cref{thm:impossibility-x-geq-12,thm:impossibility-tALmax-geq-n2}
show that accountable liveness can at best be expected
for the region circumscribed by $\netX < 1/2$ and $\tALmax < n/2$.
Furthermore, only $\tALmax \geq n/3$ is of interest,
as a protocol for $n = 3\tau+1$ with $\tau = \tL = \tS$
otherwise achieves trivial accountable liveness
(since the protocol then is always live when $f \leq \tALmax$).
In fact, 
the atomic broadcast consensus protocol of \cref{alg:tendermint-2}
together with the accountability mechanism of \cref{fig:overview1,alg:tendermint-2-alproto-multishot,alg:adjudication-criticalsubsets1}
achieves non-trivial accountable liveness with \emph{some} $\tALident > 0$
for the interior of the aforementioned region.
But is the value of $\tALident$ achieved by this combined scheme 
``good''? 
Could one hope to achieve even higher $\tALident$,
either with a different consensus protocol and/or with a different accountability mechanism?

Let us address this question in four steps.
First, a simple argument shows that \emph{no} combination of consensus protocol and accountability mechanism can achieve $\tALident > \tL + 1$.
Second, we show that for the consensus protocol of \cref{alg:tendermint-2},
\emph{every} accountability mechanism has to satisfy $\tALident < (\tL + 2) - \left\lfloor \frac{\tALmax - (\tL + 1)}{k-2} \right\rfloor$ for every $k\geq3$ and $\netX \geq 1/k$.
Third, the proof of the aforementioned bound reveals that the bound holds for a large class of PBFT-style consensus protocols, \emph{irrespective} of the liveness accountability mechanism.
As can be seen from \cref{fig:results1} (\ref{plt:results1-tALmax033-f033-Achievability1} vs.\ \ref{plt:results1-tALmax033-f033-Converse1}),
the bound closely matches the $\tALident$ achieved by the combined scheme of \cref{alg:tendermint-2,fig:overview1,alg:tendermint-2-alproto-multishot,alg:adjudication-criticalsubsets1}, and is even tight for $\netX=1/3$.
In this sense, the combined scheme of \cref{alg:tendermint-2,fig:overview1,alg:tendermint-2-alproto-multishot,alg:adjudication-criticalsubsets1}
is near optimal.
Finally, we explain why we conjecture that the bound holds for \emph{every} combination of consensus protocol (not just PBFT-style) and liveness accountability mechanism that together satisfy the design goals set out in \cref{sec:modelprelims-goal}.

For the first step,
optimal resilience
(\cref{def:optresilient},
\cf \cref{lem:existence-unliveness-crashes} 
in \cref{sec:addendum-impossibilities})
implies that for any protocol for $n = 3\tau+1$ with $\tau = \tL = \tS$,
an adversary only needs to let $\tL + 1$ of its $f$ nodes behave adversarily
(in particular, let them crash) to cause a liveness violation, while the remaining $f - (\tL + 1)$ adversary nodes can behave indistinguishably from honest nodes.
Thus, we cannot hope to guarantee to detect more than $\tL + 1$ guilty nodes 
in case of any liveness violation.
Ergo, 
no more than $\tALident = \tL + 1$ can be achieved by any protocol.

For the second step:
\begin{theorem}
    \label{thm:impossibility-tALident1-ours}
    For every $k\geq3$:
    \Cref{alg:tendermint-2},
    irrespective of the liveness accountability mechanism,
    cannot achieve accountable liveness
    for any
    $\tALmax > \tL$
    under $(\Delta, \netG, \netX)$-partial-synchrony
    for $\netX \geq 1/k$,
    unless
    $\tALident < (\tL + 2) - \left\lfloor \frac{\tALmax - (\tL + 1)}{k-2} \right\rfloor$.
\end{theorem}
The proof is relegated to \cref{sec:proof-impossibility-tALident1-ours}.

For the third step,
we note that the specifics of \cref{alg:tendermint-2}
enter into the proof of \cref{thm:impossibility-tALident1-ours}
only to argue that 
there is a timely-liveness violation
in one of the executions $E_{k,i}$ (\cf \cref{fig:tauALident-converses1} in \cref{sec:proof-impossibility-tALident1-ours}) considered in the proof.
The commonly considered classical PBFT-style consensus protocols
like 
\cref{alg:tendermint-2},
PBFT~\cite{pbft}, 
Tendermint~\cite{tendermint}, 
HotStuff~\cite{hotstuff}, 
CasperFFG~\cite{casper,casperethresearch}, 
or 
Streamlet~\cite{streamlet},
all exhibit timely-liveness violations in these executions. 
Intuitively, this is because 
the protocols never collect enough ($>2n/3$) votes to form quorum certificates, before a view change is triggered due to lack of progress.
They thus cannot confirm blocks.
More formally, this is because
the protocols satisfy the following property:
\begin{definition}
    \label{def:protocol-nowornever}
    We say a consensus protocol is \emph{now-or-never} iff
    there exists $\Delta''$ dividing $\netDeltaPeriod \netG(\netDeltaPeriod) / k$
    such that if the protocol execution is partitioned into intervals of length $\Delta''$,
    and for each interval, any $\lceil n/3 \rceil$ nodes are partitioned off temporarily until the end of the interval,
    while all other communication is next-round-delay,
    then the protocol does not confirm any transaction.
\end{definition}
This condition is in fact sufficient
to prove an analogue of \cref{thm:impossibility-tALident1-ours}:
\begin{theorem}
    \label{thm:impossibility-tALident1-pbft}
    For every $k\geq3$:
    A now-or-never consensus protocol,
    irrespective of the liveness accountability mechanism,
    cannot achieve accountable liveness
    for any
    $\tALmax > \tL$
    under $(\Delta, \netG, \netX)$-partial-synchrony
    for $\netX \geq 1/k$,
    unless
    $\tALident < (\tL + 2) - \left\lfloor \frac{\tALmax - (\tL + 1)}{k-2} \right\rfloor$.
\end{theorem}
The proof is relegated to \cref{sec:proof-impossibility-tALident1-pbft}.

Now to the final fourth step: 
Why does this proof not carry over to general atomic broadcast protocols?
The argument for \cref{thm:impossibility-tALident1-ours,thm:impossibility-tALident1-pbft}
hinges on the assumption that one of the $E_{k,i}$
(\cf \cref{fig:tauALident-converses1} in \cref{sec:proof-impossibility-tALident1-ours})
has a timely-liveness violation.
An asynchronous protocol, 
for instance,
may be able to leverage 
the limited rounds of message exchange allowed 
at the boundaries of the $\CT_i$ or $\CT^{(v)}$
to achieve timely-liveness occasionally,
independently of any $\Delta''$ and in particular for $\Delta'' = \netDeltaPeriod \netG(\netDeltaPeriod) / k$.
Nonetheless, we conjecture that even a randomized asynchronous protocol
would have a constant probability for a timely-liveness violation in one of the $E_{k,i}$,
because only a small constant number of full communication rounds can be completed.
\begin{conjecture}
    \label{thm:impossibility-tALident1-all}
    For every $k\geq3$:
    \emph{Every atomic broadcast consensus protocol},
    irrespective of the liveness accountability mechanism,
    cannot achieve accountable liveness
    for any
    $\tALmax > \tL$
    under $(\Delta, \netG, \netX)$-partial-synchrony
    for $\netX \geq 1/k$,
    unless
    $\tALident < (\tL + 2) - \left\lfloor \frac{\tALmax - (\tL + 1)}{k-2} \right\rfloor$.
\end{conjecture}

It is important to observe that, as stated, \cref{thm:impossibility-tALident1-ours,thm:impossibility-tALident1-pbft,thm:impossibility-tALident1-all} assert a bound on $\tALident$ as used in \cref{def:accountable-liveness}, \ie, where $\tALident$ is no more than the lowest number of guilty nodes identified across \emph{all executions with $f \leq \tALmax$}.
The reasoning for \cref{thm:impossibility-tALident1-ours} identifies worst-case executions where $f \approx \tALmax$ nodes act adversarily and the number of guilty nodes identified is minimized.
One may be interested in a variant of \cref{def:accountable-liveness} where $\tALident$ is a function of the actual number of adversary nodes $f$, and $\tALident(f)$ is no more than the lowest number of guilty nodes identified across \emph{all executions with $f$ adversary nodes}.
Indeed, the steps of the argument for \cref{thm:impossibility-tALident1-ours} go through with $f$ instead of $\tALmax$ to obtain
$\tALident(f) < (\tL + 2) - \left\lfloor \frac{f - (\tL + 1)}{k-2} \right\rfloor$.

\section{Related Work}
\label{sec:relatedwork} 

\paragraph{Accountable Liveness}

It appears
the term ``accountable liveness'' was first introduced by Tas \emph{et al.}~\cite[Appendix F]{DBLP:conf/sp/TasTGKMY23}, who provided a preliminary definition 
for state-machine replication protocols under synchrony.
Their work 
establishes a theorem analogous to our \cref{thm:impossibility-tALmax-geq-n2} for their 
setting.
While the technical details of their proof necessarily differ from our proof of \cref{thm:impossibility-tALmax-geq-n2}, the underlying reason both theorems hold is the same: if a \emph{certificate} implicating adversarial nodes can be constructed by an honest minority in any instance of an attack on liveness carried out by an adversarial majority, then the adversary is also able to construct certificates implicating honest nodes, leading to a failure of soundness.

An informal notion of accountable liveness appears in the Pod Network project~\cite{pod_network}, but only in the setting of partial ordering consensus---and it remains documented solely in a blog post rather than in their formal publication~\cite{alpos2025podoptimallatencycensorshipfreeaccountable}. In contrast, we provide a thorough analysis of accountable liveness 
in the context of total ordering consensus, establishing both the necessary and sufficient conditions for realizing it in our $\netX$-partially-synchronous network model. %

Ethereum~\cite{ethereum} addresses liveness violations in its consensus protocol, Gasper~\cite{gasper}---which combines LMD-GHOST for chain selection and Casper FFG~\cite{casper} for finality---through \emph{inactivity leaks}~\cite{casper,pavloff2024byzantineattacksexploitingpenalties}, a mechanism that gradually reduces the stake of 
non-participating
nodes. 
The goal is that
the remaining active nodes 
can
eventually form a supermajority of the \emph{effective} stake, restoring finality despite prolonged outages. 
This approach lacks a formal specification, and its guarantees remain unclear. The approach should be seen as a heuristic solution rather than a rigorously defined mechanism.

\paragraph{Timing Models}
In addition to the classical synchronous and partially-synchronous network models~\cite{DBLP:journals/siamcomp/DolevS83,model-psync,pbft,DBLP:journals/jacm/FischerLP85},
recent work has introduced \emph{granular synchrony}~\cite{DBLP:conf/wdag/GiridharanACN024}, a model that represents the network as a graph with communication links spanning fully synchronous, partially synchronous, and asynchronous behaviors. 
Notably, granular synchrony unifies existing timing models as specific instances within its broader framework.
Relatedly, Guo, Pass, and Shi~\cite{guopassshi} propose $\chi$-weak-synchrony, where at any time, $\chi$ fraction of nodes are honest and synchronously connected. 
Lewis-Pye and Roughgarden~\cite{lewispye2025optimalfaulttolerance} propose a timing model that serves as an interpolation between the synchronous and partially-synchronous settings. In addition to the standard parameters of the partially-synchronous model, i.e., $\Delta$ and $\GST$, they introduce an additional parameter, $\Delta^*$, with $\Delta^* \ge \Delta$ and may or may not bound message delays \emph{before} $\GST$. 
Our $\netX$-partial-synchrony model also provides a (different) interpolation between synchrony and partial synchrony, and may be of independent interest in the context of the recent interest in more fine-grained network models.

For further discussion of additional related works, see \cref{sec:addendum-relatedwork}, where we discuss papers on accountable safety, slashing, recovery procedures, and responsive and network-adaptive protocols.

\section{Discussion \& Conclusion}
\label{sec:conclusion}

In this paper, we introduced the notion of accountability for liveness in the context of atomic broadcast. By proposing the $\netX$-partially-synchronous network model, we demonstrated how to identify adversarial protocol violators using the $\funLA$ function, which combines blame accounting (\cref{sec:blaming}) with an adjudication rule (\cref{sec:adjudication}). We also proved that accountable liveness becomes impossible to achieve if the network model or adversary strength deviate beyond certain thresholds (\cref{sec:impossibilities})---underscoring the intrinsic trade-offs and additional assumptions \emph{necessary} for liveness accountability.

Beyond these theoretical foundations, our work serves as a starting point for automating responses to liveness attacks in blockchains.  
Notably, Ethereum already implements an automated response to major liveness issues. Ethereum's consensus protocol, Gasper, 
addresses liveness violations through a mechanism called \emph{inactivity leaks}, originally introduced to prevent Casper FFG~\cite{casper} from stalling indefinitely when more than one-third of nodes fail to participate.
The techniques presented in this paper provide a complementary and more general approach:  our methods enable the detection and formal identification of adversarial nodes through certificates of guilt. Once such a liveness failure is detected, offending nodes can be pinpointed and slashed---an approach akin to Ethereum’s inactivity leaks but with precise and stronger accountability guarantees.  

\begin{acks}
This work was supported in part by grant FY25-1894 from the Ethereum Foundation.
The research of Tim Roughgarden at Columbia University was supported by NSF award CNS-2212745.
\end{acks}

\nocite{fullversionofthispaper}
\nocite{accountabilityimpliesfinality}
\nocite{alpos2025podoptimallatencycensorshipfreeaccountable}
\nocite{bftforensics}
\nocite{buterinblogpost}
\nocite{casper}
\nocite{casperethresearch}
\nocite{cryptoeprint:2021/1169}
\nocite{cryptoeprint:2024/1799}
\nocite{cryptoeprint:2025/083}
\nocite{DBLP:books/daglib/0025983}
\nocite{DBLP:conf/ccs/Momose021}
\nocite{DBLP:conf/crypto/BlumZL20}
\nocite{DBLP:conf/crypto/KiayiasRDO17}
\nocite{DBLP:conf/dsn/BuchmanGK0SW22}
\nocite{DBLP:conf/dsn/RanchalPedrosaG24}
\nocite{DBLP:conf/eurocrypt/PassS18}
\nocite{DBLP:conf/fc/NeuTT22}
\nocite{DBLP:conf/icdcs/CivitGG21}
\nocite{DBLP:conf/nsdi/ShamisPC0FACCDK22}
\nocite{DBLP:conf/sigecom/BudishLR24}
\nocite{DBLP:conf/sosp/HaeberlenKD07}
\nocite{DBLP:conf/sp/AbrahamMN0Y20}
\nocite{DBLP:conf/sp/HouY23}
\nocite{DBLP:conf/sp/HouYS22}
\nocite{DBLP:conf/sp/TasTGKMY23}
\nocite{DBLP:conf/tcc/BlumKL19}
\nocite{DBLP:conf/wdag/GiridharanACN024}
\nocite{DBLP:journals/access/DeirmentzoglouP19}
\nocite{DBLP:journals/corr/abs-2002-03437}
\nocite{DBLP:journals/corr/abs-2201-07946}
\nocite{DBLP:journals/corr/abs-2304-14701}
\nocite{DBLP:journals/corr/abs-2310-06338}
\nocite{DBLP:journals/ipl/AlpernS85}
\nocite{DBLP:journals/jacm/FischerLP85}
\nocite{DBLP:journals/siamcomp/DolevS83}
\nocite{elaineshibook}
\nocite{ethereum}
\nocite{fullversionofthispaper}
\nocite{gasper}
\nocite{guopassshi}
\nocite{haberlenfaultdetectionproblem}
\nocite{hotstuff}
\nocite{lewispye2025optimalfaulttolerance}
\nocite{model-psync}
\nocite{nakamoto2008bitcoin}
\nocite{pavloff2024byzantineattacksexploitingpenalties}
\nocite{pbft}
\nocite{PikachuAzouvi}
\nocite{pod_network}
\nocite{sleepy}
\nocite{snapandchat}
\nocite{streamlet}
\nocite{sync-model}
\nocite{tendermint}

\bibliographystyle{ACM-Reference-Format}
\bibliography{references.bib}

\ifcompilefullversion\else
\section*{Appendix}
See full version for appendix:
\url{https://eprint.iacr.org/2025/693}~\cite{fullversionofthispaper}
\fi

\label{_my_appendix_start}
\appendix
\balance

\ifcompilefullversion
\section{Additional Related Work}
\label{sec:addendum-relatedwork}

We survey additional related works
beyond those discussed in \cref{sec:relatedwork}.
Achieving consensus among distributed nodes is a decades-old problem, traditionally defined by two fundamental properties: \emph{safety} and \emph{liveness}~\cite{pbft,DBLP:books/daglib/0025983}. Safety ensures that honest nodes never diverge on decided values, while liveness guarantees that decisions are eventually reached, all despite possible asynchrony and up to a threshold of adversary nodes.

\paragraph{Accountable Safety}

With the advent of blockchain technology, and in particular proof-of-stake (PoS) protocols, a stronger notion of safety called \emph{accountable safety} has emerged~\cite{casper,DBLP:conf/icdcs/CivitGG21,DBLP:conf/nsdi/ShamisPC0FACCDK22,DBLP:conf/sosp/HaeberlenKD07,DBLP:conf/dsn/BuchmanGK0SW22,bftforensics,DBLP:conf/dsn/RanchalPedrosaG24,snapandchat,DBLP:conf/fc/NeuTT22,accountabilityimpliesfinality,haberlenfaultdetectionproblem}. While accountable safety keeps the traditional requirement that decisions remain consistent under an adversarial threshold, it also allows to \emph{identify} specific misbehaving nodes in the event of a safety violation. This property is especially powerful in PoS settings, where such identification can trigger financial penalties (\emph{slashing}~\cite{casper}) on adversary nodes, thus creating economic incentives to follow the protocol~\cite{DBLP:conf/dsn/RanchalPedrosaG24,DBLP:conf/sigecom/BudishLR24}.

A key challenge arises when extending accountable safety to systems with \emph{dynamic participation}~\cite{DBLP:journals/corr/abs-2304-14701}. A recent work~\cite{DBLP:conf/fc/NeuTT22} formalizes an availability-accountability dilemma, proving that no protocol can remain fully accountably-safe while also guaranteeing liveness if the active set of nodes fluctuates like in the sleepy model~\cite{sleepy}. 
Neu \emph{et al.} design an accountability gadget that checkpoints a longest-chain protocol and can be combined with any BFT protocol providing accountable safety under static participation, thus addressing this dilemma in practical settings.

\paragraph{Slashing}

While existing PoS protocols with accountable safety can identify adversary nodes, they do not always ensure that those nodes' stakes are actually slashed---thereby falling short of providing slashable safety. In particular, adversary nodes can exploit posterior corruption attacks by reusing previously held stakes after withdrawal, making slashing ineffective~\cite{DBLP:journals/access/DeirmentzoglouP19}. To address this gap, Tas \emph{et al.}~\cite{DBLP:conf/sp/TasTGKMY23,DBLP:journals/corr/abs-2201-07946} and Azouvi and Vukolić~\cite{PikachuAzouvi} propose leveraging Bitcoin~\cite{nakamoto2008bitcoin} as a \emph{checkpointing} mechanism, anchoring critical PoS states within Bitcoin's immutable ledger. This design prevents adversary nodes from rewriting or invalidating older blocks---thereby circumventing long-range attacks---once the stake is withdrawn. Simultaneously, if a safety violation is detected in recent PoS blocks, the protocol can identify and slash the responsible nodes. 

\paragraph{Recovery Mechanisms}

Another important aspect of accountable safety involves \emph{recovery} following a slashable event. In blockchain systems like Ethereum, for instance, a major slashing often relies on social consensus to coordinate recovery---potentially including hard forks to stabilize the network. To formalize and automate this process, Lewis-Pye and Roughgarden~\cite{lewispye2025optimalfaulttolerance} propose a \emph{wrapper} approach that runs an execution of an accountably-safe, optimally-resilient state-machine replication protocol until a consistency violation occurs. When this occurs, the wrapper initiates a \emph{recovery procedure} to reach consensus on a set of adversary nodes for which proof of misbehavior exists, along with a long initial segment of the log generated by, below which no consistency violations have been detected. The wrapper then restarts the protocol with the adversary nodes removed. Gong \emph{et al.}~\cite{cryptoeprint:2025/083} further explore recovery under alive-but-corrupt nodes and partial synchrony. These works mark initial steps toward automating post-slashing recovery.

\paragraph{Responsive and Network Adaptive Protocols}

Responsive synchronous protocols~\cite{DBLP:conf/sp/AbrahamMN0Y20,DBLP:conf/eurocrypt/PassS18} operate under synchrony assumptions but 
can improve their latency if
actual
message delivery happens to be faster than the network's delay bound.
For instance, Thunderella~\cite{DBLP:conf/eurocrypt/PassS18} achieves near-instant confirmation via an asynchronous fast path under favorable conditions but falls back to a slow synchronous path if responsiveness fails, ensuring safety and liveness. These protocols leverage a \emph{fast path} to achieve low latency when the network is responsive while falling back to a conservatively timed execution when conditions degrade. While such protocols do not achieve full partial-synchrony safety (i.e., safety under arbitrarily long message delays), for any $\netX<1$, the implied synchronous model allows them to operate at the $\Delta^*$ time scale when network conditions are not optimal, benefiting from the fast path whenever possible.
On the other hand these protocols have a tradeoff between resilience on the responsive fast path and the synchronous fallback. A sequence of works~\cite{DBLP:conf/tcc/BlumKL19,DBLP:conf/crypto/BlumZL20,DBLP:journals/corr/abs-2002-03437} explored whether it is possible to design Byzantine-fault tolerant protocols that tolerate more than one-third Byzantine faults under synchrony, while still ensuring resilience to a certain fraction of faults---ideally up to one-third---in asynchrony or partial synchrony (the fast path). Their results establish that a BFT protocol can simultaneously tolerate $f_{\mathrm{a}} < \frac{n}{3}$ faults under asynchrony and $f_{\mathrm{a}} \leq f_{\mathrm{s}} < \frac{n}{2}$ faults under synchrony, if and only if the condition $2 f_{\mathrm{s}} + f_{\mathrm{a}} < n$ holds. Momose and Ren~\cite{DBLP:conf/ccs/Momose021} further demonstrate that it is possible to separate fault tolerance thresholds for different timing models and for safety and liveness. Specifically, they show that safety under synchrony can be improved while still preserving other fault thresholds, including liveness under synchrony and both safety and liveness under asynchrony.

\section{Addendum Consensus Protocol}
\label{sec:addendum-consensus}

\begin{algorithm}[tb]
    \caption{Unmodified Tendermint consensus, code for node $p$
    (based on \cite{tendermint}, \cite[Sec.\ 9.1]{DBLP:conf/sigecom/BudishLR24};
    basis for \cref{alg:tendermint-2})}
    \label{alg:tendermint-original}
    \begin{algorithmic}[1]
        \algfontsize%
        \State
            \algloclabel{loc:tendermint-original-blocks}
            Blocks:
            \begin{itemize}[itemsep=0pt,topsep=0pt,parsep=0pt,partopsep=0pt]
                \item $b \triangleq \Block(\FFp, \FFv, \FFbmI, \FFQC, \FFtxs)$ consists of:
                    creator $\FFp$,
                    view $\FFv$,
                    parent block $\FFbmI$,
                    quorum certificate $\FFQC$,
                    transactions $\FFtxs$.
                \item Genesis block: $b_0 \triangleq \Block(0, 0, \bot, \emptyset, \emptyset)$.
                \item Validity:
                    \begin{IEEEeqnarray}{rCCCl}
                        \MvalidB(\StM{}{}, b)
                        &\triangleq&
                        && (b = b_0) \nonumber\\
                        &&\lor&& ((b \in \StM{}{}) \nonumber\\ %
                        &&&\land& \MvalidQC(\StM{}{}, b.\FFQC, b.\FFbmI, 1) \nonumber\\
                        &&&\land& (b.\FFp = \Leader_{b.\FFv}) \nonumber\\
                        &&&\land& (b.\FFv > b.\FFbmI.\FFv))
                        \IEEEeqnarraynumspace
                    \end{IEEEeqnarray}
            \end{itemize}
        \State
            \algloclabel{loc:tendermint-original-votes}
            Votes:
            \begin{itemize}[itemsep=0pt,topsep=0pt,parsep=0pt,partopsep=0pt]
                \item $w \triangleq \Vote(\FFp, \FFb, \FFstage)$ consists of:
                    creator $\FFp$,
                    target block $\FFb$,
                    vote stage $\FFstage$.
                \item Validity:
                    \begin{IEEEeqnarray}{rCCCl}
                        \MvalidV(\StM{}{}, w)
                        &\triangleq&
                        && (w \in \StM{}{}) \land \MvalidB(\StM{}{}, w.\FFb)
                        \IEEEeqnarraynumspace
                    \end{IEEEeqnarray}
            \end{itemize}
        \State
            \algloclabel{loc:tendermint-original-qcs}
            Quorum certificates (QCs):
            \begin{itemize}[itemsep=0pt,topsep=0pt,parsep=0pt,partopsep=0pt]
                \item Validity:
                    \begin{IEEEeqnarray}{rCCCl}
                        \MvalidQC(\StM{}{}, Q, b, s)
                        &\triangleq&
                        && ((Q = \emptyset) \nonumber\\ 
                        &&&\land& (b = b_0)) \nonumber\\
                        &&\lor&& ((\forall w \in Q: \MvalidV(\StM{}{}, w) \land (w.\FFb = b) \land (w.\FFstage = s)) \nonumber\\
                        &&&\land& (|\{ w.\FFp \mid w \in Q \}| > 2n/3)) 
                        \IEEEeqnarraynumspace
                    \end{IEEEeqnarray}
                \item Notation: If $\MvalidQC(\StM{}{}, Q, b, s)$, then $\FFv(Q) \triangleq b.\FFv$.
            \end{itemize}
        \State $\StM{}{} \gets \{ b_0 \}$%
            \algloclabel{loc:tendermint-original-stateM}%
        \State $\StQ{}{} \gets \emptyset$
        \State
        At all times, re-broadcast all messages and transactions received from the network or as input.
        \State
        At all times, 
        add to $\StM{}{}$ any message (block or vote) received from the network, 
        upon verification and stripping of the message's signature (ensuring that message $m$ was indeed created by $m.\FFp$),
        and any transaction received from the network or as input.
        Denote the set of transactions in $\StM{}{}$ as $\StMtxs{}{}$.
        Record when elements are added to $\StM{}{}$,
        to allow access to $\StM{}{}$ and $\StMtxs{}{}$ ``as of'' time $t$
        as $\StM{t}{}$ and $\StMtxs{t}{}$.%
        \algloclabel{loc:tendermint-original-stateM-explain}
        \State
        At all times, confirm as log $\LOG_t^p$ the sequence of transactions on the path from $b_0$ to $b$ iff
        $\exists Q_1, Q_2 \subseteq \StM{}{}: \MvalidQC(\StM{}{}, Q_1, b, 1) \land \MvalidQC(\StM{}{}, Q_2, b, 2)$.%
        \algloclabel{loc:tendermint-original-confirm}
        \For{$v=1,2,3,...$}
            \At{$t=3 \Delta v + 0 \Delta$}%
                \algloclabel{loc:tendermint-original-propose}
                \If{$\Leader_v = p$}
                    \State $(b, Q) \gets \argmax_{(b, Q): \MvalidQC(\StM{}{}, Q, b, 1)} b.\FFv$%
                        \algloclabel{loc:tendermint-original-Lvchooseparent}
                    \State Sign and broadcast $\Block(\FFp \gets p, \FFv \gets v, \FFbmI \gets b, \FFQC \gets Q, \FFtxs \gets 
                        \{ \tx \in \StMtxs{}{} \mid \tx \not\in b_0.\FFtxs \| ... \| b.\FFtxs, \text{ for the path from $b_0$ to $b$} \})$%
                        \algloclabel{loc:tendermint-original-Lvproduceblock}
                \EndIf
            \EndAt
            \At{$t=3 \Delta v + 1 \Delta$}%
                \algloclabel{loc:tendermint-original-vote1}
                \If{$\exists b: \MvalidB(\StM{}{}, b) \land (b.\FFv = v)$}%
                        \algloclabel{loc:tendermint-original-validprop}%
                        \Comment{$\MvalidB(\StM{}{}, b) \implies (b.\FFp = \Leader_v)$. Proceed only with one $b$ satisfying the condition.}
                    \If{$\FFv(\StQ{}{}) \leq \FFv(b.\FFQC)$}%
                            \algloclabel{loc:tendermint-original-admissibleprop}
                        \State Sign and broadcast $\Vote(\FFp \gets p, \FFb \gets b, \FFstage \gets 1)$%
                            \algloclabel{loc:tendermint-original-stage1vote}
                    \EndIf
                \EndIf
            \EndAt
            \At{$t=3 \Delta v + 2 \Delta$}%
                \algloclabel{loc:tendermint-original-vote2}
                \If{$\exists b, Q: (b.\FFv = v) \land \MvalidQC(\StM{}{}, Q, b, 1)$}%
                    \algloclabel{loc:tendermint-original-validstage1qc}
                    \Comment{Proceed only with one $(b,Q)$ satisfying the condition.}
                    \State $\StQ{}{} \gets Q$%
                        \algloclabel{loc:tendermint-original-updatelock}
                    \State Sign and broadcast $\Vote(\FFp \gets p, \FFb \gets b, \FFstage \gets 2)$%
                        \algloclabel{loc:tendermint-original-stage2vote}
                \EndIf
            \EndAt
        \EndFor
    \end{algorithmic}
\end{algorithm}

\begin{remark}
    Note that $\MvalidQC(\StM{}{}, Q, b, s)$
    implies $\MvalidB(\StM{}{}, b)$ because $\MvalidV(\StM{}{}, w)$ implies $\MvalidB(\StM{}{}, w.\FFb)$.
\end{remark}

\begin{remark}
    \label{rem:tendermint-2-blocktree}
    Note that, through the parent block relation in \alglocref{alg:tendermint-2}{loc:tendermint-2-blocks}, valid blocks form a tree rooted at the genesis block $b_0$.
    As a result, the path from $b_0$ to any valid block $b$ exists and is unique.
    Block $b$ is an \emph{ancestor} of block $b'$, denoted as $b \preceq b'$, iff $b$ is on the path from $b_0$ to $b'$.
    Ancestors of any valid block are from strictly increasing views. 
\end{remark}

\begin{remark}
    \label{rem:tendermint-2-lockview-nondecreasing}
    Assuming $f \leq 2n/3$
    (which holds in particular if we assume $n>3f$),
    by \alglocref{alg:tendermint-2}{loc:tendermint-2-qcs},
    an honest node's vote is required to form any valid quorum certificate (QC).
    Thus, by \cref{alg:tendermint-2}, for any view $v$, no honest node can see any valid QC for $v$
    before $12 \Delta v + 4 \Delta$.
    It follows that in every honest node $p$'s view, $\FFv(\StQ{t}{p})$ as a function of $t$ is non-decreasing.
    To see this, suppose $t' = 12 \Delta v' + 7 \Delta$ for view $v'$ is a time where $p$ updates $\StQ{}{p}$
    such that $v'' \triangleq \FFv(\StQ{t}{p}) > \FFv(\StQ{t'^+}{p}) = v'$ for some time $t$ with $t < t'^+$.
    Then $p$ saw a valid QC for view $v''$ at time $t < 12 \Delta v' + 7 \Delta < 12 \Delta v'' + 4 \Delta$,
    a contradiction to the aforementioned.
\end{remark}

\subsection{Proof of \cref{thm:tendermint-2-safety}}
\label{sec:proof-tendermint-2-safety}

\begin{proof}[Proof of \cref{thm:tendermint-2-safety}]
    Towards a contradiction, suppose honest $p$ confirms 
    $b$ 
    (\ie, $p$ outputs as its log $\LOG_t^p$ the sequence of transactions on the path from $b_0$ to $b$ at some time $t$,
    as per \alglocref{alg:tendermint-2}{loc:tendermint-2-confirm}),
    and honest $p'$ confirms $b'$, but neither $b \preceq b'$ nor $b' \preceq b$.
    Without loss of generality, assume $b.\FFv \leq b'.\FFv$.
    Let $b''$ be the block such that
    $b'' \preceq b'$
    and $b''.\FFv$ is minimal,
    subject to the constraint $b''.\FFv \geq b.\FFv$.
    Note that $b''$ exists because $b'$ is a candidate,
    and $b''$ is unique due to \cref{rem:tendermint-2-blocktree}.
    Note that $b'' \neq b$ because otherwise $b \preceq b'$, which would be a contradiction to earlier assumptions.
    Note that because $p$ confirms $b$,
    $\exists Q_1, Q_2: \MvalidQC(\StM{\infty}{\cup}, Q_1, b, 1) \land \MvalidQC(\StM{\infty}{\cup}, Q_2, b, 2)$.
    Because $p'$ confirms $b'$, $p'$ views $b'$ as valid, and thus all its ancestors have valid stage-$1$ QCs, \ie,
    $\exists Q_3: \MvalidQC(\StM{\infty}{\cup}, Q_3, b'', 1)$.

    Suppose $b''.\FFv = b.\FFv$.
    Due to the assumption $n>3f$,
    and quorum intersection,
    some honest node has contributed to both $Q_1$ and $Q_3$, a contradiction because according to \cref{alg:tendermint-2}
    honest nodes send only one stage-$1$ vote per view.

    Suppose $b''.\FFv > b.\FFv$.
    Let $Q_0 \triangleq b''.\FFQC$. 
    Note that $\FFv(Q_0) < b''.\FFv$ due to \cref{rem:tendermint-2-blocktree}. 
    Note that it must be that $\FFv(Q_0) < b.\FFv$ because $b''$ was chosen with minimal $b''.\FFv$ (subject to $b''.\FFv \geq b.\FFv$).
    Due to the assumption $n>3f$,
    and quorum intersection,
    some honest node $p$ has contributed to both $Q_2$ and $Q_3$.
    Thus, $\FFv(\StQ{12 \Delta (b.\FFv + 1)}{p}) = b.\FFv$, by \alglocref{alg:tendermint-2}{loc:tendermint-2-updatelock}.
    Yet, $p$ allegedly stage-$1$ votes for $b''$ in a later $b''.\FFv \geq b.\FFv + 1$, \ie, after $12 \Delta b''.\FFv \geq 12 \Delta (b.\FFv + 1)$, specifically at $12 \Delta b''.\FFv + 4 \Delta$.
    This is even though $\FFv(b''.\FFQC) = \FFv(Q_0) < b.\FFv = \FFv(\StQ{12 \Delta (b.\FFv + 1)}{p}) \leq \FFv(\StQ{12 \Delta b''.\FFv}{p}) \leq \FFv(\StQ{12 \Delta b''.\FFv + 4 \Delta}{p})$ (recall that by \cref{rem:tendermint-2-lockview-nondecreasing}, $\FFv(\StQ{t}{p})$ is non-decreasing in $t$),
    a contradiction to \alglocref{alg:tendermint-2}{loc:tendermint-2-admissibleprop}.
\end{proof}

\subsection{Proof of \cref{thm:tendermint-2-liveness}}
\label{sec:proof-tendermint-2-liveness}

\begin{proof}[Proof of \cref{thm:tendermint-2-liveness}]
    Suppose the leader $\Leader_v$ is honest for some view $v$ with $12 \Delta v \geq \GST$.
    Because the network is synchronous,
    $\StM{12 \Delta v + 2 \Delta}{\Leader_v} \supseteq \StM{12 \Delta v}{p}$
    for any honest node $p$.
    Furthermore, honest nodes do not update their $\StQ{}{}$ in the time frame from $(12 \Delta (v-1) + 7 \Delta)^+$ to $(12 \Delta v + 7 \Delta)^-$, and thus also not in the time frame from $12 \Delta v$ to $12 \Delta v + 4 \Delta$.
    As a result, $Q$ chosen by $\Leader_v$ at $12 \Delta v + 2 \Delta$ in \alglocref{alg:tendermint-2}{loc:tendermint-2-Lvchooseparent} is such that
    for every honest node $p$, $\FFv(\StQ{12 \Delta v + 4 \Delta}{p}) \leq \FFv(Q)$
    in \alglocref{alg:tendermint-2}{loc:tendermint-2-admissibleprop}.
    Furthermore, 
    by synchrony,
    $\StM{12 \Delta v + 4 \Delta}{p} \supseteq \StM{12 \Delta v + 2 \Delta}{\Leader_v}$.
    Therefore, $b, Q$ chosen by $\Leader_v$ at $12 \Delta v + 2 \Delta$ in \alglocref{alg:tendermint-2}{loc:tendermint-2-Lvchooseparent} are viewed as
    valid by all honest nodes at $12 \Delta v + 4 \Delta$,
    and so is the resulting block $b^*$ produced by $\Leader_v$ at $12 \Delta v + 2 \Delta$ in \alglocref{alg:tendermint-2}{loc:tendermint-2-Lvproduceblock},
    which by synchrony all honest nodes receive by $12 \Delta v + 4 \Delta$.
    Thus, the condition of \alglocref{alg:tendermint-2}{loc:tendermint-2-validprop} is satisfied in all honest views,
    and all honest nodes will stage-$1$ vote for $b^*$ in \alglocref{alg:tendermint-2}{loc:tendermint-2-stage1vote}.
    By synchrony, all honest votes propagate to all honest nodes by $12 \Delta v + 7 \Delta$.
    Thus, the condition of \alglocref{alg:tendermint-2}{loc:tendermint-2-validstage1qc} is satisfied in all honest views.
    Furthermore, since $\Leader_v$ is honest and produces only one block in view $v$, $b^*$ is the only block that satisfies \alglocref{alg:tendermint-2}{loc:tendermint-2-validstage1qc} in any honest view.
    As a result,
    all honest nodes will stage-$2$ vote for $b^*$ at $12 \Delta v + 7 \Delta$.
    Again by synchrony, these votes will propagate to all honest nodes by $12 \Delta v + 9 \Delta$,
    who will thus confirm $b^*$ and with it all pending transactions that any honest node has seen
    and re-broadcast
    by $12 \Delta v + \Delta$, and which thus (by synchrony) the honest $\Leader_v$ included in $b^*$ at $12 \Delta v + 2 \Delta$.
\end{proof}

\section{Addendum Impossibility Results}
\label{sec:addendum-impossibilities}

A key step in establishing the impossibility of accountable liveness is to obtain executions in which
timely-liveness
must be violated for ``reasonable'' protocols.
In this context, a useful consequence of optimal resilience 
(\cf \cref{def:optresilient}) 
is that the liveness guarantee for such protocols is tight in the sense that there must exist an execution with $\tL+1$ \emph{crash faults} in which timely-liveness is violated:
\begin{lemma}
    \label{lem:existence-unliveness-crashes}
    If an atomic broadcast protocol is $\tS$-safe
    under partial synchrony
    and either $\tS > 0$ or $n$ is even,
    then, under
    synchrony,
    some execution
    with $\lceil(n-\tS)/2\rceil$ crash faults
    violates 
    timely-liveness.
\end{lemma}
\begin{proof}
    The argument proceeds analogously to the classical ``split-brain converse''~\cite{model-psync}.
    Towards a contradiction, suppose $\Pi$ 
    satisfies the conditions of \cref{lem:existence-unliveness-crashes},
    \ie, $\Pi$
    is $\tS$-safe under partial synchrony,
    but under
    $(\Delta, \netG, \netX)$-partial-synchrony
    $\Pi$
    preserves timely-liveness
    (in fact it suffices to consider traditional liveness)
    for all except at most $(\lceil(n-\tS)/2\rceil-1)$ honest nodes
    in all executions with $\lceil(n-\tS)/2\rceil$ crash faults.
    Note that $\tL \leq (\lceil(n-\tS)/2\rceil-1)$ is the maximum liveness resilience that can be expected due to \cref{def:optresilient}.

    Note that
    we can partition $\CP$ into $\CP_1, \CP_2, \CP_3$ with
    $|\CP_1| = |\CP_3| = \lceil(n-\tS)/2\rceil$
    and $|\CP_2| \leq \tS$.
    Let $\tx_1, \tx_2$ be two high-entropy transactions,
    so that $\Pi$ cannot guess them, and they cannot have been hard-coded into $\Pi$. 

    \underline{Execution $E_1$:}
    The nodes of $\CP_1$ crash.
    The nodes of $\CP_2 \cup \CP_3$ are honest.
    The network is synchronous.
    The environment inputs $\tx_1$ into all nodes,
    and does not input $\tx_2$ into any node.
    Since $\Pi$ is assumed to preserve 
    liveness 
    for all except at most $(\lceil(n-\tS)/2\rceil-1)$ honest nodes
    in all executions with $\lceil(n-\tS)/2\rceil$ crash faults
    under synchrony
    and $|\CP_1| = \lceil(n-\tS)/2\rceil$
    and $|\CP_3| = \lceil(n-\tS)/2\rceil > \lceil(n-\tS)/2\rceil-1$,
    by some time $t_1$,
    some (honest) node $p_3\in\CP_3$ will output
    a log containing $\tx_1$ and not containing $\tx_2$.

    \underline{Execution $E_2$:}
    The nodes of $\CP_3$ crash.
    The nodes of $\CP_1 \cup \CP_2$ are honest.
    The network is synchronous.
    The environment inputs $\tx_2$ into all nodes,
    and does not input $\tx_1$ into any node.
    Since $\Pi$ is assumed to preserve 
    liveness 
    for all except at most $(\lceil(n-\tS)/2\rceil-1)$ honest nodes
    in all executions with $\lceil(n-\tS)/2\rceil$ crash faults
    under synchrony %
    and $|\CP_3| = \lceil(n-\tS)/2\rceil$
    and $|\CP_1| = \lceil(n-\tS)/2\rceil > \lceil(n-\tS)/2\rceil-1$,
    by some time $t_2$,
    some (honest) node $p_1\in\CP_1$ will output
    a log containing $\tx_2$ and not containing $\tx_1$.
    
    \underline{Execution $E_3$:}
    Before round $\max(t_1, t_2)$,
    communication among the nodes of $\CP_1 \cup \CP_2$ is synchronous,
    and communication among the nodes of $\CP_2 \cup \CP_3$ is synchronous,
    but any nodes $p \in \CP_1, q \in \CP_3$ cannot communicate (asynchrony).
    Such a delay satisfies partial synchrony with $\GST = \max(t_1, t_2)$.
    The nodes of $\CP_1 \cup \CP_3$ are honest.
    The nodes of $\CP_2$ behave like the ``split-brain'' adversary:
    To the nodes of $\CP_1$, they behave like the nodes $\CP_2$ in $E_2$.
    To the nodes of $\CP_3$, they behave like the nodes $\CP_2$ in $E_1$.
    The environment inputs $\tx_1$ into all nodes of $\CP_2 \cup \CP_3$,
    and $\tx_2$ to all nodes of $\CP_1 \cup \CP_2$.
    Since $E_3$ is indistinguishable from $E_1$ until round $\max(t_1, t_2)$
    for the nodes of $\CP_3$,
    and $E_3$ is indistinguishable from $E_2$ until round $\max(t_1, t_2)$
    for the nodes of $\CP_1$,
    (honest) node $p_3\in\CP_3$ (see $E_1$) will output a log containing $\tx_1$ and not containing $\tx_2$,
    and (honest) node $p_1\in\CP_1$ (see $E_2$) will output a log containing $\tx_2$ and not containing $\tx_1$.
    This is a safety violation that contradicts the assumption that $\Pi$ is $\tS$-safe under partial synchrony,
    since $|\CP_2| \leq \tS$.
\end{proof}

\subsection{Proof of \cref{thm:impossibility-tALmax-geq-n2}}
\label{sec:proof-impossibility-tALmax-geq-n2}

\begin{proof}[Proof of \cref{thm:impossibility-tALmax-geq-n2}]
    Towards a contradiction, suppose $\Pi$ satisfies the conditions of \cref{thm:impossibility-tALmax-geq-n2} for period length $\netDeltaPeriod$
    and is accountably live with appropriate parameters.
    Pick any even $n$.

    \underline{Execution $E_1$:}
    Consider an execution 
    under synchrony %
    where the crash of the nodes in $\CP_1$ with $|\CP_1|=n/2$
    causes a 
    timely-liveness violation.
    Such an execution exists according to \cref{lem:existence-unliveness-crashes}.
    Let $\CP_2 \triangleq \CP \setminus \CP_1$ behave honestly.
    Due to $\tALmax \geq n/2$, $\tALident > 0$, 
    $\Pi$ purportedly being accountably live, and the timely-liveness violation, honest nodes will eventually produce
    a certificate of guilt for one of the nodes in $\CP_1$.
    
    \underline{Execution $E_2$:}
    The roles of $\CP_1$ and $\CP_2$ are swapped compared to $E_1$:
    $\CP_1$ are now honest, and $\CP_2$ are adversary.
    The adversary nodes do not communicate with the honest nodes,
    but otherwise behave like the honest nodes in $E_1$.

    Because the views (on the protocol execution) of the adversary nodes in $E_2$
    are identical to those of the honest nodes in $E_1$,
    the adversary nodes 
    will eventually produce
    a certificate of guilt for one of the nodes in $\CP_1$.
    But all $\CP_1$ are honest in $E_2$.
    This is a contradiction to the definition of accountable liveness, as desired.
\end{proof}

\subsection{Proof of \cref{thm:impossibility-x-geq-12}}
\label{sec:proof-impossibility-x-geq-12}

\begin{proof}[Proof of \cref{thm:impossibility-x-geq-12}]
    \ifcompilewithoutheavyfigures\else
        \begin{figure*}[tb]%
    \centering%
    \begin{tikzpicture}[%
            nodesHon/.style={thick,Green},
            nodesAdv/.style={thick,Red},
            viewsHon/.style={thick,Green},
            viewsAdv/.style={thick,Red},
            commsBdY/.style={{Straight Barb[angle=60:1.5pt 3]}-{Straight Barb[angle=60:1.5pt 3]},Green},
            commsBdN/.style={{Straight Barb[angle=60:1.5pt 3]}-{Straight Barb[angle=60:1.5pt 3]},Red},
            commsUdY/.style={-{Straight Barb[angle=60:1.5pt 3]},Green},
            commsUdN/.style={-{Straight Barb[angle=60:1.5pt 3]},Red},
            commsP/.style={Red},
        ]
        \footnotesize

        \begin{scope}[x=2cm,y=1.5cm,xshift=-7.2cm,yshift=4.25cm]
            \myDrawScenario{P/0.4/{$\CP_1(\tx)$}/{nodesHon},Q/0.4/{$\CP_2(\tx)$}/{nodesAdv},R/0.2/{$\CP_3(\tx)$}/{nodesHon}}{V1/0.5/{$\CT_1$}/{viewsHon},V2/0.5/{$\CT_2$}/{viewsAdv}};

            \node [commsP] at (QV1mid) {\myIconAdvPartition};
            \node [commsP] at (PV2mid) {\myIconHonPartition};
            \draw [commsBdY] (PV1botlet1) -- (RV1toplet1);
            \draw [commsBdY] (QV2botrit1) -- (RV2toprit1);
            \draw [commsUdN] (QV1toprit1) -- (PV2botlet1);
            \draw [commsUdN] (PV1botrit1) -- (QV2toplet1);

            \node [anchor=south] (EB) at ([yshift=2em]0.5,1) {Execution $E_B$};
            \node [anchor=north,align=center] (EBlabel) at ([yshift=-2.5em]0.5,0) {\emph{Assume for contradiction:}\\Certif.\ of guilt for $p'\in\CP_2$};

            \node [above,xshift=-0.75em] at (toplet) {\tiny\scriptspacing$T_0=0$};
            \node [above,xshift=-0.5em] at (V1toprit) {\tiny\scriptspacing$T_1=\netDeltaPeriod \netG(\netDeltaPeriod)/2$};
            \node [above,xshift=1.25em] at (toprit) {\tiny\scriptspacing$T_2=\netDeltaPeriod \netG(\netDeltaPeriod)$};
        \end{scope}

        \begin{scope}[x=2cm,y=1.5cm,xshift=-3.7cm,yshift=4.25cm]
            \myDrawScenario{P/0.4/{$\CP_1(\tx)$}/{nodesAdv},Q/0.4/{$\CP_2(\tx)$}/{nodesHon},R/0.2/{$\CP_3(\tx)$}/{nodesHon}}{V1/0.5/{$\CT_1$}/{viewsAdv},V2/0.5/{$\CT_2$}/{viewsHon}};

            \node [commsP] at (QV1mid) {\myIconHonPartition};
            \node [commsP] at (PV2mid) {\myIconAdvPartition};
            \draw [commsBdY] (PV1botlet1) -- (RV1toplet1);
            \draw [commsBdY] (QV2botrit1) -- (RV2toprit1);
            \draw [commsUdN] (QV1toprit1) -- (PV2botlet1);
            \draw [commsUdN] (PV1botrit1) -- (QV2toplet1);

            \node [anchor=south] (EC) at ([yshift=2em]0.5,1) {Execution $E_C$};
            \node [anchor=north,align=center] (EClabel) at ([yshift=-2.5em]0.5,0) {Certif.\ of guilt for $p'\in\CP_2$\\\emph{Contradiction, $\CP_2 \subseteq \CPh$!}};

            \draw [{Straight Barb[angle=60:1.5pt 3]}-{Straight Barb[angle=60:1.5pt 3]}] ([xshift=0.5em]Ptoprit) -- ([xshift=0.5em]Pbotrit) node [midway,right] {$=\tL+1$};
            \draw [{Straight Barb[angle=60:1.5pt 3]}-{Straight Barb[angle=60:1.5pt 3]}] ([xshift=0.5em]Qtoprit) -- ([xshift=0.5em]Qbotrit) node [midway,right] {$=\tL+1$};
            \draw [{Straight Barb[angle=60:1.5pt 3]}-{Straight Barb[angle=60:1.5pt 3]}] ([xshift=0.5em]Rtoprit) -- ([xshift=0.5em]Rbotrit) node [midway,right] {$\leq\tS$};

            \node [above,xshift=-0.75em] at (toplet) {\tiny\scriptspacing$T_0=0$};
            \node [above,xshift=-0.5em] at (V1toprit) {\tiny\scriptspacing$T_1=\netDeltaPeriod \netG(\netDeltaPeriod)/2$};
            \node [above,xshift=1.25em] at (toprit) {\tiny\scriptspacing$T_2=\netDeltaPeriod \netG(\netDeltaPeriod)$};
        \end{scope}

        \node [anchor=south] at ($(EB.south)!0.5!(EC.south)$) {$\approx$};
        
        \begin{scope}[x=2cm,y=1.5cm,xshift=3.7cm,yshift=4.25cm,opacity=0.5]
            \myDrawScenario{P/0.4/{$\CP_1(\tx')$}/{nodesHon},Q/0.4/{$\CP_2(\tx')$}/{nodesAdv},R/0.2/{$\CP_3(\tx')$}/{nodesHon}}{V1/0.5/{$\CT_1$}/{viewsAdv},V2/0.5/{$\CT_2$}/{viewsHon}};

            \node [commsP] at (PV1mid) {\myIconHonPartition};
            \node [commsP] at (QV2mid) {\myIconAdvPartition};
            \draw [commsBdY] (QV1botlet1) -- (RV1toplet1);
            \draw [commsBdY] (PV2botrit1) -- (RV2toprit1);
            \draw [commsUdN] (QV1toprit1) -- (PV2botlet1);
            \draw [commsUdN] (PV1botrit1) -- (QV2toplet1);

            \node [anchor=south] (ECp) at ([yshift=2em]0.5,1) {Execution $E_{C'}$};
            \node [anchor=north,align=center] (ECplabel) at ([yshift=-2.5em]0.5,0) {Certif.\ of guilt for $p'\in\CP_1$\\\emph{Contradiction, $\CP_1 \subseteq \CPh$!}};

            \node [above] at (toplet) {\tiny\scriptspacing$T_0$};
            \node [above] at (V1toprit) {\tiny\scriptspacing$T_1$};
            \node [above] at (toprit) {\tiny\scriptspacing$T_2$};
        \end{scope}
        
        \begin{scope}[x=2cm,y=1.5cm,xshift=7.2cm,yshift=4.25cm,opacity=0.5]
            \myDrawScenario{P/0.4/{$\CP_1(\tx')$}/{nodesAdv},Q/0.4/{$\CP_2(\tx')$}/{nodesHon},R/0.2/{$\CP_3(\tx')$}/{nodesHon}}{V1/0.5/{$\CT_1$}/{viewsHon},V2/0.5/{$\CT_2$}/{viewsAdv}};

            \node [commsP] at (PV1mid) {\myIconAdvPartition};
            \node [commsP] at (QV2mid) {\myIconHonPartition};
            \draw [commsBdY] (QV1botlet1) -- (RV1toplet1);
            \draw [commsBdY] (PV2botrit1) -- (RV2toprit1);
            \draw [commsUdN] (QV1toprit1) -- (PV2botlet1);
            \draw [commsUdN] (PV1botrit1) -- (QV2toplet1);

            \node [anchor=south] (EBp) at ([yshift=2em]0.5,1) {Execution $E_{B'}$};
            \node [anchor=north,align=center] (EBplabel) at ([yshift=-2.5em]0.5,0) {\emph{Assume for contradiction:}\\Certif.\ of guilt for $p'\in\CP_1$};

            \node [above] at (toplet) {\tiny\scriptspacing$T_0$};
            \node [above] at (V1toprit) {\tiny\scriptspacing$T_1$};
            \node [above] at (toprit) {\tiny\scriptspacing$T_2$};
        \end{scope}

        \node [anchor=south,opacity=0.5] at ($(EBp.south)!0.5!(ECp.south)$) {$\approx$};

        \begin{scope}[x=2cm,y=1.5cm,xshift=-7.2cm]
            \myDrawScenario{P/0.4/{$\CP_1(\tx)$}/{nodesHon},Q/0.4/{$\CP_2(\tx)$}/{nodesHon},R/0.2/{$\CP_3(\tx)$}/{nodesHon}}{V1/0.5/{$\CT_1$}/{viewsAdv},V2/0.5/{$\CT_2$}/{viewsAdv}};

            \node [commsP] at (QV1mid) {\myIconHonPartition};
            \node [commsP] at (PV2mid) {\myIconHonPartition};
            \draw [commsBdY] (PV1botlet1) -- (RV1toplet1);
            \draw [commsBdY] (QV2botrit1) -- (RV2toprit1);
            \draw [commsUdN] (QV1toprit1) -- (PV2botlet1);
            \draw [commsUdN] (PV1botrit1) -- (QV2toplet1);

            \node [anchor=south] (E1) at ([yshift=1em]0.5,1) {Execution $E_A = E_1$};
            \node [anchor=north,align=center] (E1label) at ([yshift=-2.5em]0.5,0) {\emph{Assume for contradiction:}\\Some $p_1\in\CP_1$ confirms $\tx$\\and does not confirm $\tx'$};
        \end{scope}
        
        \begin{scope}[x=2cm,y=1.5cm,xshift=-3.7cm]
            \myDrawScenario{P/0.4/{$\CP_1(\tx)$}/{nodesHon},Q/0.4/{$\CP_2(\tx')$}/{nodesHon},R/0.2/{$\CP_3(\tx)$}/{nodesHon}}{V1/0.5/{$\CT_1$}/{viewsAdv},V2/0.5/{$\CT_2$}/{viewsAdv}};

            \node [commsP] at (QV1mid) {\myIconHonPartition};
            \node [commsP] at (PV2mid) {\myIconHonPartition};
            \draw [commsBdY] (PV1botlet1) -- (RV1toplet1);
            \draw [commsBdY] (QV2botrit1) -- (RV2toprit1);
            \draw [commsUdN] (QV1toprit1) -- (PV2botlet1);
            \draw [commsUdN] (PV1botrit1) -- (QV2toplet1);

            \node [anchor=south] (E3) at ([yshift=1em]0.5,1) {Execution $E_3$};
            \node [anchor=north,align=center] (E3label) at ([yshift=-2.5em]0.5,0) {Some $p_1\in\CP_1$ confirms $\tx$\\and does not confirm $\tx'$};
        \end{scope}
        
        \begin{scope}[x=2cm,y=1.5cm,xshift=0cm]
            \myDrawScenario{P/0.4/{$\CP_1(\tx)$}/{nodesHon},Q/0.4/{$\CP_2(\tx')$}/{nodesHon},R/0.2/{$\CP_3(\tx,\tx')$}/{nodesAdv}}{V1/0.5/{$\CT_1$}/{viewsAdv},V2/0.5/{$\CT_2$}/{viewsAdv}};

            \node [commsP] at (PV2mid) {\myIconHonPartition};
            \node [commsP] at (QV2mid) {\myIconHonPartition};
            \draw [commsBdY] (PV1botlet1) -- (RV1toplet1);
            \draw [commsBdY] (QV1botlet2) -- (RV1toplet2);
            \draw [commsBdN] (PV1botlet2) -- (QV1toplet2);
            \draw [commsUdN] (QV1toprit1) -- (PV2botlet1);
            \draw [commsUdN] (PV1botrit1) -- (QV2toplet1);

            \node [anchor=south] (E5) at ([yshift=1em]0.5,1) {Execution $E_5$};
            \node [anchor=north,align=center] (E5label) at ([yshift=-2.5em]0.5,0) {Some $p_1\in\CP_1$ confirms $\tx$\\and does not confirm $\tx'$\\Some $p_2\in\CP_2$ confirms $\tx'$\\and does not confirm $\tx$\\\emph{Safety violation, contradiction!}};
        \end{scope}
        
        \begin{scope}[x=2cm,y=1.5cm,xshift=3.7cm]
            \myDrawScenario{P/0.4/{$\CP_1(\tx)$}/{nodesHon},Q/0.4/{$\CP_2(\tx')$}/{nodesHon},R/0.2/{$\CP_3(\tx')$}/{nodesHon}}{V1/0.5/{$\CT_1$}/{viewsAdv},V2/0.5/{$\CT_2$}/{viewsAdv}};

            \node [commsP] at (PV1mid) {\myIconHonPartition};
            \node [commsP] at (QV2mid) {\myIconHonPartition};
            \draw [commsBdY] (QV1botlet1) -- (RV1toplet1);
            \draw [commsBdY] (PV2botrit1) -- (RV2toprit1);
            \draw [commsUdN] (QV1toprit1) -- (PV2botlet1);
            \draw [commsUdN] (PV1botrit1) -- (QV2toplet1);

            \node [anchor=south] (E4) at ([yshift=1em]0.5,1) {Execution $E_4$};
            \node [anchor=north,align=center] (E4label) at ([yshift=-2.5em]0.5,0) {Some $p_2\in\CP_2$ confirms $\tx'$\\and does not confirm $\tx$};
        \end{scope}
        
        \begin{scope}[x=2cm,y=1.5cm,xshift=7.2cm]
            \myDrawScenario{P/0.4/{$\CP_1(\tx')$}/{nodesHon},Q/0.4/{$\CP_2(\tx')$}/{nodesHon},R/0.2/{$\CP_3(\tx')$}/{nodesHon}}{V1/0.5/{$\CT_1$}/{viewsAdv},V2/0.5/{$\CT_2$}/{viewsAdv}};

            \node [commsP] at (PV1mid) {\myIconHonPartition};
            \node [commsP] at (QV2mid) {\myIconHonPartition};
            \draw [commsBdY] (QV1botlet1) -- (RV1toplet1);
            \draw [commsBdY] (PV2botrit1) -- (RV2toprit1);
            \draw [commsUdN] (QV1toprit1) -- (PV2botlet1);
            \draw [commsUdN] (PV1botrit1) -- (QV2toplet1);

            \node [anchor=south] (E2) at ([yshift=1em]0.5,1) {Execution $E_2$};
            \node [anchor=north,align=center] (E2label) at ([yshift=-2.5em]0.5,0) {\emph{Assume for contradiction:}\\Some $p_2\in\CP_2$ confirms $\tx'$\\and does not confirm $\tx$};
        \end{scope}

        \node [anchor=south] at ($(E1.south)!0.5!(E3.south)$) {$\overset{\CP_1}{\approx}$};
        \node [anchor=south] at ($(E3.south)!0.5!(E5.south)$) {$\overset{\CP_1}{\approx}$};
        \node [anchor=south] at ($(E5.south)!0.5!(E4.south)$) {$\overset{\CP_2}{\approx}$};
        \node [anchor=south] at ($(E4.south)!0.5!(E2.south)$) {$\overset{\CP_2}{\approx}$};

        \node [anchor=center] at ($(E1.north)!0.5!(EBlabel.south)$) {$\approx$};
        \node [anchor=center,opacity=0.5] at ($(E2.north)!0.5!(EBplabel.south)$) {$\approx$};
        
    \end{tikzpicture}%
    \caption[]{%
        Illustration of indistinguishable executions $E_B, E_C, E_A = E_1, E_2, E_3, E_4, E_5$ used in the proof of \cref{thm:impossibility-x-geq-12}.
        Nodes (with their respective transaction input at round $0$) are indicated on the vertical axis, time (from the beginning of round $0$ to the beginning of round $\netDeltaPeriod \netG(\netDeltaPeriod)$) is indicated on the horizontal axis.
        Green/red bars at the edge of the rectangles indicate honest/adversary nodes and synchronous/asynchronous rounds, respectively.
        Green/red arrows indicate communication between groups of nodes that does/doesn't occur.
        Red ``\myIconHonPartition{}'' indicates a network partition,
        ``\myIconAdvPartition{}'' indicates an adversary-emulated network partition.
    }%
    \label{fig:impossibility-x-geq-12}%
\end{figure*}%
    \fi
    Without loss of generality, we consider $\netX=1/2$.
    If $\tALmax \geq n/2$, we are done by \cref{thm:impossibility-tALmax-geq-n2}.
    So we assume $\tALmax < n/2$.
    Towards a contradiction, suppose protocol $\Pi$ satisfies the conditions of \cref{thm:impossibility-x-geq-12} for some period length $\netDeltaPeriod$.
    We require the mild regularity condition that $\netDeltaPeriod \netG(\netDeltaPeriod)$ is an even number.

    We proceed in two steps:
    (1)~We describe three executions $E_A$, $E_B$, and $E_C$ 
    that are indistinguishable in terms of which and when messages are received by honest nodes. 
    We assume that 
    timely-liveness
    is violated 
    (with a particular set of nodes that do not confirm)
    in $E_A$
    (and due to indistinguishability also in $E_B$ and $E_C$).
    We then show that if $\Pi$ is accountably live with $\tALident > 0$, then a 
    certificate of guilt must eventually be produced in $E_B$ for an adversary node,
    but since $E_B$ and $E_C$ are indistinguishable, a 
    certificate of guilt is eventually produced in $E_C$ for the same node,
    but that node is honest in $E_C$.
    This is the contradiction, and thus such $\Pi$ cannot exist.
    (2)~We show, using five executions $E_1, E_2, E_3, E_4, E_5$, where $E_A = E_1$, that due to the protocol being optimally-resilient, 
    timely-liveness
    must indeed be violated 
    (with a particular set of nodes that do not confirm)
    in $E_A$, discharging the earlier assumption used in~(1).
    
    We start with~(1).
    Partition $\CP$ into $\CP_1, \CP_2, \CP_3$ with
    $|\CP_1| = \tL + 1 \leq \tALmax$,
    $|\CP_2| = \tL + 1 \leq \tALmax$,
    $|\CP_3| = n - 2(\tL + 1) \leq \tS$. 
    See \cref{fig:impossibility-x-geq-12}, which illustrates all subsequently detailed executions from the beginning of round $0$ to the beginning of round $\netDeltaPeriod \netG(\netDeltaPeriod)$.
    For easy reference to sets of rounds of the executions,
    let $T_i \triangleq i \netDeltaPeriod \netG(\netDeltaPeriod)/2$ for $i\in\{0,1,2\}$,
    $\CT_i \triangleq [T_{i-1},T_i)$ for $i\in\{1,2\}$,
    and $\CT \triangleq \CT_1 \cup \CT_2$.

    \underline{Execution~$E_A$ (\cf \cref{fig:impossibility-x-geq-12}, ``Execution $E_A = E_1$''):}
    All nodes are input $\tx$ 
    at time $0$. No other transactions are input.
    All nodes are honest.
    The network is asynchronous 
    during $\CT$ (see \cref{fig:impossibility-x-geq-12}),
    and next-round-delay afterwards for all nodes (not shown in \cref{fig:impossibility-x-geq-12}).
    (By ``next-round-delay'' we mean every message arrives in the next round of the model. Recall, the network delay bound is $\Delta$ rounds.)
    During asynchrony, communication between any nodes $p\in \CP_1, q\in \CP_2$ is delayed until $T_2$.
    During $\CT_1$,
    communication in $\CP_1 \cup \CP_3$ is next-round-delay. All other communication is delayed until $T_1$.
    During $\CT_2$,
    communication in $\CP_2 \cup \CP_3$ is next-round-delay. All other communication is delayed until $T_2$.

    Note that we are \emph{not invoking accountable liveness} in $E_A$,
    and therefore $E_A$ does not have to be $\netX$-partially-synchronous,
    and we may (and do) assume that the network is asynchronous for the entire $\CT$.
    We only assume (for the purposes of part~(1), and this assumption is subsequently discharged in part~(2)) that there is a timely-liveness
    violation in $E_A$.
    More specifically, we assume that $\tx$ is not confirmed by any honest node $p_1 \in \CP_1$ by the end of $\CT$ in $E_A$.

    \underline{Execution~$E_B$ (\cf \cref{fig:impossibility-x-geq-12}):}
    All nodes are input $\tx$ 
    at time $0$. No other transactions are input.
    Nodes in $\CP_1 \cup \CP_3$ are honest.
    Nodes in $\CP_2$ are adversary.
    The network is asynchronous during 
    $\CT_2$,
    and otherwise has next-round-delay for all nodes.
    During 
    $\CT$,
    adversary nodes in $\CP_2$ delay the sending to and receiving from nodes in $\CP_1$ until 
    $T_2$.
    During 
    $\CT_1$,
    adversary nodes in $\CP_2$ also delay the sending to and receiving from nodes in $\CP_3$ until 
    $T_1$.
    During 
    $\CT_2$,
    adversary nodes in $\CP_2$ communicate with nodes in $\CP_3$ using next-round-delay (and the network asynchrony allows this).
    During network asynchrony, communication between any node $p\in \CP_1$ and any node outside $\CP_1$ is delayed until 
    $T_2$.
    After 
    $T_2$,
    the adversary nodes behave honestly.

    \underline{Execution~$E_C$ (\cf \cref{fig:impossibility-x-geq-12}):}
    All nodes are input $\tx$ 
    at time $0$. No other transactions are input.
    Nodes in $\CP_2 \cup \CP_3$ are honest.
    Nodes in $\CP_1$ are adversary.
    The network is asynchronous during 
    $\CT_1$,
    and otherwise has next-round-delay for all nodes.
    During 
    $\CT$,
    adversary nodes in $\CP_1$ delay the sending to and receiving from nodes in $\CP_2$ until 
    $T_2$.
    During 
    $\CT_2$,
    adversary nodes in $\CP_1$ also delay the sending to and receiving from nodes in $\CP_3$ until 
    $T_2$.
    During 
    $\CT_1$,
    adversary nodes in $\CP_1$ communicate with nodes in $\CP_3$ using next-round-delay (and the network asynchrony allows this).
    During network asynchrony, communication between any node $p\in \CP_2$ and any node outside $\CP_2$ is delayed until 
    $T_1$.
    After 
    $T_2$,
    the adversary nodes behave honestly.
    
    Note that $E_A, E_B, E_C$ are indistinguishable to honest nodes 
    because they receive the same messages at the same rounds
    during each of the executions.
    Recall that we have assumed a timely-liveness
    violation in $E_A$, where for every $p\in\CP_1$,
    $\tx\not\in\LOG_{\netDeltaPeriod \netG(\netDeltaPeriod)-1}^p$, \ie,
    $p$ has not confirmed $\tx$ by the end of $\CT$ in $E_A$.
    So there must be a timely-liveness
    violation in $E_B$,
    and eventually a message is produced in $E_B$ that identifies an adversary node $p' \in \CP_2$ as guilty for the timely-liveness
    violation,
    because $\Pi$ is assumed to be accountably live with $\tALident > 0$.
    Then, due to $E_B$ and $E_C$ being indistinguishable 
    in terms of which and when messages are produced,
    eventually a message is produced in $E_C$ that identifies $p' \in \CP_2$ as guilty for a timely-liveness
    violation.
    (Note that there was no timely-liveness
    violation in $E_C$ because $\CP_1$ are adversary in $E_C$. Regardless, due to the indistinguishability of $E_B$ and $E_C$, honest nodes go along with the adversary nodes to eventually produce a certificate of guilt for $p' \in \CP_2$ in $E_C$.)
    But that is a contradiction because $\CP_2$ are honest in $E_C$.

    We now proceed to~(2), showing that $E_A$ must indeed have a timely-liveness
    violation (namely where none of $\CP_1$ confirms) if $\Pi$ is optimally-resilient.
    We consider five executions $E_1$, $E_2$, $E_3$, $E_4$, $E_5$, where $E_A = E_1$,
    see \cref{fig:impossibility-x-geq-12}.

    \underline{Execution~$E_1$ (\cf \cref{fig:impossibility-x-geq-12}, ``Execution $E_A = E_1$''):}
    All nodes are input $\tx$ 
    at time $0$. No other transactions are input.
    All nodes are honest.
    The network is asynchronous during 
    $\CT$,
    and next-round-delay afterwards for all nodes.
    During asynchrony, communication between any nodes $p\in \CP_1, q\in \CP_2$ is delayed until 
    $T_2$.
    During 
    $\CT_1$,
    communication in $\CP_1 \cup \CP_3$ is next-round-delay. All other communication is delayed until 
    $T_1$.
    During 
    $\CT_2$,
    communication in $\CP_2 \cup \CP_3$ is next-round-delay. All other communication is delayed until 
    $T_2$.

    \underline{Execution~$E_2$ (\cf \cref{fig:impossibility-x-geq-12}):}
    All nodes are input $\tx'$ 
    at time $0$. No other transactions are input.
    All nodes are honest.
    The network is asynchronous during 
    $\CT$,
    and next-round-delay afterwards for all nodes.
    During asynchrony, communication between any nodes $p\in \CP_1, q\in \CP_2$ is delayed until 
    $T_2$.
    During 
    $\CT_1$,
    communication in $\CP_2 \cup \CP_3$ is next-round-delay. All other communication is delayed until 
    $T_1$.
    During 
    $\CT_2$,
    communication in $\CP_1 \cup \CP_3$ is next-round-delay. All other communication is delayed until 
    $T_2$.

    \underline{Execution~$E_3$ (\cf \cref{fig:impossibility-x-geq-12}):}
    Same as $E_1$, except nodes $\CP_2$ are input $\tx'$ 
    at time $0$.

    \underline{Execution~$E_4$ (\cf \cref{fig:impossibility-x-geq-12}):}
    Same as $E_2$, except nodes $\CP_1$ are input $\tx$ 
    at time $0$.

    \underline{Execution~$E_5$ (\cf \cref{fig:impossibility-x-geq-12}):}
    Nodes $\CP_1$ are input $\tx$, nodes $\CP_2$ are input $\tx'$, and nodes $\CP_3$ are input $\tx, \tx'$ 
    at time $0$.
    Nodes $\CP_1 \cup \CP_2$ are honest.
    Nodes $\CP_3$ are adversary.
    The network is asynchronous during 
    $\CT$,
    and next-round-delay afterwards for all nodes.
    During asynchrony, communication between any nodes $p\in \CP_1, q\in \CP_2$ is delayed until 
    $T_2$.
    During 
    $\CT_1$,
    communication in $\CP_2 \cup \CP_3$ is next-round-delay and communication in $\CP_1 \cup \CP_3$ is next-round-delay.
    All other communication is delayed until 
    $T_2$.
    Nodes $\CP_3$ perform a split-brain attack, behaving to $\CP_1$ like in $E_3$ and to $\CP_2$ like in $E_4$.

    Towards a contradiction, assume that there is no timely-liveness
    violation in $E_1$ and in $E_2$
    where none of $\CP_1$ and none of $\CP_2$ confirm $\tx$ and $\tx'$, respectively,
    \ie, by time $\netDeltaPeriod \netG(\netDeltaPeriod)$,
    some honest node $p_1 \in \CP_1$
    has 
    confirmed $\tx$ 
    and not confirmed $\tx'$ in $E_1$ (because they had no way of knowing $\tx'$ which is assumed to be high-entropy),
    and some honest node $p_2 \in \CP_2$
    has 
    confirmed $\tx'$ 
    and not confirmed $\tx$ in $E_2$ (because they had no way of knowing $\tx$ which is assumed to be high-entropy).
    (If either $E_1$ or $E_2$ has a timely-liveness
    violation where either none of $\CP_1$ or none of $\CP_2$ confirm, we can take that as $E_A$ and adjust~(1) accordingly and we are done---which is hinted in \cref{fig:impossibility-x-geq-12} with executions $E_{B'}, E_{C'}$.)
    Note that from the perspective of the nodes $\CP_1$,
    $E_1$ and $E_3$ are indistinguishable until $T_2$,
    thus 
    $p_1 \in \CP_1$
    confirms $\tx$ in $E_3$ and does not confirm $\tx'$ 
    in $E_3$ 
    by time $\netDeltaPeriod \netG(\netDeltaPeriod)$.
    Note that from the perspective of the nodes $\CP_2$,
    $E_2$ and $E_4$ are indistinguishable until $T_2$,
    thus 
    $p_2 \in \CP_2$
    confirms $\tx'$ in $E_4$ and does not confirm $\tx$ 
    in $E_4$ by time $\netDeltaPeriod \netG(\netDeltaPeriod)$.
    Note also that from the perspective of the nodes $\CP_1$,
    $E_3$ and $E_5$ are indistinguishable until $T_2$,
    and from the perspective of the nodes $\CP_2$,
    $E_4$ and $E_5$ are indistinguishable until $T_2$.
    Thus, 
    $p_1 \in \CP_1$
    confirms $\tx$ in $E_5$ and does not confirm $\tx'$
    in $E_5$ 
    by time $\netDeltaPeriod \netG(\netDeltaPeriod)$.
    Likewise, 
    $p_2 \in \CP_2$
    confirms $\tx'$ in $E_5$ and does not confirm $\tx$ 
    in $E_5$ 
    by time $\netDeltaPeriod \netG(\netDeltaPeriod)$.
    Since both $p_1$ and $p_2$ are honest in $E_5$,
    this constitutes a safety violation, even though $|\CP_3| \leq \tS$ in $E_5$, giving the required
    contradiction.
    Thus, there must be a timely-liveness
    violation in either $E_1$ or $E_2$
    where either none of $\CP_1$ or none of $\CP_2$ confirm,
    and we can use that as $E_A$ in~(1).
\end{proof}

\subsection{Proof of \cref{thm:impossibility-tALident1-ours}}
\label{sec:proof-impossibility-tALident1-ours}

\ifcompilewithoutheavyfigures\else
    \begin{figure*}[tbp]%
    \centering%
    \begin{tikzpicture}[%
            nodesHon/.style={thick,Green},
            nodesAdv/.style={thick,Red},
            viewsHon/.style={thick,Green},
            viewsAdv/.style={thick,Red},
            commsBdY/.style={{Straight Barb[angle=60:1.5pt 3]}-{Straight Barb[angle=60:1.5pt 3]},Green},
            commsBdN/.style={{Straight Barb[angle=60:1.5pt 3]}-{Straight Barb[angle=60:1.5pt 3]},Red},
            commsUdY/.style={-{Straight Barb[angle=60:1.5pt 3]},Green},
            commsUdN/.style={-{Straight Barb[angle=60:1.5pt 3]},Red},
            commsP/.style={Red},
            nodeC/.style={Red},
        ]
        \footnotesize

        \begin{scope}[x=3cm,y=2cm]
            \myDrawScenario{%
                P/0.1667/{$\CP_0$}/{nodesAdv},%
                Q1/0.1667/{$\CP_1$}/{nodesHon},%
                Q2/0.1666/{$\CP_2$}/{nodesAdv},%
                Q3/0.1666/{$\CP_3$}/{nodesAdv},%
                R/0.3334/{$\CP_4$}/{nodesHon}%
            }{%
                V1/0.3334/{$\CT_1$}/{viewsHon},%
                V2/0.3333/{$\CT_2$}/{viewsHon},%
                V3/0.3333/{$\CT_3$}/{viewsAdv}%
            };

            \node [nodeC] at (PV1mid) {\myIconAdvCrash};
            \node [nodeC] at (PV2mid) {\myIconAdvCrash};
            \node [nodeC] at (PV3mid) {\myIconAdvCrash};
            \node [commsP] at (Q3V1mid) {\myIconAdvPartition};
            \node [commsP] at (Q2V2mid) {\myIconAdvPartition};
            \node [commsP] at (Q1V3mid) {\myIconHonPartition};

            \node [anchor=south] (E31) at ([yshift=1em]0.5,1) {Execution $E_{3,1}$};
        \end{scope}

        \begin{scope}[x=3cm,y=2cm,xshift=4cm]
            \myDrawScenario{%
                P/0.1667/{$\CP_0$}/{nodesAdv},%
                Q1/0.1667/{$\CP_1$}/{nodesAdv},%
                Q2/0.1666/{$\CP_2$}/{nodesHon},%
                Q3/0.1666/{$\CP_3$}/{nodesAdv},%
                R/0.3334/{$\CP_4$}/{nodesHon}%
            }{%
                V1/0.3334/{$\CT_1$}/{viewsHon},%
                V2/0.3333/{$\CT_2$}/{viewsAdv},%
                V3/0.3333/{$\CT_3$}/{viewsHon}%
            };

            \node [nodeC] at (PV1mid) {\myIconAdvCrash};
            \node [nodeC] at (PV2mid) {\myIconAdvCrash};
            \node [nodeC] at (PV3mid) {\myIconAdvCrash};
            \node [commsP] at (Q3V1mid) {\myIconAdvPartition};
            \node [commsP] at (Q2V2mid) {\myIconHonPartition};
            \node [commsP] at (Q1V3mid) {\myIconAdvPartition};

            \node [anchor=south] (E32) at ([yshift=1em]0.5,1) {Execution $E_{3,2}$};
        \end{scope}

        \begin{scope}[x=3cm,y=2cm,xshift=8cm]
            \myDrawScenario{%
                P/0.1667/{$\CP_0$}/{nodesAdv},%
                Q1/0.1667/{$\CP_1$}/{nodesAdv},%
                Q2/0.1666/{$\CP_2$}/{nodesAdv},%
                Q3/0.1666/{$\CP_3$}/{nodesHon},%
                R/0.3334/{$\CP_4$}/{nodesHon}%
            }{%
                V1/0.3334/{$\CT_1$}/{viewsAdv},%
                V2/0.3333/{$\CT_2$}/{viewsHon},%
                V3/0.3333/{$\CT_3$}/{viewsHon}%
            };

            \node [nodeC] at (PV1mid) {\myIconAdvCrash};
            \node [nodeC] at (PV2mid) {\myIconAdvCrash};
            \node [nodeC] at (PV3mid) {\myIconAdvCrash};
            \node [commsP] at (Q3V1mid) {\myIconHonPartition};
            \node [commsP] at (Q2V2mid) {\myIconAdvPartition};
            \node [commsP] at (Q1V3mid) {\myIconAdvPartition};

            \node [anchor=south] (E33) at ([yshift=1em]0.5,1) {Execution $E_{3,3}$};

            \draw [{Straight Barb[angle=60:1.5pt 3]}-{Straight Barb[angle=60:1.5pt 3]}] ([xshift=0.5em]Ptoprit) -- ([xshift=0.5em]Q2botrit) node [pos=0.95,right] {$\leq\tALmax$};
            \draw [{Straight Barb[angle=60:1.5pt 3]}-{Straight Barb[angle=60:1.5pt 3]}] ([xshift=1em]Ptoprit) -- ([xshift=1em]Q1botrit) node [pos=0.8,right] {$>\tL$};
            \draw [{Straight Barb[angle=60:1.5pt 3]}-{Straight Barb[angle=60:1.5pt 3]}] ([xshift=1.5em]Ptoprit) -- ([xshift=1.5em]Pbotrit) node [pos=0.4,right] {$<\tALident$};
        \end{scope}

        \node [anchor=south] at ($(E31.south)!0.5!(E32.south)$) {$\approx$};
        \node [anchor=south] at ($(E32.south)!0.5!(E33.south)$) {$\approx$};

        \begin{scope}[x=3cm,y=3cm,yshift=-5cm]
            \myDrawScenario{%
                P/0.2500/{$\CP_0$}/{nodesAdv},%
                Q1/0.0833/{$\CP_1$}/{nodesHon},%
                Q2/0.0833/{$\CP_2$}/{nodesAdv},%
                Q3/0.0834/{$\CP_3$}/{nodesAdv},%
                Q4/0.0833/{$\CP_4$}/{nodesAdv},%
                R/0.4167/{$\CP_5$}/{nodesHon}%
            }{%
                V1/0.25/{$\CT_1$}/{viewsHon},%
                V2/0.25/{$\CT_2$}/{viewsHon},%
                V3/0.25/{$\CT_3$}/{viewsHon},%
                V4/0.25/{$\CT_4$}/{viewsAdv}%
            };

            \node [nodeC] at (PV1mid) {\scriptsize\myIconAdvCrash};
            \node [nodeC] at (PV2mid) {\scriptsize\myIconAdvCrash};
            \node [nodeC] at (PV3mid) {\scriptsize\myIconAdvCrash};
            \node [nodeC] at (PV4mid) {\scriptsize\myIconAdvCrash};
            \node [commsP] at (Q4V1mid) {\scriptsize\myIconAdvPartition};
            \node [commsP] at (Q3V2mid) {\scriptsize\myIconAdvPartition};
            \node [commsP] at (Q2V3mid) {\scriptsize\myIconAdvPartition};
            \node [commsP] at (Q1V4mid) {\scriptsize\myIconHonPartition};

            \node [anchor=south] (E41) at ([yshift=1em]0.5,1) {Execution $E_{4,1}$};
        \end{scope}

        \begin{scope}[x=3cm,y=3cm,yshift=-5cm,xshift=4cm]
            \myDrawScenario{%
                P/0.2500/{$\CP_0$}/{nodesAdv},%
                Q1/0.0833/{$\CP_1$}/{nodesAdv},%
                Q2/0.0833/{$\CP_2$}/{nodesHon},%
                Q3/0.0834/{$\CP_3$}/{nodesAdv},%
                Q4/0.0833/{$\CP_4$}/{nodesAdv},%
                R/0.4167/{$\CP_5$}/{nodesHon}%
            }{%
                V1/0.25/{$\CT_1$}/{viewsHon},%
                V2/0.25/{$\CT_2$}/{viewsHon},%
                V3/0.25/{$\CT_3$}/{viewsAdv},%
                V4/0.25/{$\CT_4$}/{viewsHon}%
            };

            \node [nodeC] at (PV1mid) {\scriptsize\myIconAdvCrash};
            \node [nodeC] at (PV2mid) {\scriptsize\myIconAdvCrash};
            \node [nodeC] at (PV3mid) {\scriptsize\myIconAdvCrash};
            \node [nodeC] at (PV4mid) {\scriptsize\myIconAdvCrash};
            \node [commsP] at (Q4V1mid) {\scriptsize\myIconAdvPartition};
            \node [commsP] at (Q3V2mid) {\scriptsize\myIconAdvPartition};
            \node [commsP] at (Q2V3mid) {\scriptsize\myIconHonPartition};
            \node [commsP] at (Q1V4mid) {\scriptsize\myIconAdvPartition};

            \node [anchor=south] (E42) at ([yshift=1em]0.5,1) {Execution $E_{4,2}$};
        \end{scope}

        \begin{scope}[x=3cm,y=3cm,yshift=-5cm,xshift=8cm]
            \myDrawScenario{%
                P/0.2500/{$\CP_0$}/{nodesAdv},%
                Q1/0.0833/{$\CP_1$}/{nodesAdv},%
                Q2/0.0833/{$\CP_2$}/{nodesAdv},%
                Q3/0.0834/{$\CP_3$}/{nodesHon},%
                Q4/0.0833/{$\CP_4$}/{nodesAdv},%
                R/0.4167/{$\CP_5$}/{nodesHon}%
            }{%
                V1/0.25/{$\CT_1$}/{viewsHon},%
                V2/0.25/{$\CT_2$}/{viewsAdv},%
                V3/0.25/{$\CT_3$}/{viewsHon},%
                V4/0.25/{$\CT_4$}/{viewsHon}%
            };

            \node [nodeC] at (PV1mid) {\scriptsize\myIconAdvCrash};
            \node [nodeC] at (PV2mid) {\scriptsize\myIconAdvCrash};
            \node [nodeC] at (PV3mid) {\scriptsize\myIconAdvCrash};
            \node [nodeC] at (PV4mid) {\scriptsize\myIconAdvCrash};
            \node [commsP] at (Q4V1mid) {\scriptsize\myIconAdvPartition};
            \node [commsP] at (Q3V2mid) {\scriptsize\myIconHonPartition};
            \node [commsP] at (Q2V3mid) {\scriptsize\myIconAdvPartition};
            \node [commsP] at (Q1V4mid) {\scriptsize\myIconAdvPartition};

            \node [anchor=south] (E43) at ([yshift=1em]0.5,1) {Execution $E_{4,3}$};
        \end{scope}

        \begin{scope}[x=3cm,y=3cm,yshift=-5cm,xshift=12cm]
            \myDrawScenario{%
                P/0.2500/{$\CP_0$}/{nodesAdv},%
                Q1/0.0833/{$\CP_1$}/{nodesAdv},%
                Q2/0.0833/{$\CP_2$}/{nodesAdv},%
                Q3/0.0834/{$\CP_3$}/{nodesAdv},%
                Q4/0.0833/{$\CP_4$}/{nodesHon},%
                R/0.4167/{$\CP_5$}/{nodesHon}%
            }{%
                V1/0.25/{$\CT_1$}/{viewsAdv},%
                V2/0.25/{$\CT_2$}/{viewsHon},%
                V3/0.25/{$\CT_3$}/{viewsHon},%
                V4/0.25/{$\CT_4$}/{viewsHon}%
            };

            \node [nodeC] at (PV1mid) {\scriptsize\myIconAdvCrash};
            \node [nodeC] at (PV2mid) {\scriptsize\myIconAdvCrash};
            \node [nodeC] at (PV3mid) {\scriptsize\myIconAdvCrash};
            \node [nodeC] at (PV4mid) {\scriptsize\myIconAdvCrash};
            \node [commsP] at (Q4V1mid) {\scriptsize\myIconHonPartition};
            \node [commsP] at (Q3V2mid) {\scriptsize\myIconAdvPartition};
            \node [commsP] at (Q2V3mid) {\scriptsize\myIconAdvPartition};
            \node [commsP] at (Q1V4mid) {\scriptsize\myIconAdvPartition};

            \node [anchor=south] (E44) at ([yshift=1em]0.5,1) {Execution $E_{4,4}$};

            \draw [{Straight Barb[angle=60:1.5pt 3]}-{Straight Barb[angle=60:1.5pt 3]}] ([xshift=0.5em]Ptoprit) -- ([xshift=0.5em]Q3botrit) node [pos=0.9,right] {$\leq\tALmax$};
            \draw [{Straight Barb[angle=60:1.5pt 3]}-{Straight Barb[angle=60:1.5pt 3]}] ([xshift=1em]Ptoprit) -- ([xshift=1em]Q1botrit) node [pos=0.85,right] {$>\tL$};
            \draw [{Straight Barb[angle=60:1.5pt 3]}-{Straight Barb[angle=60:1.5pt 3]}] ([xshift=1.5em]Ptoprit) -- ([xshift=1.5em]Pbotrit) node [pos=0.4,right] {$<\tALident$};
        \end{scope}

        \node [anchor=south] at ($(E41.south)!0.5!(E42.south)$) {$\approx$};
        \node [anchor=south] at ($(E42.south)!0.5!(E43.south)$) {$\approx$};
        \node [anchor=south] at ($(E43.south)!0.5!(E44.south)$) {$\approx$};
        
    \end{tikzpicture}%
    \caption[]{%
        Illustration of indistinguishable executions used in \cref{thm:impossibility-tALident1-ours}, for $k=3$ and $k=4$.
        See caption of \cref{fig:impossibility-x-geq-12} for legend.
        Red ``\myIconAdvCrash{}'' indicates adversary-emulated crashed nodes.%
    }%
    \label{fig:tauALident-converses1}%
\end{figure*}%
\fi

\begin{proof}[Proof of \cref{thm:impossibility-tALident1-ours}]
    Towards a contradiction,
    suppose for some $k\geq3$,
    \Cref{alg:tendermint-2}
    with some liveness accountability mechanism
    satisfies the conditions of \cref{thm:impossibility-tALident1-ours}
    for $\netX = 1/k$ (which is without loss of generality)
    with
    $(\tALident - 1) \geq (\tL + 1) - \left\lfloor \frac{\tALmax - (\tL + 1)}{k-2} \right\rfloor$.
    Then $\CP$ can be partitioned into $\CP_0, \CP_1, ..., \CP_k, \CP_{k+1}$,
    such that
    $|\CP_0| = \tALident - 1$,
    $|\CP_i| = \left\lfloor \frac{\tALmax - (\tL + 1)}{k-2} \right\rfloor$ for $i\in\{1,...,k\}$,
    $|\CP_{k+1}| = n - \sum_{i=0}^{k}|\CP_i|$.
    Note that for every $i\in\{1,...,k\}$,
    $|\CP_0| + |\CP_i| \geq \tL + 1 > \tL$,
    and
    $|\CP_0| + \sum_{i=0}^{k-1}|\CP_i| \leq \tALmax$,
    $|\CP_0| < \tALident$.

    We require the mild regularity condition that $k$ divides $\netDeltaPeriod \netG(\netDeltaPeriod)$.
    Partition the first $\netDeltaPeriod \netG(\netDeltaPeriod)$ rounds
    into $k$ equally-sized intervals $\CT_i$ for $i\in\{1,...,k\}$:
    Let $T_i \triangleq i \netDeltaPeriod \netG(\netDeltaPeriod)/k$ for $i\in\{0,...,k\}$,
    $\CT_i \triangleq [T_{i-1},T_i)$ for $i\in\{1,...,k\}$,
    and $\CT \triangleq \bigcup_{i=1}^k \CT_i$.

    Now consider the executions $E_{k,i}$ for $i\in\{1,...,k\}$
    (see \cref{fig:tauALident-converses1} for an illustration),
    where in the $i$-th execution the following holds:
    $\CP_0$ are crashed,
    $\CP_{k+1}$ are honest,
    $\CP_i$ are honest,
    and all other nodes are adversary.
    All nodes are input a high-entropy transaction $\tx$ at round $0$.
    Furthermore,
    $\CT_{k-i+1}$ is asynchronous,
    the remaining rounds are next-round-delay.
    During asynchrony, messages to, from, and within $\CP_i$ are delayed until the end of asynchrony;
    all other messages have next-round-delay.
    For every $j\in\{1,...,k\}\setminus\{i\}$,
    the non-crashed adversary nodes in $\CP_j$
    behave during $\CT_{k-j+1}$ as if messages to, from, and within $\CP_j$ are delayed until the end of $\CT_{k-j+1}$, \ie, they delay processing incoming messages and sending messages until the end of $\CT_{k-j+1}$.
    Other than the extra delay, adversary nodes follow the protocol.
    Observe that all executions $E_{k,i}$ for $i\in\{1,...,k\}$ are indistinguishable from the perspective of honest nodes in terms of which messages they receive and when.
    
    Observe that in every execution $E_{k,i}$,
    according to \cref{alg:tendermint-2}, $\tx$ is not confirmed by any of the honest nodes
    by $T_k$, \ie, by the beginning of round $\netDeltaPeriod \netG(\netDeltaPeriod)$.
    This is because for every view $v$ before $\netDeltaPeriod \netG(\netDeltaPeriod)$,
    the following holds (let $t_v \triangleq 12 \Delta v$):
    The block $b_v$ proposed by $\Leader_v$ at $t_v + 2 \Delta$
    (\alglocref{alg:tendermint-2}{loc:tendermint-2-Lvproduceblock})
    reaches, by $t_v + 4 \Delta$, 
    only $\Leader_v$ (if for some $j\in\{1,...,k\}$, $t_v \in \CT_{k-j+1}$ and $\Leader_v \in \CP_0 \cup \CP_j$)
    or less than $n - \tL$ nodes (otherwise).
    In any case,
    $b_v$ reaches at most $2n/3$ nodes (since $\tL = \lfloor (n-1)/3 \rfloor$ for \cref{alg:tendermint-2}, \cf \cref{thm:tendermint-2-liveness}) by $t_v + 4 \Delta$.
    But no node votes stage-$1$ for $b_v$ after $t_v + 4 \Delta$
    (\alglocref{alg:tendermint-2}{loc:tendermint-2-stage1vote}).
    As a result, 
    at most $2n/3$ stage-$1$ votes are ever produced for $b_v$ 
    according to \cref{alg:tendermint-2},
    and thus
    no stage-$1$ QC is ever produced for $b_v$,
    and thus $b_v$ is never confirmed
    (\alglocref{alg:tendermint-2}{loc:tendermint-2-confirm}).
    This implies a timely-liveness
    violation.

    Recall that \cref{alg:tendermint-2} was assumed
    to be equipped with a liveness accountability mechanism
    that would render it
    accountably live with sensitivity $\tALident$,
    so in all of the executions $E_{k,i}$ 
    (for our purposes here it suffices that this is the case in one of the executions), eventually, a proof of guilt is produced for $\tALident$ adversary nodes. 
    Since $|\CP_0| = \tALident - 1 < \tALident$, 
    it follows that there must be
    a proof of guilt for some node not in $\CP_0$.
    But then by indistinguishability of the executions,
    there is an execution among the $E_{k,i}$ where that node is honest,
    yet a certificate of guilt is produced for that node.
    This is the desired contradiction.
\end{proof}

\subsection{Proof of \cref{thm:impossibility-tALident1-pbft}}
\label{sec:proof-impossibility-tALident1-pbft}

\begin{proof}[Proof of \cref{thm:impossibility-tALident1-pbft}]
    We follow the steps of the proof of \cref{thm:impossibility-tALident1-ours}.
    Only the network delay (both real and adversary-emulated) in the executions $E_{k,i}$
    is slightly adjusted to leverage the now-or-never property.

    Partition the first $\netDeltaPeriod \netG(\netDeltaPeriod)$ rounds
    into equally-sized intervals $\CT^{(v)}$ of length $\Delta''$.
    This is possible because $\Delta''$ divides $\netDeltaPeriod \netG(\netDeltaPeriod) / k$ by assumption.
    Consider the executions $E_{k,i}$ of the proof of \cref{thm:impossibility-tALident1-ours}.
    The adversary nodes in $\CP_0$ no longer crash.
    During asynchrony, within each $\CT^{(v)}$,
    messages to, from, and within $\CP_i \cup \CP_0$ are delayed until the end of $\CT^{(v)}$.
    For every $j\in\{1,...,k\}\setminus\{i\}$,
    the non-crashed adversary nodes in $\CP_j \cup \CP_0$
    behave during $\CT_{k-j+1}$, within each $\CT^{(v)}$, as if messages to, from, and within $\CP_j \cup \CP_0$ are delayed until the end of $\CT^{(v)}$.
    Observe that all executions $E_{k,i}$ for $i\in\{1,...,k\}$ are still indistinguishable from the perspective of honest nodes in terms of which messages they receive and when.
    
    In every execution $E_{k,i}$ and every $\CT^{(v)}$,
    $\lceil n/3 \rceil$ nodes are partitioned off temporarily until the end of $\CT^{(v)}$.
    Thus, by the now-or-never property,
    $\tx$ is not confirmed by any of the honest nodes
    by $T_k$, \ie, by the beginning of round $\netDeltaPeriod \netG(\netDeltaPeriod)$.
    This implies a timely-liveness
    violation. 
    The rest of the proof proceeds as in the proof of \cref{thm:impossibility-tALident1-ours}.
\end{proof}

\deferredsection{proofsblaming}{Addendum Blame Accounting}
\deferredsection{proofsadjudication}{Addendum Adjudication Rule}
\fi

\end{document}